\documentclass[12pt,a4paper]{elsarticle}
\usepackage[top=25mm,bottom=37mm,left=2.5cm,right=2.5cm,asymmetric]{geometry}
\usepackage{xcolor}
\newcommand{\eref}[1]{(\ref{#1})}

\newcommand{\vecsym}[1]{\boldsymbol{#1}}

\newcommand{\ie}{\textit{i.e.}}
\newcommand{\eg}{\emph{e.g.\ }}

\usepackage{amssymb}
\usepackage{amsmath}
\usepackage{bbm}
\usepackage{bm}
\usepackage{dsfont}
\usepackage[nodots]{numcompress}
\usepackage{url}

\usepackage{natbib}
\biboptions{authoryear}

\usepackage{units}
\usepackage{graphicx}
\usepackage[caption = false]{subfig}
\usepackage{caption}
\usepackage{float}
\usepackage{color}
\usepackage[titletoc,toc,title]{appendix}

\journal{International Journal}
\makeatletter
\def\ps@pprintTitle{%
\let\@oddhead\@empty
\let\@evenhead\@empty
\let\@oddfoot\@empty
\let\@evenfoot\@oddfoot
}
\makeatother

\begin{document}

\begin{frontmatter}
	
	\title{Analytic prediction of yield stress and strain hardening in a strain gradient plasticity material reinforced by small elastic particles}
	
	\author{Philip Cron\'{e}}
	\author{Peter Gudmundson}
	\author{Jonas Faleskog \corref{CORR}}
	
	\address{Solid Mechanics, Department of Engineering Mechanics, KTH Royal Institute of Technology, 10044 Stockholm, Sweden}
	
	\cortext[CORR]{Corresponding author. E-mail: faleskog@kth.se}
	
	\begin{abstract}
		The influence on macroscopic work hardening of small, spherical, elastic particles dispersed within a matrix is studied using an isotropic strain gradient plasticity framework. An analytical solution, based on a recently developed yield strength model is proposed. The model accounts for random variations in particle size and elastic properties, and is numerically validated against FE solutions in 2D/3D material cell models. Excellent agreement is found as long as the typical particle radius is much smaller than the material length scale, given that the particle volume fraction is not too large ($<10\%$) and that the particle/matrix elastic mismatch is within a realistic range. Finally, the model is augmented to account for strengthening contribution from shearable particles using classic line tension models and successfully calibrated against experimental tensile data on an $Al-2.8wt\%Mg-0.16wt\%Sc$ alloy.
	\end{abstract}
	
	
\end{frontmatter}

\section{Introduction}
The effects of precipitated particles distributed within a bulk material is a well studied phenomenon. Pioneering work in the development of its theory are due to \cite{orowan1948discussion}, \cite{mott1940attempt}, who explained the increase in yield strength using the concept of dislocations interacting with the particles. An excellent review on the subject of precipitation strengthening has been given by \cite{ardell1985precipitation}. In essence, there are two types of interactions between a dislocation and a particle: Either, the particle is small enough that the dislocation will cut right through it, or the dislocation will bypass the particle without plastically deforming it.

To model the strengthening and hardening of complex metallic material systems in general, a wide range of mechanisms have to be taken into account. These mechanisms contribute to the material flow stress by impeding dislocation motion and are typically due to grain boundaries, secondary phases, forest dislocations, solute atoms, and intrinsic dislocation glide resistance. Depending on the specific material system and processing, some of the above mechanisms will most likely dominate over the others which simplifies the modelling. How to superimpose contributions of relevance to flow stress is not trivial \citep{queyreau2010orowan,de2013superposition}. During the last 35 years, micro mechanical modelling of hardening based on dislocation-obstacle interaction has heavily relied on the work of Kocks, Mecks and Estrin (KME) \citep{mecking1981kinetics,estrin1984unified,kocks2003physics}, who propose a model for the evolution of dislocation density as a function of plastic strain. In essence, their model postulates that the accumulation rate of dislocation density is governed by the difference in rate of storage and annihilation. Given its versatile nature, this model can be modified to include a wide range of micro structure dependent mechanisms relevant to hardening. Once the governing equation for the dislocation density is established, its output typically enters the classic Taylor model of dislocation hardening. Some notable work on modelling the strengthening and hardening in metal systems containing precipitates include \cite{russell1972dispersion,fazeli2008modeling,myhr2010combined,cheng2003influence,deschamps1998influence,fribourg2011microstructure}.

Another method of modelling the work hardening is to directly model the movement of each individual dislocation within a large group of dislocations under an applied macroscopic stress. Contributions using this framework, which is commonly referred to as Discrete Dislocation Dynamics (DDD), include \cite{arsenlis2007enabling,koslowski2002phase,santos2020multiscale,vattre2009dislocation,zbib20003d}.

Because the apparent size effect in particle strengthening, cf.  \citet{kouzeli2002size}, conventional continuum plasticity theories fall short in their predictive power due to their lack of an intrinsic length scale. However, strain gradient enriched continuum plasticity models offer a mean to resolve these small scale size effects by recognizing the close connection between the density of geometrically necessary dislocations (GNDs) and gradients in plastic strain. By adding the contributions from plastic strain gradients to the internal work and to the strain based measure of total dislocation density, a higher order theory framework with an intrinsic length scale can be constructed from conventional local theories of continuum plasticity, see \eg \cite{fleck1993phenomenological,fleck1994strain,hutchinson1997strain,fleck2001reformulation,gudmundson2004unified,gurtin2004gradient,gurtin2005theory,fleck2009mathematical,fleck2009mathematical2}. For a thorough review on the subject of Strain Gradient Plasticity (SGP) theories, the reader is referred to the recent article by \cite{voyiadjis2019strain}. Further, by also adding the work performed at an internal interface to the internal work, interface phenomena such as dislocation pile-ups and other grain boundary type features can be modelled. Relating dislocation movement and production to plastic strain, the higher order theories allows for restrictions, including a complete blockage, to the flow of dislocations through an internal boundary. Following the second law of thermodynamics, constitutive laws for the forces due to dislocation pile ups at internal interfaces, \ie, interface models, can be established by relating plastic strain to the moment stresses present in higher order SGP theories. These constitutive laws may be purely dissipative, purely energetic or a mix of both. The latter can be motivated through the fact that dislocation networks does store some elastic energy as they distort the lattice. Several thermodynamically consistent interface models within the framework of higher order isotropic SGP theories have been proposed over the years, \eg \cite{fleck2009mathematical,fredriksson2005size,Asgharzadeh2021a,dahlberg2013improved,voyiadjis2009formulation}.

Early attempts at studying particulate reinforced metals using conventional $J_2$ plasticity continuum mechanics include \cite{christman1989experimental,bao1991particle}. The authors in the latter reference found a simple relation for the elevation in yield stress for a metal reinforced by impenetrable rigid particles given by $\sigma_{\text{p}}=\beta\sigma_{0}f$ where $\sigma_{0}$ is the yield stress in the matrix material, $f$ is the volume fraction of particles and $\beta$ is a proportionality parameter dependent on particle shape.

Investigations using isotropic SGP theories to study above mentioned metals include \cite{chen2002size,liu2003strain,zhang2007three,yueguang2001particulate,qu2005study,dai1999strain,chen2002size,zhu1997strain,xue2002particle}. In a recent work by \cite{Asgharzadeh2021a}, numerical simulations were performed on a metal matrix reinforced with impenetrable particles using the higher order SGP theory proposed by \cite{gudmundson2004unified}. A contribution to the composite yield stress proportional to the total surface area of the particles was identified and a model was proposed. Good agreement between model and experimental data was achieved. An upper bound solution to predict the yield stress based on Eshelbys tensors for the inhomogeneous inclusion problem was later provided by \cite{faleskog2021analytical} on the following form:
\begin{equation}
\sigma_{\rm y}^{\rm UBS} =  \dfrac{\sigma_0}{1-\Gamma f} \biggl(1 - f + 3 f \alpha \dfrac{\ell}{a}\biggl),
\end{equation}
where $\ell$ denotes the material length scale of the matrix, $a$ the particle radius, and $\alpha$ a parameter related to the particle/matrix interface. This upper bound estimate is in close agreement with corresponding numerical full field solutions as long as the volume fraction of particles is kept in the lower range of roughly $<10\%$.

The objective of the current work is to extend the theory proposed by \cite{faleskog2021analytical} to also include work hardening.

The paper is structured in the following manner: In Section 2, the continuum model used for bulk and interface respectively is presented. In Section 3, an upper bound solution to the particle reinforced composite problem is presented. Section 4 contains simulation results validating the upper bound solution for a wide range of parameters. Results for a homogeneous distribution of spherical particles using an axi-symmetric model as well as a non uniform distribution in a 3D model are presented, and model limitations at high strain levels are discussed. The model is augmented to account for particle shearing in Section 5, and Section 6 contains comparisons between computations and experimental data. Finally, the article is summarized in Section 7.

\section{Problem description}
\subsection{Model}
\begin{figure}[!htb]
	\begin{center}
		\includegraphics[width=0.7\textwidth]{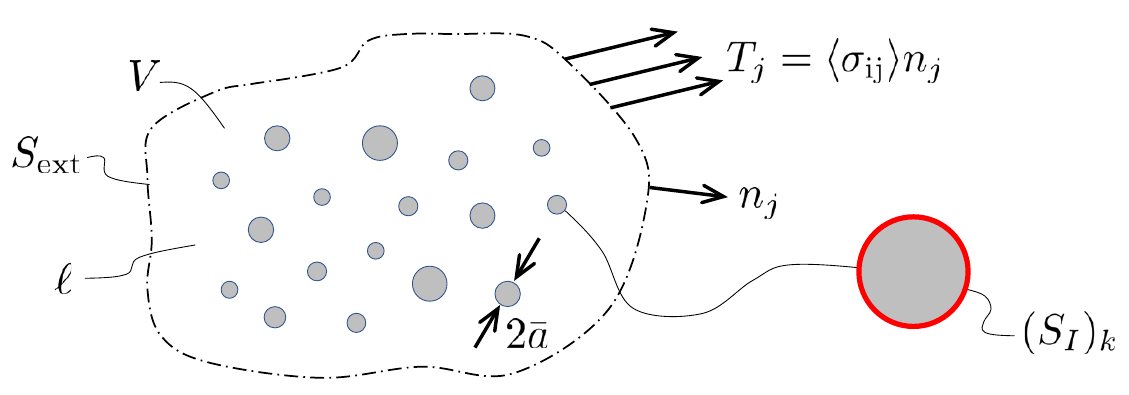}
		\caption{Schematic of a micro-structure in volume $V$ containing a distribution of particles, with the key parameters indicated; effective particle radius $\bar{a}$, and matrix material characteristic length scale $\ell$. The traction acting on the external boundary $S_{\rm ext}$ is given by the volume average of the stresses $\langle \sigma_{ij} \rangle$ in $V$. On internal interfaces, highligthed in color red at particle $k$, a special interface description is used.}
		\label{fig1}
	\end{center}
\end{figure}

The current investigation aims at predicting both yield strength and post yield strain hardening in a particle reinforced metal-matrix composite as measured in a standard tensile test. Essentially, the material dealt with is assumed to consist of small, smooth particles having a distribution of size and spatial arrangement, see Fig. \ref{fig1}. The particles are assumed to be linearly elastic and will thus block plastic slip and act as obstacles to dislocation motion. The pileup of dislocations around the particles will have implications concerning the constitutive modelling of the region in close proximity to a particle. A separate constitutive law for this region will therefore be proposed in terms of a sharp interface between a particle and the matrix. The matrix is assumed to be elastic-plastic and governed by the higher order SGP theory proposed by \cite{gudmundson2004unified}, with a single material length scale $\ell$. 

In the volume considered $V$, particles occupy volume $V_{\rm p}$ and the matrix material $V_{\rm m} = V - V_{\rm p}$. Sub-indices 'm' and 'p' will henceforth be used to distinguish between quantities in the matrix and particles, respectively. It should also be noted that a super-script 'p' will in the following denote plastic quantities. In addition to the volume fraction of particles $f = V_{\rm p}/V$, a representative (effective) value of particle size $\bar{a}$ suffice to characterize the microstructure of the material. The influence of particles on strengthening depends on their size relative to the length parameter $\ell$ of the matrix material and on their volume fraction.

\subsection{Variational formulation of strain gradient plasticity theory}
The primary kinematic variables in the higher order, isotropic, small strain SGP theory by \cite{gudmundson2004unified} are displacements $u_{i}$ and plastic strains $\varepsilon_{ij}^{\rm p}$. These are related as
\begin{equation}\label{eqn:kinematic}
\varepsilon_{ij} = \frac{1}{2}(u_{i,j}+u_{j,i}), \quad \varepsilon_{ij} = \varepsilon_{ij}^{\rm e} + \varepsilon_{ij}^{\rm p},  \quad  \varepsilon_{kk}^{\rm p} = 0,
\end{equation}
where $\varepsilon_{ij}$ are the components of total strain tensor and $\varepsilon_{ij}^{\rm e}$ are the elastic strain components. As in conventional $J_2$-plasticity theory, plastic incompressibility is assumed.

Mechanical equilibrium in absence of body forces in a solid of volume $V$, with external/internal surfaces $S_{\rm ext}$/$S_I$, is expressed through the principle of virtual work as
\begin{equation}\label{eqn:VirtualWork}
	\begin{split}
		\int_{V}\left[ \sigma_{ij}\delta\varepsilon_{ij} + (q_{ij}-s_{ij})\delta\varepsilon_{ij}^{\rm p} + m_{ijk}\delta\varepsilon_{ij,k}^{\rm p} \right]{\rm d}V +  \int_{S_I}\left[M^I_{ij}\delta\varepsilon^{\rm p}_{ij}  \right]{\rm d}S & \\ = \int_{S_{\rm ext}}\left[ T_{i}\delta u_{i} + M_{ij}\delta\varepsilon_{ij}^{\rm p} \right]{\rm d}S,
	\end{split}
\end{equation}
where, $ \sigma_{ij} $ and $ s_{ij} = \sigma_{ij} - \delta_{ij}\sigma_{kk}/3 $ denote the components of the Cauchy stress tensor and its respective deviator while $ \delta_{ij} $ denotes the Kronecker delta symbol. Including plastic strain and its gradients in Eq. \eref{eqn:VirtualWork} brings out their work conjugate stresses in this tensorial higher order theory. These are the micro stresses $q_{ij}$ and the moment stresses  $m_{ijk}$, where the latter give rise to moment tractions  $M_{ij}$ on boundaries. Thus, on the internal and external surfaces, the higher order moment tractions $M^I_{ij}$ and $M_{ij}$ naturally arise, both being work conjugate to $\varepsilon_{ij}^{\rm p}$. By integration by parts of Eq. \eref{eqn:VirtualWork}, the strong form of the equilibrium condition along with the natural boundary conditions on external and internal boundaries are obtained as
\begin{equation}\label{eqn:equilibrium}
\begin{array}{lcll}
\sigma_{ij,j} = 0        & {\rm and} & m_{ijk,k} + s_{ij} - q_{ij} = 0 & \textrm{ in $V = V_m + V_p$}, \vspace{2mm} \\
\sigma_{ij}n_{j} = T_{i} & {\rm and} & m_{ijk}n_{k} = M_{ij}           & \textrm{ on $ S_{\rm ext} $}, \vspace{2mm} \\
& &  M^I_{ij} + m_{ijk}n^I_{k} = 0 & \textrm{ on $ S_I $}.
\end{array}
\end{equation}
Here, $n_{k}$ denotes the outward normal vector to surface $S_{\rm ext}$, and $n^I_{k}$ denotes the normal vector to the interface surface $S_I$ in the direction from the matrix into the particle. It should be noted from Eq. \eref{eqn:equilibrium} that in absence of moment stresses, \ie, $m_{ijk}=0$, the micro stresses coincide with the deviatoric Cauchy stresses.

\subsection{Constitutive equations}
In both matrix and particles, the elastic response is assumed to be linear and isotropic with Young's modulus $E_{\rm m}$, $E_{\rm p}$ and Poisson's ratios $\nu_{\rm m}$, $\nu_{\rm p}$, with shear modulus $G_{\rm m} = E_{\rm m} / [2(1+\nu_{\rm m})]$ and $G_{\rm p} = E_{\rm p} / [2(1+\nu_{\rm p})]$.

Assuming a purely dissipative bulk, \ie, the free energy in the bulk does not depend on plastic strains nor on the gradients thereof, the rate of dissipation can be written as
\begin{equation}\label{eqn:dissipation}
\dot{D} = q_{ij} \dot{\varepsilon}_{ij}^{\rm p} + m_{ijk}\dot{\varepsilon}_{ij,k}^{\rm p} = \Sigma \dot{E}^{\rm p} \geq 0,
\end{equation}
where the effective work conjugate measures of stress and plastic strain rate, respectively, has been introduced as   
\begin{equation}\label{eqn:Effective}
	\Sigma = \sqrt{\frac{3}{2}\left( q_{ij}q_{ij} + \frac{m_{ijk}m_{ijk}}{\ell^{2}} \right)}, \quad 
	\dot{E}^{\rm p} = \sqrt{\frac{2}{3}\left( \dot{\varepsilon}_{ij}^{\rm p}\dot{\varepsilon}_{ij}^{\rm p}   +  \ell^{2}\dot{\varepsilon}_{ij,k}^{\rm p}\dot{\varepsilon}_{ij,k}^{\rm p} \right) }.
\end{equation}
These measures reduce to those of standard $ J_2 $-plasticity quantities in absence of plastic strain gradients. Note that the inclusion of higher order terms in Eq. \eref{eqn:Effective} necessitates a length parameter $ \ell $, here viewed as a material constant. In the context of the current problem, $ \ell $ sets the scale on which plastic strain gradients may influence the strengthening due to the build up of plastic strain gradients at the particle/matrix interface. Note that a length scale $ \ell $ much smaller than other problem specific relevant length scales will remove the effects of strain gradients.

Constitutive rate equations satisfying Eqs. \eref{eqn:dissipation} and \eref{eqn:Effective} are employed and defined as
\begin{equation} \label{eqn:consteq1}
\dot{\varepsilon}_{ij}^{\rm p}=\dot{E}^{\rm p}\frac{3q_{ij}}{2\Sigma} , \quad \dot{\varepsilon}_{ij,k}^{\rm p}=\dot{E}^{\rm p}\frac{3m_{ijk}}{2l^2\Sigma}.
\end{equation}
For a rate independent material, Eq. \eref{eqn:consteq1} is valid provided that the yield condition $\Sigma = \sigma_{\rm m} $ is satisfied, where $\sigma_{\rm m}$ is the flow stress (of the matrix material) which depends on the accumulated plastic effective strain. Here, the matrix material will be assumed to obey a power law hardening function according to
\begin{equation}\label{eqn:flowfcn}
	\sigma_{\rm m} = \sigma_{0} \left( 1 + \frac{E^{\rm p}}{\varepsilon_{0}}\right)^{N}, \quad  E^{\rm p} = \int{\dot{E}^{\rm p} {\rm{d}}t}, 
\end{equation}
where $\sigma_{0}$ is the initial yield stress of the matrix material, $N$ is a plastic work hardening exponent and $\varepsilon_0 = \sigma_0 / E_{\rm m}$.

\subsection{Interface description}

The constitutive behaviour of the particle/matrix interface must satisfy the requirement of positive rate of dissipation, \ie,
\begin{equation}\label{eqn:dissInt}
   \dot{D}_I = \left( M^{I}_{ij} - \frac{\partial \psi}{\partial \varepsilon_{ij}^{\rm p}} \right)\dot{\varepsilon}_{ij}^{\rm p} \ge 0, 
\end{equation}
where $\psi$ is the free energy per unit surface area at the interface $S_I$. From a mechanical perspective in the case of proportional loading, it is immaterial whether $M^{I}_{ij}$ is evaluated on the basis of a purely energetic or a purely dissipative formulation, respectively \citep{fredriksson2005size,fredriksson2007modelling}. In the present study a purely energetic formulation will be employed and $\dot{D}_I$ is taken to be zero. Regarding the specific form of $\psi$, \cite{Asgharzadeh2021a} and \cite{faleskog2021analytical} find that it is possible to obtain precipitation strengthening in line with existing experimental results using a $\psi$-function that depends linearly on the current value of effective plastic strain $\varepsilon_{\rm e}^{\rm p} = \sqrt{\frac{2}{3}\varepsilon_{ij}\varepsilon_{ij}}$ at the interface $S^I$. This suggests that the moment traction can be obtained as
\begin{equation} 	\label{eqn:consteq2}
	M_{ij}^I = \frac{\partial \psi}{\partial \varepsilon_{ij}^{\rm p}} = \frac{{\rm d} \psi}{{\rm d} \varepsilon_{\rm e}^{\rm p}} \frac{\partial \varepsilon_{\rm e}^{\rm p}}{\partial \varepsilon_{ij}^{\rm p}} =  \psi^{\,\prime}  \frac{2}{3} \frac{\varepsilon_{ij}^{\rm p}}{\varepsilon_{\rm e}^{\rm p}}.
\end{equation}
A convenient and appropriate definition of the proportionality factor is $\psi^{\,\prime} = \sigma_0 \ell \alpha$, when prediction of yield strength is of primary interest, cf. \cite{Asgharzadeh2021a}, \cite{faleskog2021analytical}. Here, $\alpha$ is a non-dimensional parameter in the range $[0,1]$ that determines the plastic constraint at an interface: a value of zero puts no constraint, whereas a value of one puts full constraint on the plastic strains. The latter is identical to a micro-hard interface where plastic strains are constrained to zero.

\begin{figure}
	\begin{center}
		\includegraphics[width=0.5\textwidth]{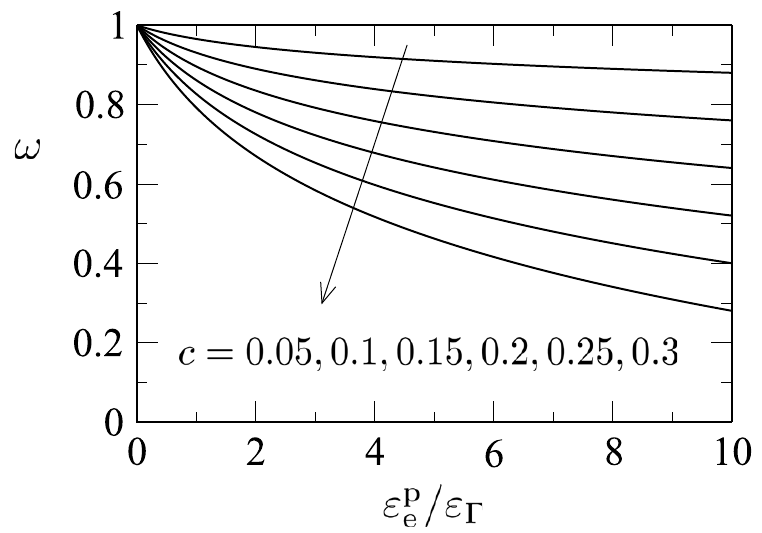}
		\caption{Interface decay function $\omega = \psi^{\,\prime} / (\sigma_0 \ell \alpha) $ plotted versus normalized effective plastic strain at the interface for different values of parameter $c$.}
		\label{fig2}
	\end{center}
\end{figure}

In the current work, post yield strain hardening is of primary interest. Therefore, the surface free energy model will be extended to account for the evolution of plastic straining. From a physical point of view, $\psi$ should correlate with the stored energy associated with dislocation networks in the close vicinity of $S_I$. These networks may be described as a surface dislocation density, \ie, the total length of dislocations per unit area $\hat{\rho}$ (hat accent is used to separate it from standard dislocation density denoted $\rho$). In Appendix A, a model for the free energy $\psi$ per unit surface area of the interface in terms of $\hat{\rho}$ is presented as (see Eq. \eref{appeq:A04})
\begin{equation} \label{eqn:intfenergy}
	\psi =  \dfrac{G_{\rm m} b^2}{4\pi} \hat{\rho} \ln(\dfrac{10}{b\hat{\rho}}),
\end{equation}
where $G_{\rm m}$ denotes the shear modulus and $b$ the magnitude of Burgers vector. Furthermore, a relation between surface dislocation density $\hat{\rho}$ and accumulated plastic $\varepsilon_{\rm e}^{\rm p}$ is derived (see Eq. \eref{appeq:A07}) according to
\begin{equation}  \label{eqn:rho-peeq}
	\hat{\rho} = \hat{\rho}_0 + \dfrac{k}{b} \varepsilon_{\rm e}^{\rm p}.
\end{equation}
Here, $\hat{\rho}_0$ is of order $\sqrt{\rho_0}$, where $\rho_0$ represents the dislocation density in the matrix at the onset of plastic deformation, and $k$ is a constant of order one. By use of Eqs. \eref{eqn:intfenergy} and \eref{eqn:rho-peeq}, and requiring that $\psi^{\,\prime}(\varepsilon_{\rm e}^{\rm p}=0) = \sigma_0 \ell \alpha$, a modified and physically based expression for $\psi^{\,\prime}$ may after some manipulations be obtained as
\begin{equation} \label{eqn:intfstr}
	\psi^{\,\prime} = \sigma_0 \ell \alpha \omega, \quad \quad \omega = \omega(\varepsilon_{\rm e}^{\rm p}) = \left[ 1 - c\,\ln(1+\varepsilon_{\rm e}^{\rm p}/\varepsilon_{\Gamma}) \right].
\end{equation}
The parameters in the nondimensional function $\omega$ can readily be identified as $c = 1 / [\ln(10/(b\hat\rho_0))-1]$ and $\varepsilon_{\Gamma} = b \hat\rho_0 / k$. Note that, $\omega(0) = 1$ at the onset of overall plastic deformation. In additon, the heuristic derivation in Appendix A suggest that $\alpha \sigma_0 \ell = G_{\rm m} b k /(4\pi)$.  The behaviour of function $\omega(\varepsilon_{\rm e}^{\rm p})$ in Eq. \eref{eqn:intfstr} is illustrated in Fig. \ref{fig2}.

\section{Limit Analysis-Upper bound solution}

Consider a volume containing a large number of elastic particles as sketched in Fig. \ref{fig1} with a matrix material described by the constitutive relationships presented above. Accurate analytic solutions for yield stress predictions for such materials is proposed in \cite{faleskog2021analytical}, where the authors use perturbation analysis of the governing equations on non-dimensional form to show that the plastic strain field in the matrix is to zeroth order constant for sufficiently small values of $f$ and $a / \ell$. This is exploited in \cite{faleskog2021analytical} to derive an upper bound solution for the yield stress of the composite material. The objective here is to extend their upper bound solution and develop an analytical expression in the post yield regime for the volume average stress $\langle \sigma_{ij} \rangle$ and how it evolves with plastic straining in the matrix. For an additional discussion of upper bound solutions for strain gradient plasticity models, see \cite{fleck2009mathematical,fleck2009mathematical2,polizzotto2010strain,reddy2020bounds}.

Noting that the prescribed traction $\langle \sigma_{ij} \rangle n_j$ on the external boundary to $V$ corresponds to the volume average stress, the principle of virtual work in Eq. \eref{eqn:VirtualWork} can be evaluated as
\begin{equation} \label{eqn:vWUB1}
	V \langle \sigma_{ij} \rangle \langle \varepsilon_{ij}^{*} \rangle = \int_{V}\left[ \sigma_{ij}\varepsilon^{*\rm e}_{ij} + q_{ij}\varepsilon^{*\rm p}_{ij} + m_{ijk}\varepsilon_{ij,k}^{*\rm p} \right]{\rm d}V +  \int_{S_{\rm I}}M^I_{ij}\varepsilon^{*\rm p}_{ij}{\rm d}S,
\end{equation}
where, $\langle \varepsilon_{ij}^{*} \rangle$ is the volume average virtual strain resulting from the arbitrary virtual displacements $u_i^{*}$ and $\varepsilon_{ij}^{*\rm e} = \varepsilon_{ij}^{*} - \varepsilon_{ij}^{*\rm p}$, where $\varepsilon_{ij}^{*\rm p}$ is independent of $u_i^{*}$ according to \eref{eqn:vWUB1}. Note that it is assumed that the moment tractions vanish at the external boundary.

The principle of maximum plastic dissipation, cf. Eq. \eref{eqn:dissipation}, may now be formulated by the following inequality
\begin{equation} \label{eqn:PMdiss}
	q_{ij}\varepsilon^{*\rm p}_{ij} + m_{ijk}\varepsilon_{ij,k}^{*\rm p} \le q_{ij}^*\varepsilon^{*\rm p}_{ij} + m_{ijk}^*\varepsilon_{ij,k}^{*\rm p} = \sigma_{\rm m} E^{* \rm p}.
\end{equation}
Here, $q_{ij}^*$ and $m_{ijk}^*$ are the micro stresses and moment stresses corresponding to $\varepsilon^{*\rm p}_{ij}$ and $\varepsilon_{ij,k}^{*\rm p}$, respectively, according to Eq. \eref{eqn:consteq1}, and $E^{* \rm p}$ is the virtual effective plastic strain. Utilizing the constitutive relation \eref{eqn:consteq2} for $M_{ij}^I$ at the interface, the following inequality apply for the last term in Eq. \eref{eqn:vWUB1}
\begin{equation} \label{eqn:InterfaceInequality}
	M_{ij}^I\varepsilon^{*\rm p}_{ij} = \psi^{\,\prime} \frac{2}{3} \frac{\varepsilon^{\rm p}_{ij}\varepsilon^{*\rm p}_{ij}}{\sqrt{\frac{2}{3}\varepsilon^{\rm p}_{ij}\varepsilon^{\rm p}_{ij}}} \le \psi^{\,\prime} \frac{2}{3} \frac{\varepsilon^{*\rm p}_{ij}\varepsilon^{*\rm p}_{ij}}{\sqrt{\frac{2}{3}\varepsilon^{*\rm p}_{ij}\varepsilon^{*\rm p}_{ij}}} = \psi^{\,\prime} \varepsilon^{*\rm p}_{\rm e}.
\end{equation}

An inequality may now be formulated from Eqs. \eref{eqn:vWUB1}-\eref{eqn:InterfaceInequality} for the volume average stress,
\begin{equation} \label{eqn:StressInequality}
	V \langle \sigma_{ij} \rangle \langle \varepsilon_{ij}^{*} \rangle \le \int_{V_{\rm p}} \sigma_{ij}\varepsilon^{*\rm e}_{ij} {\rm d}V + \int_{V_{\rm m}}  \left[ \sigma_{ij}\varepsilon^{*\rm e}_{ij} + \sigma_{\rm m} E^{* \rm p}  \right]{\rm d}V + \int_{S_{\rm I}} \psi^{\,\prime} \varepsilon^{*\rm p}_{\rm e}{\rm d}S,
\end{equation}
where the integration over $V$ has been split up into sub-volumes $V_{\rm p}$ and $V_{\rm m}$. 

Supported by the numerical studies by \cite{Asgharzadeh2021a} and \cite{Asgharzadeh2021b}, 
the assumption that plastic deformation develops in the whole matrix simultaneously as soon as the condition for initiation of plastic strain is met is successfully employed in \cite{faleskog2021analytical}. Furthermore, these authors note that the plastic strain field evolving in the matrix is essentially homogeneous provided that $a / \ell$ and $f$ are sufficiently small. Based on these observations, a homogeneous virtual strain field $\varepsilon_{ij}^0$ corresponding to virtual displacements $u_i^{*} = \varepsilon_{ij}^0 x_j$, will be assumed in the present analysis. Thus, in the elastic particles residing in $V_{\rm p}$, $\varepsilon^{*\rm e}_{ij} = \varepsilon_{ij}^0$. In the matrix $V_{\rm m}$, it will be assumed that $\varepsilon^{*\rm p}_{ij} = \varepsilon_{ij}^0$ and $\varepsilon^{*\rm e}_{ij} = 0$. With these assumptions, \eref{eqn:StressInequality} may be evaluated as
\begin{equation} \label{eqn:StressInequality2}
	\left[\langle \sigma_{ij} \rangle - f \langle  \sigma_{ij} \rangle_{\rm p} \right] \frac{\varepsilon_{ij}^0}{\varepsilon_{\rm e}^0} \le (1-f) \sigma_{\rm m}  + \frac{1}{V} \int_{S_{\rm I}} \psi^{\,\prime} {\rm d}S,
\end{equation}
where $\varepsilon^0_{\rm e}$ is the effective plastic strain corresponding to $\varepsilon_{ij}^0$ and thus $E^{* \rm p} = \varepsilon_{\rm e}^0$, and $\sigma_{\rm m}$ represents the flow stress of the matrix in absence of strain gradients. Furthermore, $\langle  \sigma_{ij} \rangle_{\rm p}$ denotes the volume average stress in all the particles, and $f = V_{p}/V$ is the total volume fraction of particles. It is noted that the inequality only depends on the direction of $\varepsilon_{ij}^0$ and not on its magnitude. Moreover, since the virtual strains are volume preserving, only the deviatoric parts $\langle s_{ij} \rangle$ and $\langle s_{ij} \rangle_{\rm p}$ of $\langle \sigma_{ij} \rangle$ and $\langle  \sigma_{ij}\rangle_{\rm p}$, respectively, will contribute to Eq. \eref{eqn:StressInequality2}. Hence,
\begin{equation} \label{eqn:StressInequality3}
	\left[\langle s_{ij} \rangle - f \langle s_{ij} \rangle_{\rm p} \right] \frac{\varepsilon_{ij}^0}{\varepsilon_{\rm e}^0} \le (1-f) \sigma_{\rm m}  + \frac{1}{V} \int_{S_{\rm I}} \psi^{\,\prime} {\rm d}S.
\end{equation}
The right-hand-side of Eq. \eref{eqn:StressInequality3} does not depend on either the applied stress nor on the assumed virtual strain $\varepsilon_{ij}^0$. To obtain the lowest possible bound for the external loading, the left-hand-side of Eq. \eref{eqn:StressInequality3} should be maximized for a given $\langle s_{ij} \rangle$. This is achieved when $\varepsilon_{ij}^0$ is co-linear with $\langle s_{ij} \rangle - f \langle s_{ij} \rangle_{\rm p} $. Furthermore, the average stress deviator in particles is related to the applied deviatoric stress and the plastic strain evolving in the matrix in the post yield regime. This relation can be established by introducing a known eigenstrain in the particles that reflects the homogeneous plastic strain in the matrix, and analysing the composite as an Eshelby problem, cf. \cite{eshelby1957determination} and \cite{Mura87}. Details of this analysis is given in Appendix B, and for particles of spherical shape this relation can be expressed as
\begin{equation} \label{eqn:EshelbyRes}
	\langle s_{ij} \rangle_{\rm p} =  \bar{\Gamma} \langle s_{ij} \rangle + 2 \bar{G} \varepsilon_{ij}^{\rm p},
\end{equation}
where $\bar{\Gamma}$ and $\bar{G}$ represent effective values of $\Gamma$ and $G$ for the composite in case the elastic properties differ among particles, and $\varepsilon_{ij}^{\rm p}$ denotes the homogeneous plastic strain assumed to prevail in the matrix, \ie, in volume $V_{\rm m} = V - V_{\rm p}$. For an individual spherical particle (cf. Appendix B)
\begin{equation} \label{eqn:EshelbyParam}
	\begin{array}{l}
		\Gamma = \dfrac{15(1-\nu_{\rm m})g}{7 - 5\nu_{\rm m} + 2(4 - 5\nu_{\rm m})g} = \{\nu_{\rm m} = 0.3 \} = \dfrac{21 g}{11 + 10g},  \vspace{2 mm} \\ G  = \dfrac{G_{\rm m} \cdot (7-5\nu_{\rm m})g}{7-5\nu_{\rm m} + 2(4-5\nu_{\rm m})g}  = \{ \nu_{\rm m} = 0.3 \} = \dfrac{G_{\rm m} \cdot 11 g}{11 + 10g},
	\end{array}
\end{equation}
where $g = G_{\rm p}/G_{\rm m}$. It is noteworthy that $G / (\Gamma G_{\rm m}) = (7-5\nu_{\rm m})/[15(1-\nu_{\rm m})]$, \ie, this ratio is independent of $g$ and only exhibits a weak dependence on $\nu_{\rm m}$.

By applying a virtual strain field $\varepsilon_{ij}^0$ that is co-linear with $\langle s_{ij} \rangle - f \langle s_{ij} \rangle_{\rm p} $, the left-hand-side of Eq. \eref{eqn:StressInequality3} may be simplified as
\begin{equation} \label{eqn:StressIneqLHS}
	\begin{array}{l}
	\sqrt{\dfrac{3}{2} \left(\langle s_{ij} \rangle - f \langle s_{ij} \rangle_{\rm p} \right)\left(\langle s_{ij} \rangle - f \langle s_{ij} \rangle_{\rm p} \right) } \vspace{2 mm} \\ \qquad = \sqrt{\left( \langle\sigma_{\rm e}\rangle (1-f\bar{\Gamma})\right)^2 - 2 (1-f\bar{\Gamma})\langle\sigma_{\rm e}\rangle \,  3\bar{G} f \varepsilon_{\rm e}^{\rm p} + \left( 3\bar{G} f \varepsilon_{\rm e}^{\rm p} \right)^2}  \vspace{2 mm} \\ \qquad =  \, \sigma_{\rm e} (1-f\bar{\Gamma}) - 3\bar{G} f \varepsilon_{\rm e}^{\rm p}.
	\end{array}
\end{equation}
To arrive at the second step in Eq. \eref{eqn:StressIneqLHS}, equalities  $3 \langle s_{ij}\rangle \langle s_{ij}\rangle / 2 = \langle\sigma_{\rm e}^2\rangle$, $\varepsilon_{ij}^{\rm p} \varepsilon_{ij}^{\rm p} = 3 (\varepsilon_{\rm e}^{\rm p})^2 / 2$ and $\langle s_{ij}\rangle \varepsilon_{ij}^{\rm p} = \langle \sigma_{\rm e}\rangle  \varepsilon_{\rm e}^{\rm p}$ has been utilized. The latter equality rests on the assumption of proportinal loading, which is justified in this case. Also, the bracket symbol $\langle \rangle$ has been omitted in the third step in Eq. \eref{eqn:StressIneqLHS} and will henceforth not be used.

Returning to Eq. \eref{eqn:StressInequality3}, the second term on the right-hand-side  may for spherical particles be evaluated as
\begin{equation} \label{eqn:InterfaceTerm}
	\frac{1}{V} \int_{S_{\rm I}} \psi^{\,\prime} {\rm d}S = 3 f \sigma_0 \bar{\alpha}\omega \frac{\ell}{\bar{a}},
\end{equation}
where $\bar{\alpha}$ and $\bar{a}$ represents effective values of $\alpha$ and $a$, respectively, which will be defined below.

Finally, by substition of Eqs. \eref{eqn:StressIneqLHS} and \eref{eqn:InterfaceTerm} into Eq. \eref{eqn:StressInequality3}, the upper bound solution for the effective stress (flow stress) of a composite containing a random distribution of elastic spherical particles can be formulated as
\begin{equation} \label{eqn:UBS}
	\sigma_{\rm e} \le \frac{(1-f)}{(1-\bar{\Gamma}f)} \sigma_{\rm m}(\varepsilon_{\rm e}^{\rm p})  + \frac{3 f \sigma_0 \bar{\alpha}\,\omega(\varepsilon_{\rm e}^{\rm p})}{(1-\bar{\Gamma}f)}\frac{\ell}{\bar{a}} + \frac{3 \bar{G} f }{(1-\bar{\Gamma}f)} \varepsilon_{\rm e}^{\rm p} = \sigma_{\rm e}^{\rm UBS},
\end{equation}
where
\begin{equation} \label{eqn:UBS_peeq}
	\varepsilon_{\rm e}^{\rm p} = \frac{\bar{\varepsilon}_{\rm e}^{\rm p}}{1-f}.
\end{equation}
Here, $\bar{\varepsilon}_{\rm e}^{\rm p}$ denotes the volume average of the effective plastic strain in the composite material, \ie, in $V$. This is a more convenient measure for practical purposes, and will be used in Section 4 where model predictions are compared with full field FEM results. Note that the first term in Eq. \eref{eqn:UBS} is the flow stress from the matrix material adjusted for a volume fraction of particles and that the second and third term correspond to particles' contributions to yield stress and strain hardening, respectively. Also, notice that the third term in Eq. \eref{eqn:UBS} corresponds to a particle induced linear strain hardening, similar to what has been derived in \cite{tanaka1970hardening,brown1971work}. However, the linear hardening may be counteracted by the second term in Eq. \eref{eqn:UBS} if the interface suffers from a logarithmic decay with increasing effective plastic strain.

The effective values in Eq. \eref{eqn:UBS} are evaluated as
\begin{equation} \label{eqn:EffectiveVals}
	\bar{\Gamma} = \frac{\langle \Gamma(a) a^3 \rangle}{\langle a^3 \rangle}, \quad \bar{G} = \frac{\langle G(a) a^3 \rangle}{\langle a^3 \rangle},  \quad \bar{\alpha} = \frac{\langle \alpha(a) a^2 \rangle}{\langle a^2 \rangle}, \quad \bar{a} = \frac{\langle a^3 \rangle}{\langle a^2 \rangle},
\end{equation}
where, $\langle (\bullet) \rangle = \sum_{i=1}^{N_{\rm p}} (\bullet)_i / N_{\rm p}$ denotes the expected mean value taken over all the $N_{\rm p}$ particles in $V$. Notice that only the interface parameter $\alpha$ is subject to averaging whereas the parameters associated with the non-dimensional function $\omega$ are assumed to belong to the matrix material. If all particles possess the same properties, the effective properties in Eq. \eref{eqn:UBSave_val} simplify to
\begin{equation} \label{eqn:UBSave_val}
	\bar{\Gamma} = \Gamma, \quad \bar{G} = G, \quad \bar{\alpha} = \alpha.
\end{equation}
Moreover, if all particles have the same size, $\bar{a} = a$.

\section{Numerical validation of upper bound solution}

A numerical examination and validation of the upper bound solution (UBS) extended for post yield strain hardening is presented in this Section. Two different unit cell models will be employed for this purpose, where full-field solutions are obtained by finite element analysis. General model features and overall accuracy will be discussed first on the basis of a 2D-axisymmetric model, where both perfectly plastic and strain hardening matrix materials are considered, with or without a decaying interface strength. This is followed by three-dimensional analyses, where variations of interface strength and shear modulus with particle size are investigated in the post yield regime.

\subsection{Unit cell models} \label{sec4.1}

\begin{figure}[H]
	\begin{center}
		\includegraphics[width=0.8\textwidth]{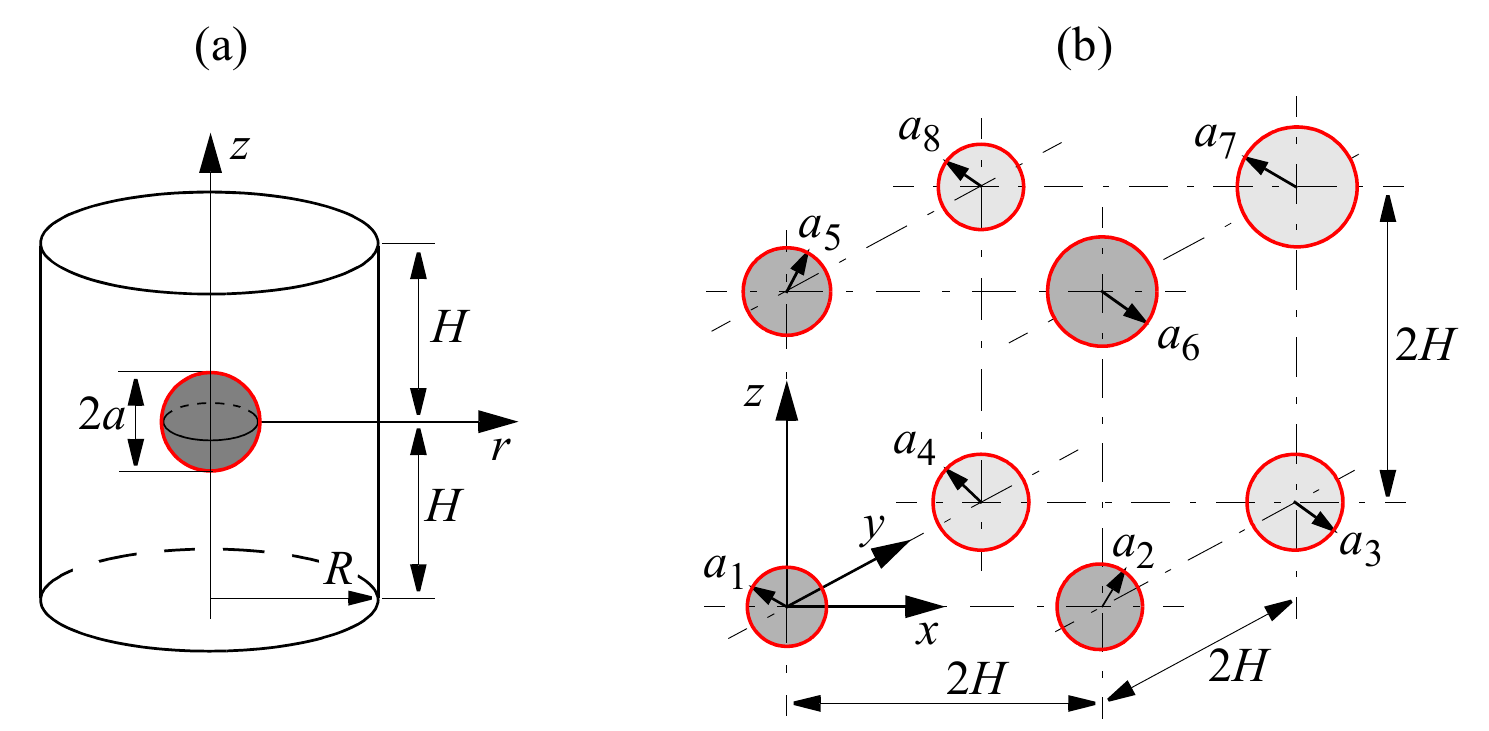}
		\caption{Unit cell models: (a) 2D axisymmetric model containing a spherical particle, and (b) 3D cubic model containing 8 spherical particles of different size ($a_1, ..., a_8$) with spacing $2H$.}
		\label{fig3}
	\end{center}
\end{figure}

The unit cell models employed in the numerical analysis are displayed in Fig. \ref{fig3}, showing a 2D axisymmetric model in (a) and a 3D cubic model in (b). The axisymmetric model is representative of a microstructure with particles of equal size positioned in hexagonal patterns in planes stacked on top of each other, whereas the cubic model represents a microstructure with a repetitive unit containing 8 particles of different size in the material. The radii of the particles $a_1, \ldots, a_8$ were chosen to follow a geometric series such that the surface area ratio between the largest and smallest particle becomes 10. Thus, $a_i=a_1\lambda^{i-1}$ with $\lambda=10^{1/14}$, where $a_1$ is determined by the volume fraction $f$ as $a_1 = H \left[ 48f \, (\lambda^3-1)\,/\,(\pi(\lambda^{24}-1))\right]^{1/3}$.

\begin{figure}[H]
	\begin{center}
		\includegraphics[width=0.85\textwidth]{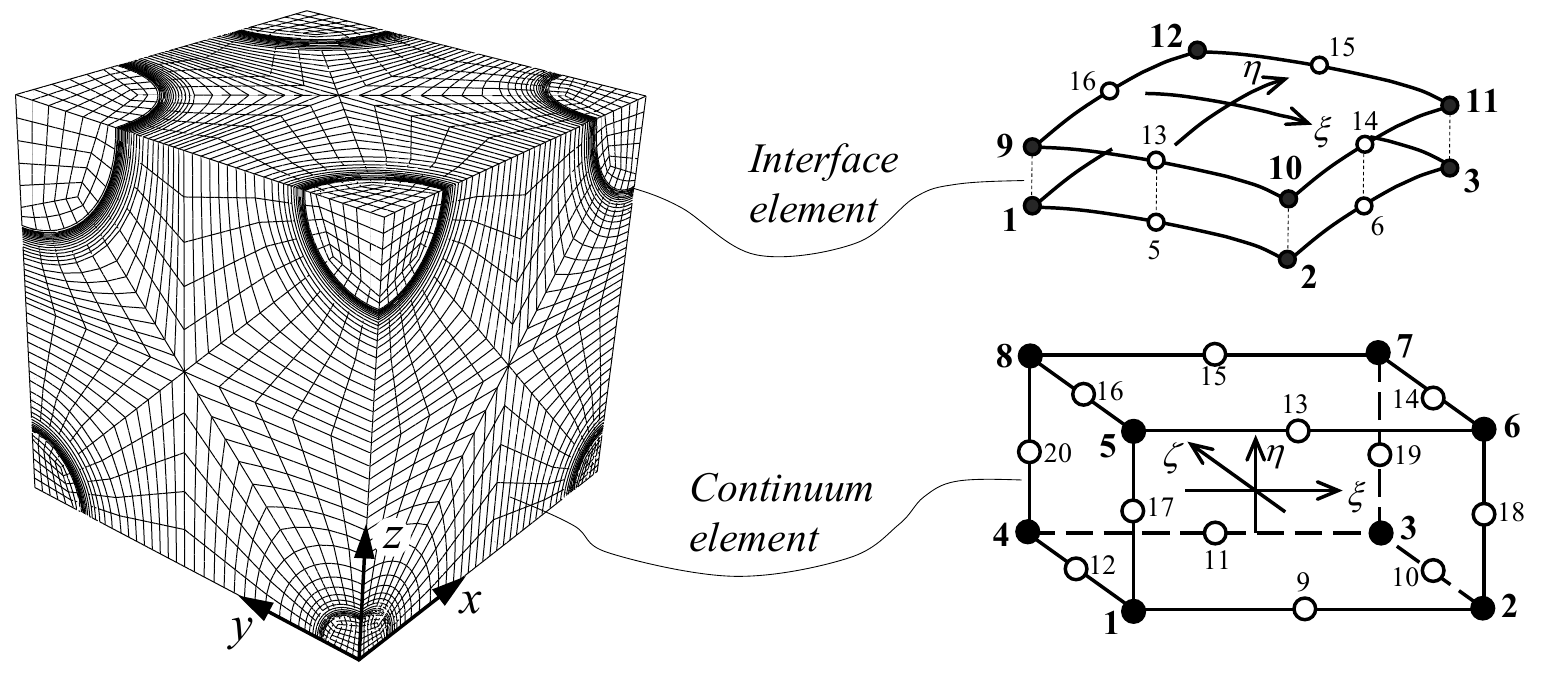}
		\caption{A representative FEM discretization of the 3D cubic model ($f = 0.04$), containing 17200 20-noded hexahedral elements with tri-quadratic interpolation for displacements (60 dof) and tri-linear interpolation for plastic strains (40 dof), and 600 12-noded interface elements (88 dof in total).}
		\label{fig4}
	\end{center}
\end{figure}

 These unit cell models have proven to be suitable for validation of the upper bound model for yield stress predictions (cf. \cite{Asgharzadeh2021a}, \cite{Asgharzadeh2021b}, \cite{faleskog2021analytical}), and were numerically analysed by the FEM formulation developed in \cite{dahlberg2013improved} and \cite{dahlberg2013deformation}, augmented with an axisymmetric element in \cite{Asgharzadeh2021a} and a tetrahedral element in \cite{Asgharzadeh2021b}. The FEM code was further developed in this work by implementation of a hexahedral 20 node element with an associated 16 node interface element. A typical cubic mesh is shown in Fig. \ref{fig4}, where the new elements are illustrated. All primary variables, \ie, displacements and plastic strains, are described at vertex nodes (solid circles in Fig. \ref{fig4}), whereas only displacements are described at the mid nodes (open circles in Fig. \ref{fig4}). 

To circumvent the problem of the indeterminacy of $q_{ij}$, $m_{ijk}$ in the elastic regime, a rate dependent visco-plastic constitutive formulation was used, cf. \cite{dahlberg2013improved}. Specifically, the effective plastic strain rate $\dot{E}^{\rm p}$ in Eqs. \eref{eqn:dissipation}-\eref{eqn:flowfcn} is governed by a viscosity law $\dot{E}^{\rm p} = \dot{\varepsilon}_{0} \Phi(\Sigma,\sigma_{\rm m})$ where $\Phi(\Sigma,\sigma_{\rm m}) = \kappa(\Sigma/\sigma_{\rm m}) + (\Sigma/\sigma_{\rm m})^{n}$ (visco-plastic response function). However, close to rate independent results can effectively be achieved provided that $\kappa$ and $n$ are small, respectively, large enough. For a detailed discussion on appropriate values for the current application, see \cite{Asgharzadeh2021a}. Furthermore, an Euler-backward algorithm was used to establish the linearised equation system, which was solved in an iterative manner in the FEM code. A sparse solver and OpenMP parallelization was utilized to speed up the computations. For details the reader is referred to the papers referenced above.

Symmetry displacement boundary conditions were applied on the unit cells in a standard manner. The higher order symmetric boundary conditions were enforced as
\begin{equation} \label{eqn:BC}
	\begin{array}{ll}
	{\textbf{3D-cubic:}}\vspace{1 mm} \\
	M_{xx}=M_{yy}=M_{zz}=M_{yz}=\varepsilon^p_{xy}=\varepsilon^p_{xz}=0\text{ on }x=0\text{ and }x=A \vspace{1 mm} \\
	M_{xx}=M_{yy}=M_{zz}=M_{xz}=\varepsilon^p_{xy}=\varepsilon^p_{yz}=0\text{ on }y=0\text{ and }y=A \vspace{1 mm} \\
	M_{xx}=M_{yy}=M_{zz}=M_{xy}=\varepsilon^p_{xz}=\varepsilon^p_{yz}=0\text{ on }z=0\text{ and }z=A \vspace{3 mm} \\

	{\textbf{2D-axisymmetric:}}\vspace{1 mm} \\
	M_{rr}=M_{zz}=\varepsilon^p_{rz}\text{ on }r=0, R \text{ and }z=0, H \vspace{1 mm} \\
	\end{array}
\end{equation}
\noindent In addition, $\varepsilon^p_{ij} = 0$ are prescribed in all particles. All unit cells were subjected to loading in uniaxial tension, which suffices as the flow stress in Eq. \eref{eqn:UBS} only depends on the second deviatoric stress invariant, not on the third or on the mean stress. Symmetric boundary conditions in combination with applied uniaxial tension were enforced by use of Lagrange equations.

The volume average of the Cauchy stress (the volume integral is recast into a surface integral by use of the divergence theorem) and the volume average of the effective Cauchy stress in a unit cell were calculated as 
\begin{equation} \label{eqn:EffStress}
\bar{\sigma}_{ij} = \frac{1}{V} \int_{S_{ext}} \frac{1}{2} \left(\sigma_{ik} x_j n_k + \sigma_{jk} x_i n_k \right) {\rm d}S, \quad \bar{\sigma}_e = \sqrt{\frac{3}{2}\bar{s}_{ij}\bar{s}_{ij}}, \quad \bar{s}_{ij}=\bar{\sigma}_{ij}-\frac{1}{3}\bar{\sigma}_{kk}\delta_{ij},
\end{equation}
and the volume average of strain and the effective strain were evaluated as
\begin{equation}\label{eqn:EffStrain}
\bar{\varepsilon}_{ij}=\frac{1}{V}\int_{S_{ext}}\frac{1}{2}(u_in_j+u_jn_i){\rm d}S, \quad \bar{\varepsilon}_e=\sqrt{\frac{3}{2}\bar{\varepsilon}_{ij}^{\:\prime}\bar{\varepsilon}_{ij}^{\:\prime}}, \quad  \bar{\varepsilon}_{ij}^{\:\prime}=\bar{\varepsilon}_{ij}-\frac{1}{3}\bar{\varepsilon}_{kk}\delta_{ij}.
\end{equation}

The results from the unit cell analyses presented below will be referred to as 'FEM' if obtained by finite element analysis or as 'UBS' if obtained by the upper bound solution in Eq. \eref{eqn:UBS}. Key parameters in the analysis are $f$, the ratio $\ell / \bar{a}$, and the ratio between the shear modulus in particles and matrix, respectively, as this ratio enters in parameters $\bar{\Gamma}$ and $\bar{G}$. Poisson's ratio play a minor role, and $\nu_{\rm p} = \nu_{\rm m} = 0.3$ will be used throughout the analysis.

Below, the upper bound solution will be compared with the outcome from numerical analyses of the unit cell models. Volume average effective stress-strain curves may then be constructed for the upper bound solution as
\begin{equation}\label{eqn:UBS_Effectve_se_eps}
		\sigma_{\rm e} = \left\{ \begin{array}{ll}
			3 G_{\rm eff} \bar{\varepsilon}_{e}, &  \bar{\varepsilon}_{e} \le \varepsilon_{y0}  \vspace{2 mm}\\
			\sigma_{\rm e}^{\rm UBS},  &  \bar{\varepsilon}_{e} > \varepsilon_{y0}, \quad{\rm where} \quad \bar{\varepsilon}_{e} = \sigma_{\rm e}^{\rm UBS}(\bar{\varepsilon}_{\rm e}^{\rm p})/3G_{\rm eff} + \bar{\varepsilon}_{\rm e}^{\rm p}.
		 \end{array} \right.
\end{equation}
Here, $\varepsilon_{\text{y}0} = \sigma_{\text{y}0}/3G_{\rm eff}$ represents the strain at the onset of overall plastic deformation with $\sigma_{\text{y}0} = \sigma_{\rm e}^{\rm UBS}(0)$ being the yield stress, and $G_{\rm eff}$ is the effective shear modulus of the composite material. The latter may be determined by use of Eshelby's equivalent inclusion method, which for spherical particles become (see pp. 428-430 in \cite{Mura87})
\begin{equation} \label{eqn:EffG_composite}
	G_{\rm eff} = \dfrac{G_{\rm m}}{ 1 + f \bar{\gamma}},
\end{equation}
where 
\begin{equation} \label{eqn:EffG_gamma}
	\bar{\gamma} = \dfrac{\langle \gamma a^3 \rangle}{\langle a^3 \rangle} \qquad \textrm{with} \qquad \gamma = \dfrac{15(1-\nu_{\rm m})(1-g)}{7 - 5\nu_{\rm m} + 2(4 - 5\nu_{\rm m})g}.
\end{equation}

The tangent modulus of the effective stress-strain curve \eref{eqn:UBS_Effectve_se_eps} will be compared with that obtained from unit cell analysis. For a perfectly plastic matrix combined with an interface energy independent of plastic straining, the tangent modulus takes the simple form
\begin{equation} \label{eqn:UBS_slope}
	\dfrac{{\rm d}\sigma_{\rm e}^{\rm UBS}}{{\rm d}\bar{\varepsilon}_{\rm e}} = \dfrac{h_{\rm UBS}}{1+h_{\rm UBS}/3G_{\rm eff}} \quad {\rm with} \quad h_{\rm UBS} =\dfrac{3\bar{G}f}{(1-\Gamma f)(1-f)}.
\end{equation}

\subsection{Strain hardening and hardening saturation: A perfectly plastic matrix} \label{sec4.2}
\begin{figure}[H]
	\begin{center}
		\subfloat[]{\includegraphics[width=0.40\textwidth]{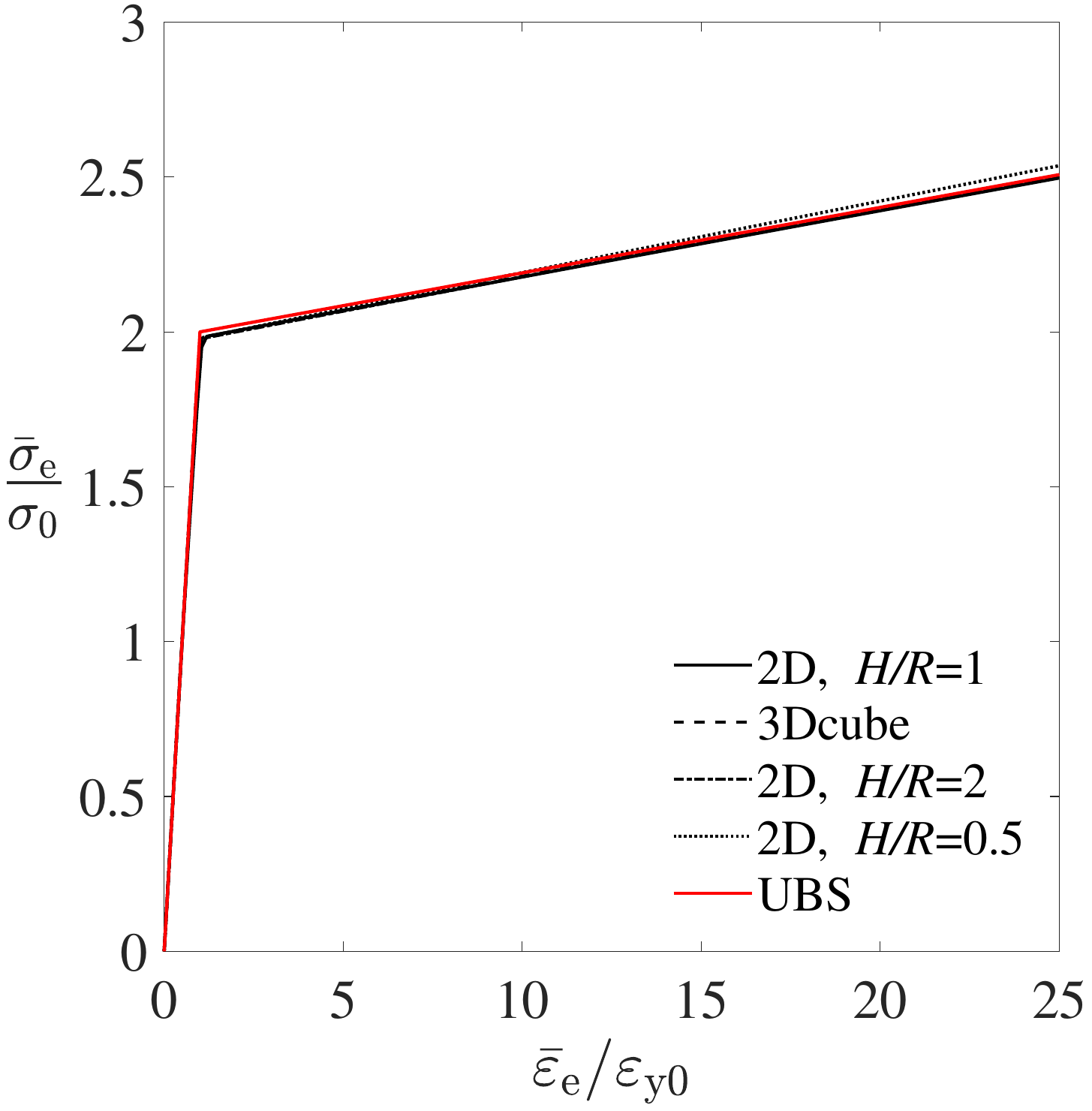}} \\
		\subfloat[]{\includegraphics[width=0.392\textwidth]{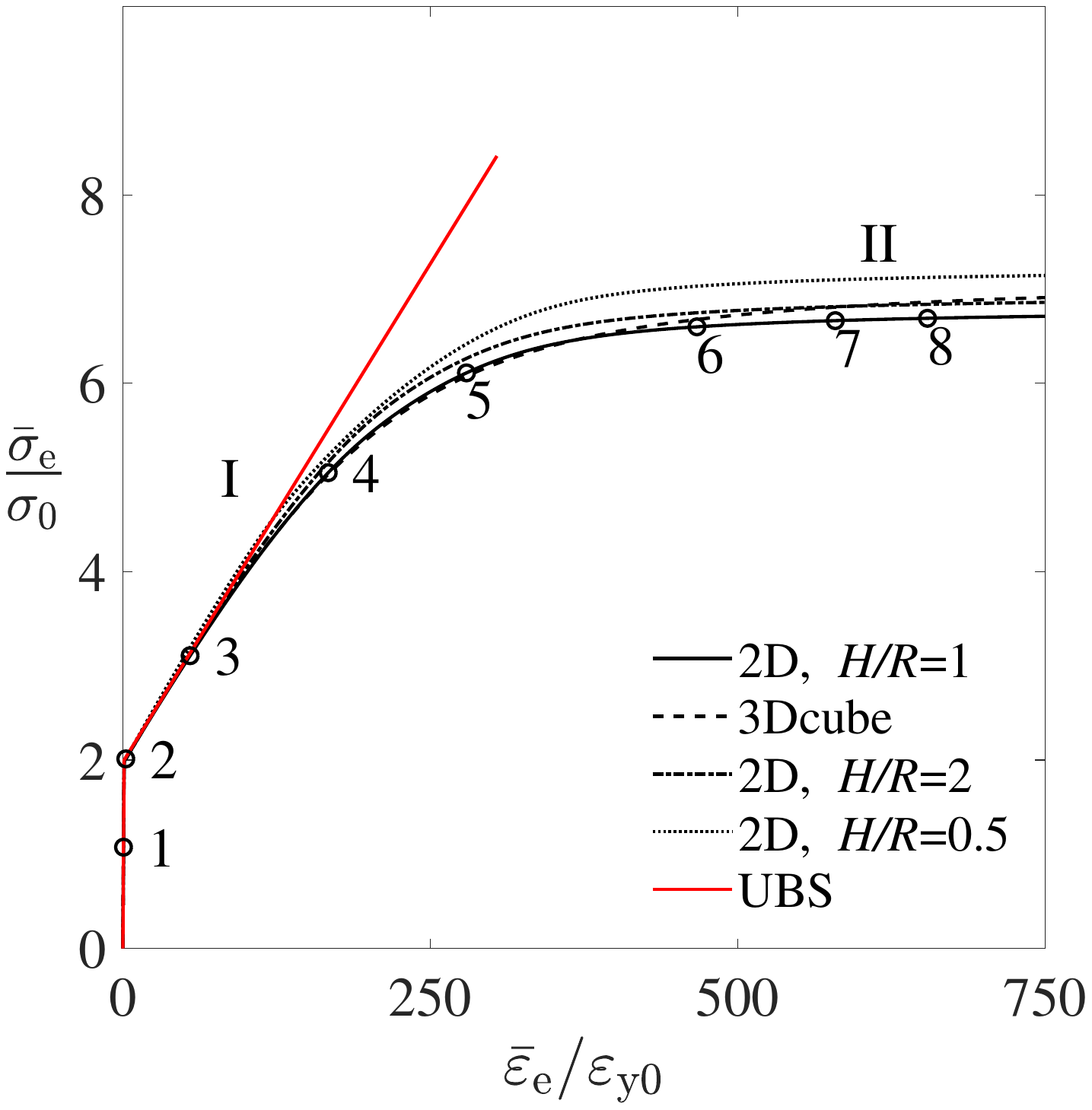}} \hspace{1 cm}
		\subfloat[]{\includegraphics[width=0.40\textwidth]{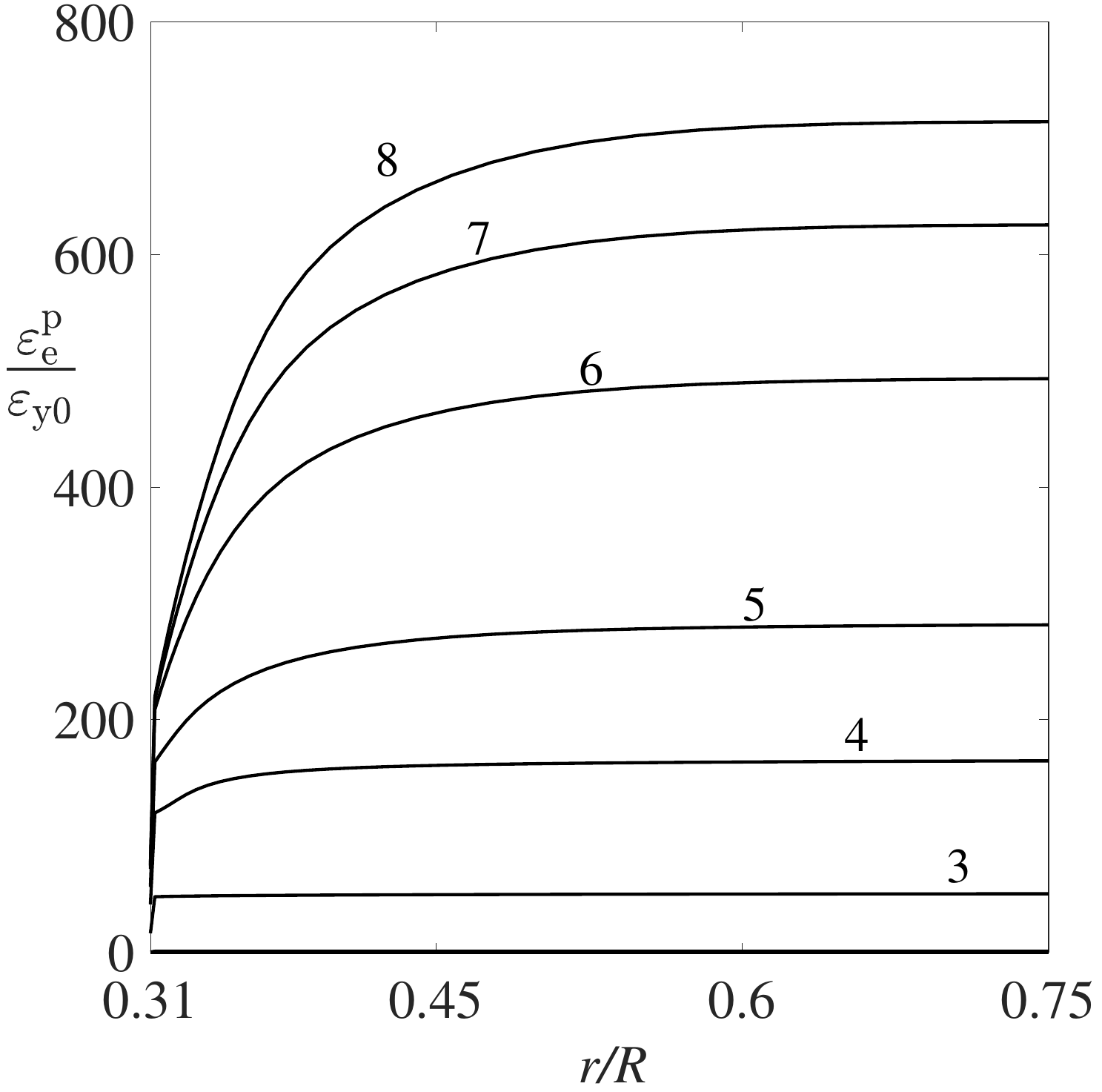}}
		\caption{A material with a perfectly plastic matrix material reinforced by 2\% particles with interface strength $\alpha = 0.99$ and relative size $\bar{a}/\ell = 1/16.498$. Normalized volume average effective stress-strain curves for different particle arrangements; (a) linear regime I and (b) saturation regime II. (c) Effective plastic strain plotted versus radius at $z=0$ resulting from the 2D axisummetric model with $H=R$.}
		\label{fig5}
	\end{center}
\end{figure}

The upper bound solution \eref{eqn:UBS} suggests that the flow stress is independent of the spatial distribution of particles. It depends on the size distribution of particles through the effective value $\bar{a}$, and also through $\bar{\alpha}$, $\bar{\Gamma}$ and $\bar{G}$ if the interface properties respectively shear modulus vary with particle size. To demonstrate this, numerical results were generated by use of the 2D axisymmetric model with three different cell ratios, $H/R = 0.5, 1, 2$, and use of the 3D cube model containing particles with the size variation described in Section 4.1. A material with a perfectly plastic matrix ($N=0$) containing a particle with the same shear modulus as the matrix ($g = 1$) and volume fraction $f = 0.02$, a close to micro-hard particle/matrix interface $\alpha = 0.99$ with $\omega = 1$ ($c = 0$), was considered. In addition, a ratio $\ell/\bar{a} = 16.498$ was chosen, which according to \eref{eqn:UBS} gives an initial yield stress $\sigma_{\rm y0} = 2\sigma_0$. The resulting overall effective stress-strain curves are shown in Fig. \ref{fig5}(a). The curves are normalized by $\sigma_0$ and $\varepsilon_{\rm y0}$, respectively. As noted, the results from the unit cell analysis coincide and are well captured by the UBS (curve in color red).

As the applied strain increases and becomes sufficiently large, the numerical results reveal that the UBS cease to be valid, see region II in Fig. \ref{fig5}(b). At this instance in the loading, the stress saturates (recall that the matrix is perfectly plastic, i.e. no hardening is provided by the matrix) and particles no longer contribute with additional strain hardening. The transition from linear hardening to zero hardening occurs at a strain that seem to scale with $\varepsilon_{0} \ell/a \cdot (1+1/g)$ and also depends on the spatial arrangement of particles as can be concluded from the spread in results noticed in region II in Fig. \ref{fig5}(b). This will be further elaborated on in Section \ref{Sect.Saturation}.

Some insight to this phenomenon can be gained by the characteristics of the distribution of effective plastic strain $\varepsilon_e^p$, in the matrix material at increasing levels of the overall strain. Radial distributions of $\varepsilon_e^p$ at $z=0$ obtained from the axisymmetric analysis with $H=R$ are presented in Fig. \ref{fig5}(c). At sufficiently low strain levels (curve labelled 3), it is noted that $\varepsilon_e^p$ is essentially homogeneous in the matrix material. This is in agreement with the analytical predictions for $a/\ell \ll 1$ presented in \cite{faleskog2021analytical}. However, as the overall strain increases, a noticeable strain gradient develops near the interface of the particle (curves labelled 4 and 5). Eventually the interface becomes incrementally micro-hard, and further development of plastic strain at the interface vanishes, which leads to even stronger plastic strain gradients in the matrix (curves labelled 6 to 8). At this stage in the loading, the overall stress level saturates.

Nevertheless, the UBS appears to be valid up to rather large strain levels (region I) as noted in the examples shown in Fig. \ref{fig5}(b). This suggests that the proposed UBS may be used for quite a wide range of physically relevant model parameters.

\subsection{Assessment of post yield response of the upper bound solution} \label{sec4.3}
A selective parametric study will now be presented based on the 2D axisymmetric model to explore the influence of the volume fraction $f$, modulus mismatch $g = G_{\rm p}/G_{\rm p}$, matrix strain hardening $N$, and the interface strength decay coefficients $c$ and $\varepsilon_{\Gamma}$, on strain hardening and the accuracy of UBS predictions. In all cases the ratio $\ell/a$ was chosen such that $\sigma_{\rm e}^{\rm UBS}(0) = 2\sigma_0$ in a microstructure with a medium strong particle/matrix interface $\alpha = 0.5$. All FEM calculations were progressed until an effective strain $\bar{\varepsilon}_e = 11 \varepsilon_{\rm y0}$ was reached.

\begin{figure}[H]
	\begin{center}
		\subfloat[]{\includegraphics[width=0.4\textwidth]{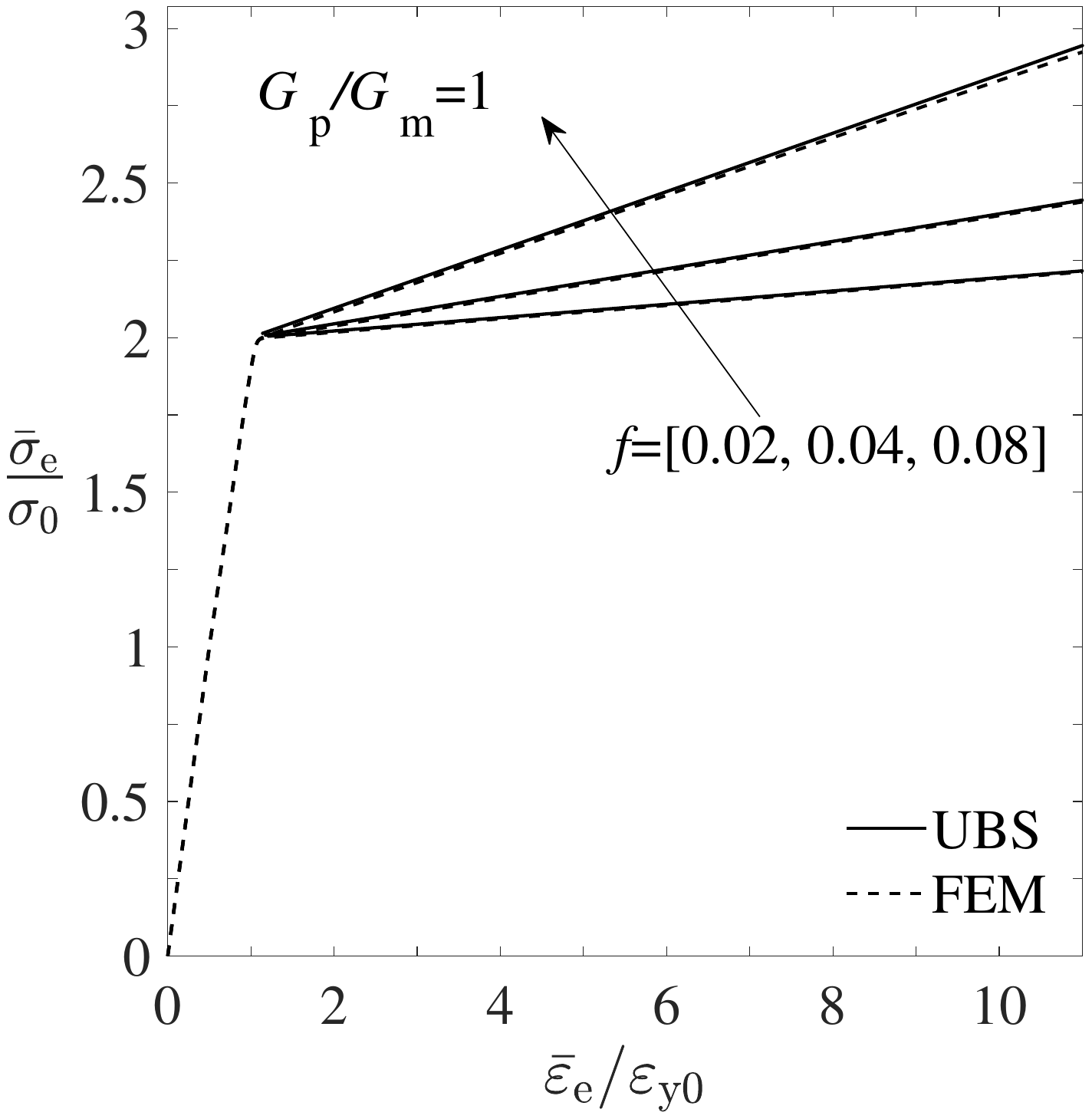}} \hspace{1 cm}
		\subfloat[]{\includegraphics[width=0.4\textwidth]{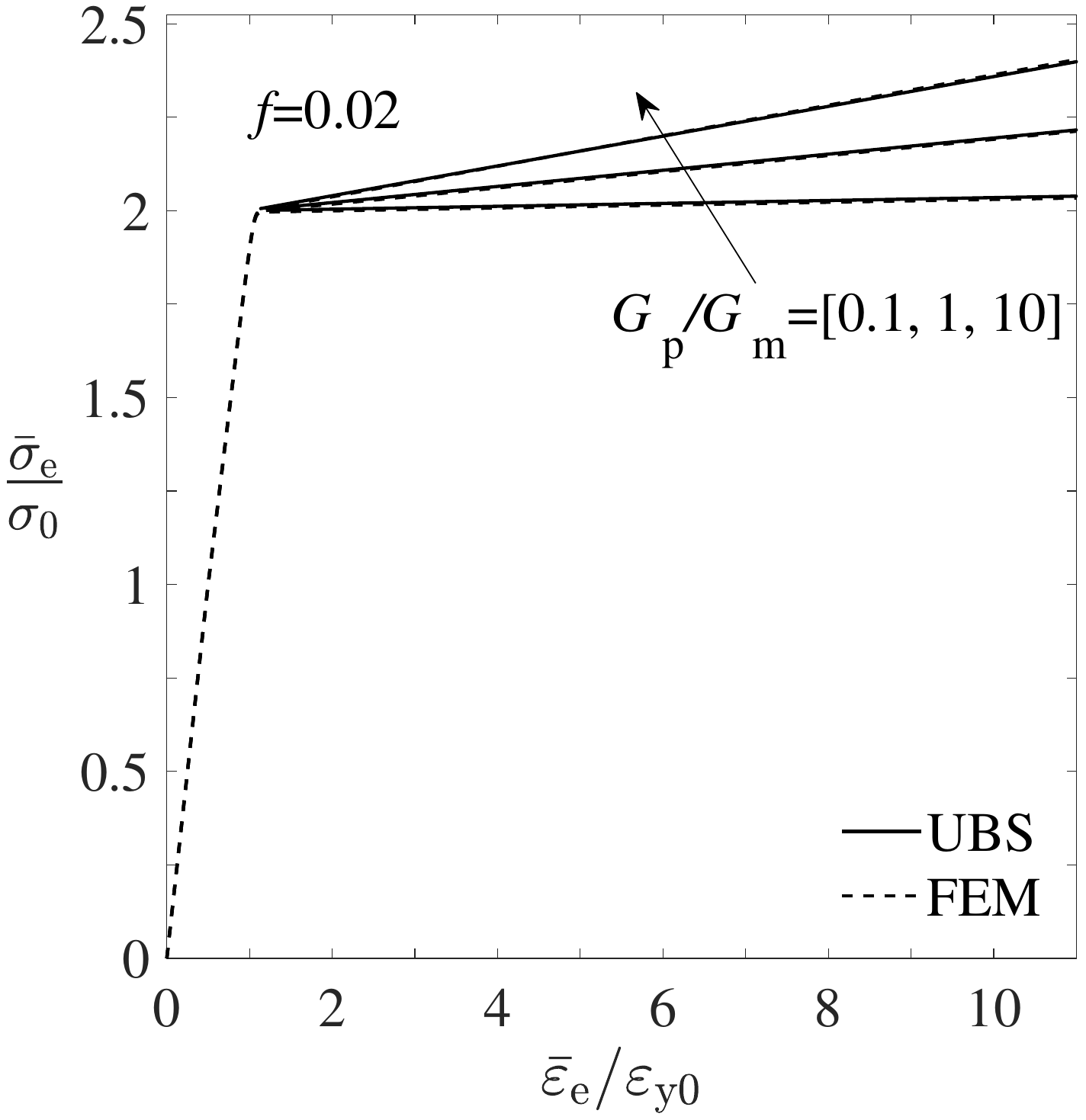}}\\
		\subfloat[]{\includegraphics[width=0.4\textwidth]{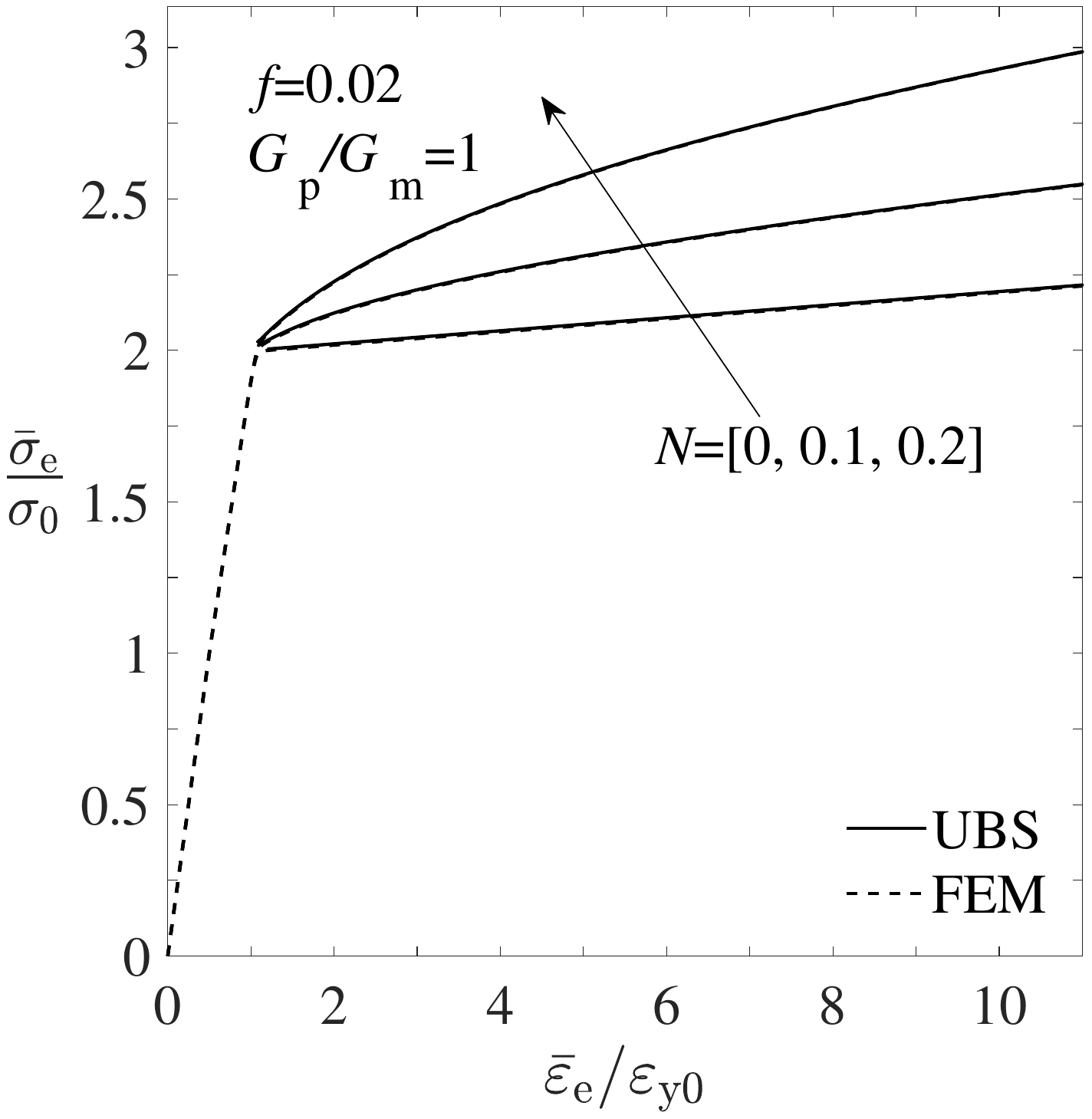}}
		\caption{Comparison of normalized volume average effective stress-strain curves resulting from UBS and FEM analysis of the 2D axisymmetric model. (a) Variation of volume fraction of particles keeping $G_{\rm p}=G_{\rm m}$ and $N=0$ constant. (b) Variation of shear modulus ratio $g=G_{\rm p}/G_{\rm m}$ keeping $f = 0.02$ and $N=0$ constant. (c) Variation of matrix strain hardening keeping $f = 0.02$ and $G_{\rm p}=G_{\rm m}$ constant.}
		\label{fig6}
	\end{center}
\end{figure}

Particles with an interface strength $\psi'$ that is independent on plastic strain (\ie, $c=0$ in Eq. \eref{eqn:intfstr}) and embedded in a perfectly plastic matrix ($N = 0$) are considered first. Figure \ref{fig6}(a) shows effective stress-strain responses for $f = \{0.02, 0.04, 0.08\}$ with $G_{\rm p}=G_{\rm m}$. Solid lines pertain to UBS predictions and dashed lines to FEM results. As observed, the UBS predictions agree very well with the FEM results, but some deviation is noticed for the material with the highest volume fraction ($f = 0.08$). This trend may be expected, as the accuracy of the UBS is closely related to the accuracy of the Eshelby solution, which assumes that particles are remote from each other. The effects of a mismatch in shear modulus is shown in Fig. \ref{fig6}(b) for materials with $G_{\rm p}/G_{\rm m} = \{0.1, 1, 10\}$, all having $f = 0.02$. Again the agreement between the UBS predictions and FEM results are very good.

Now, the perfectly plastic matrix is exchanged with a strain hardening matrix. The interaction between a strain hardening matrix and hardening due to particles is displayed in Fig. \ref{fig6}(c). Here, stress-strain curves for materials with hardening exponents $N = \{0, 0.1, 0.2 \}$, with $f = 0.02$ and $G_{\rm p}=G_{\rm m}$ are shown. As noticed, the UBS predictions fully captures the stress-strain curves resulting from the FEM analysis.

If the interface suffers from a logarithmic decay evolving with plastic straining, softening may follow upon the onset of overall plastic yielding. The degree of softening depends on the interplay between matrix strain hardening with additional contribution from particles, and the interface decay governed by parameters $c$ and $\varepsilon_{\Gamma}$ (illustrated in Fig. \ref{fig2}). 

\begin{figure}[H]
	\begin{center}
		\subfloat[]{\includegraphics[width=0.4\textwidth]{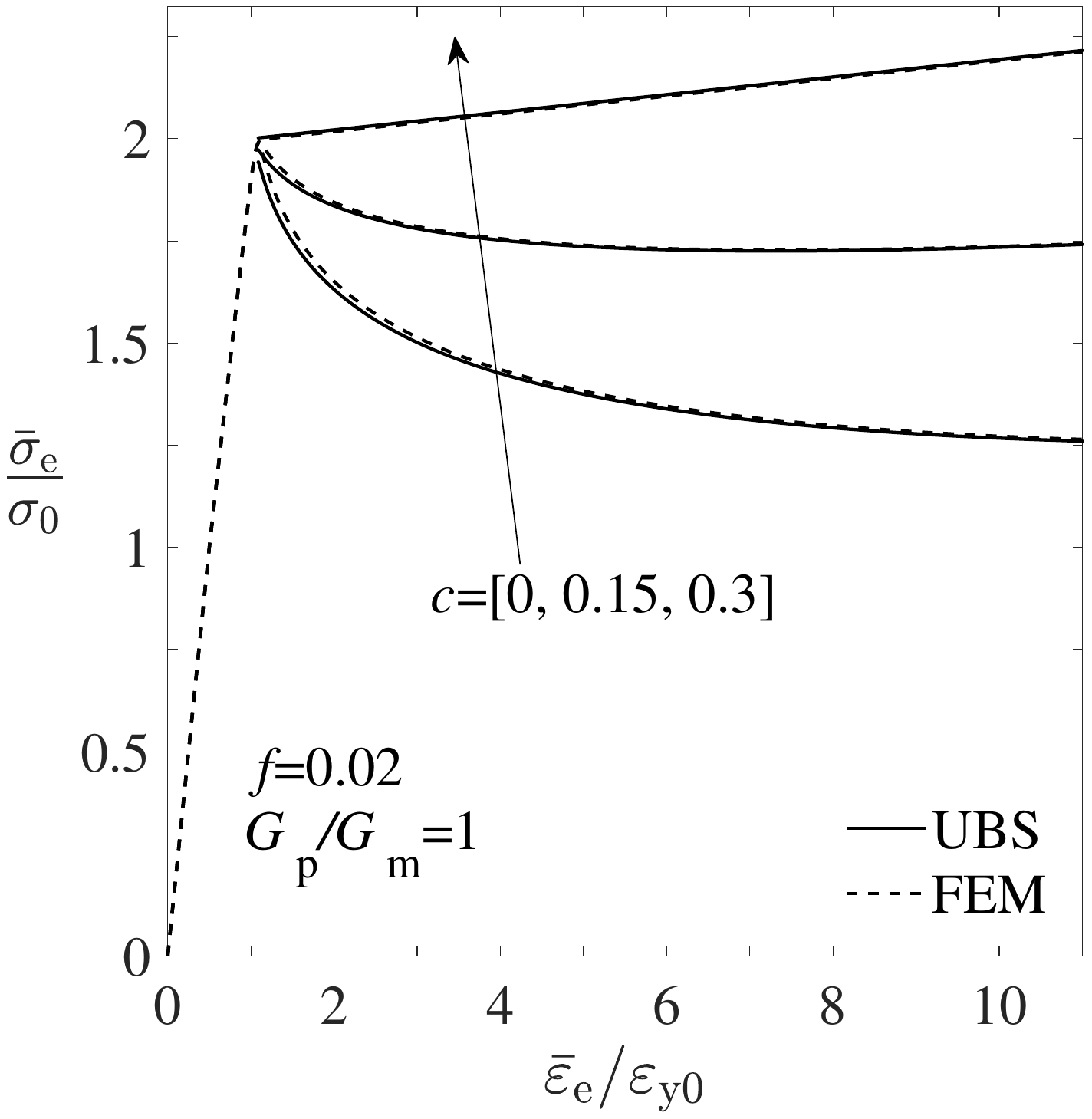}}  \hspace{1 cm}
		\subfloat[]{\includegraphics[width=0.4\textwidth]{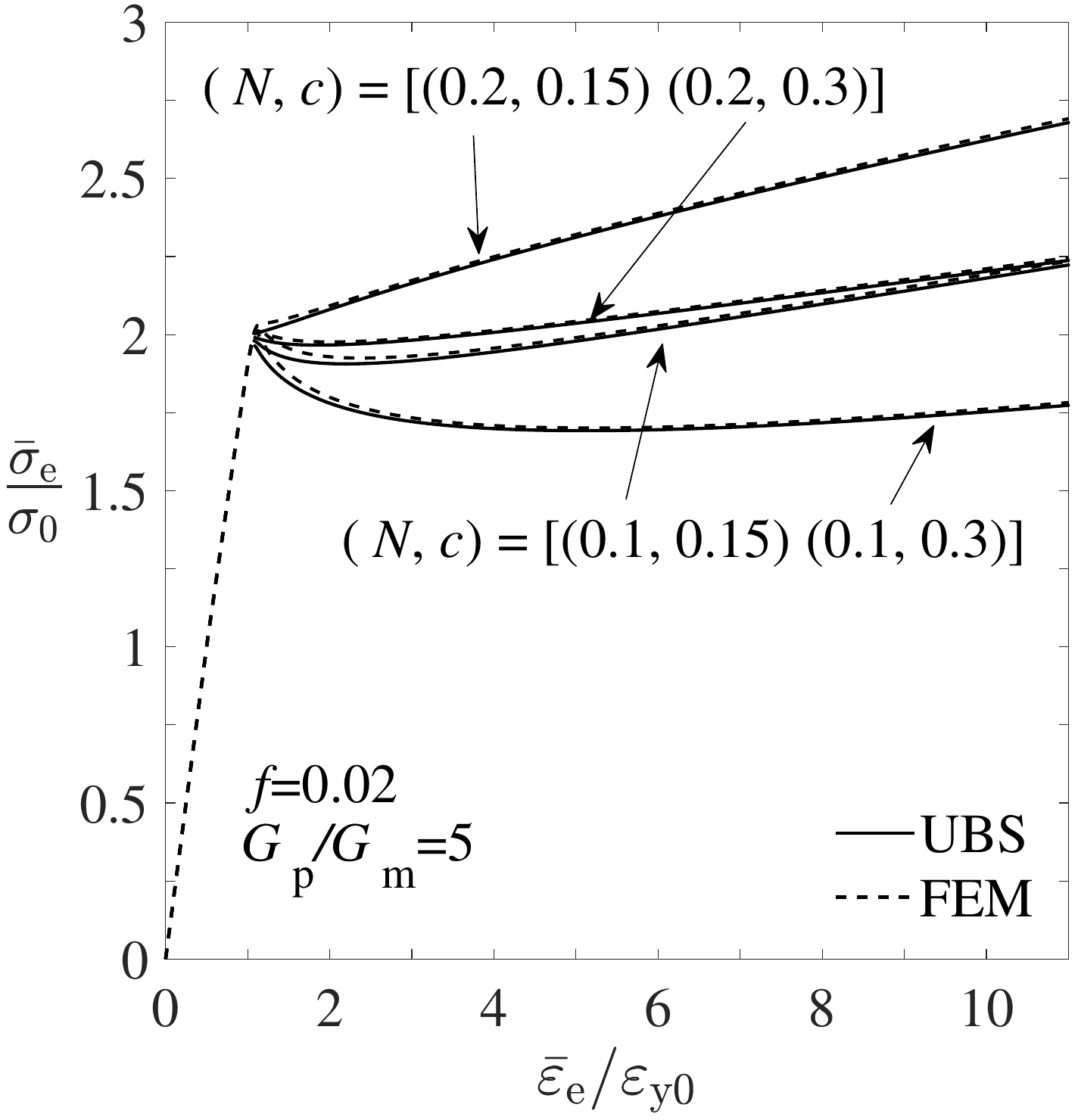}}
		\caption{Comparison of normalized volume average effective stress-strain curves resulting from UBS and FEM analysis of the 2D axisymmetric model, for a variation of the decay parameter $c$ in an initially strong interface ($\alpha=0.99$).	(a) A material with perfectly plastic matrix and 2\% particles with $G_{\rm p}=G_{\rm m}$. (b) Materials with strain hardening matrix, $N = \{0.1, 0.2\}$, and 2\% stiff particles, $G_{\rm p}=5G_{\rm m}$.}
		\label{fig7}
	\end{center}
\end{figure}

Thus, softening is most pronounced in materials with a perfectly plastic matrix ($N=0$), as can be observed from the examples shown in Fig. \ref{fig7}(a). Here, stress-strain curves are plotted for three different interfaces: $c = \{0, 0.15, 0.3\}$ with $\varepsilon_{\Gamma} = 0.88 \sigma_0/3G_{\rm m}$ and volume fraction $f=0.02$ of particles with matching shear modulus ($G_{\rm p}=G_{\rm m}$). The ensuing softening is significant, and here, the linear hardening contribution from particles observed for $c = 0$ is completely wiped out by the decaying interface strength, as seen by the curves with $c>0$.

The interplay between matrix strain hardening and interface decay is shown in Fig. \ref{fig7}(b), where four stress-strain curves generated by combinations of $N = \{0.1, 0.2\}$ and $c = \{0.15, 0.3\}$ ($\varepsilon_{\Gamma} = 0.88 \sigma_0/3G_{\rm m}$) are plotted. Particles have volume fraction $f=0.02$ and a shear modulus $G_{\rm p} = 5G_{\rm m}$. In these examples, only materials with $N=0.1$ exhibit some initial softening at onset of plastic deformation. It is noteworthy that all UBS predictions virtually lie on top of the FEM results in Fig. \ref{fig7}. 

\subsection{Overall accuracy of the upper bound solution} \label{sec4.4}
The volume fraction of particles $f$ and the shear modulus ratio $G_{\rm p}/G_{\rm m}$ play key roles for the flow properties (see Eq. \eref{eqn:UBS}) of the composite, as both affect yield stress and strain hardening. In addition, the yield stress is strongly affected by the ratio $\ell/\bar{a}$. In \cite{faleskog2021analytical} it is suggested that intervals of practical relevance for these parameters are $0.001 \le f \le 0.1$, $1 \le \ell/\bar{a} \le 100$ and $0.1 \le G_{\rm p}/G_{\rm m} \le 10$, where error maps for UBS predictions of yield stress are presented. Here, an analogous FEM analysis was carried out by use of the 2D axisymmetric model to investigate the accuracy of UBS predictions well into the post yield regime. This is accomplished by loading the unit cell until an effective strain $\bar{\varepsilon}_{\rm e*}$, defined where $\bar{\varepsilon}_{\rm e}^{\rm p} = 10\bar{\varepsilon}_{\rm y0}$ in \eref{eqn:UBS_Effectve_se_eps}, was reached. Calculations were performed on materials with a medium strong interface ($\alpha=0.5$) and a micro-hard interface ($\alpha=1$), both with $\omega = 1$. Three modulus mismatch ratios were considered; an under-matched $G_{\rm p}=0.1G_{\rm m}$, a matched $G_{\rm p}=G_{\rm m}$, and an over-matched $G_{\rm p}=10G_{\rm m}$, respectivly.

\begin{figure}[H]
	\begin{center}
		\includegraphics[width=0.95\textwidth]{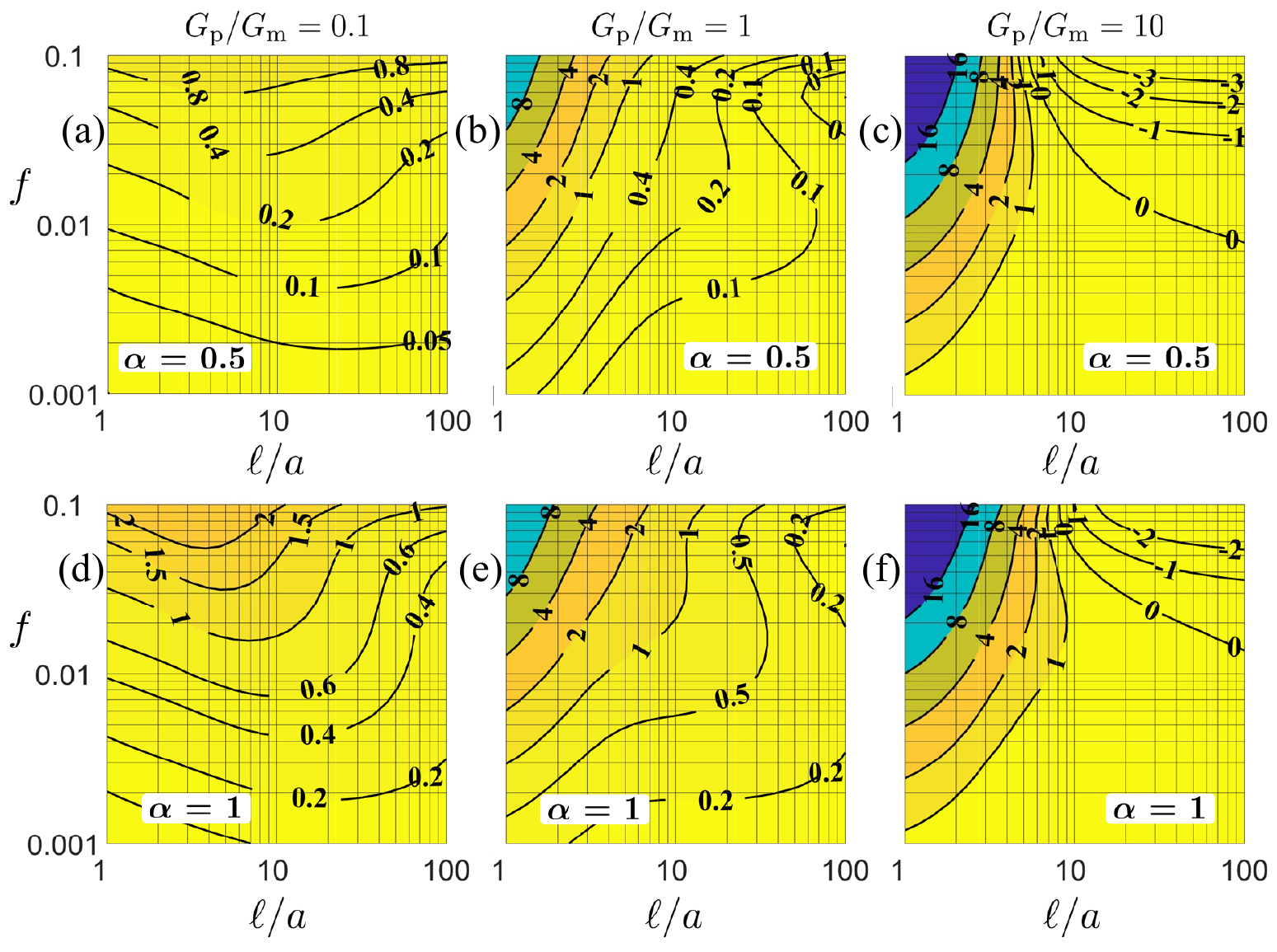}
		\caption{Error maps showing relative error in percent defined as $(\bar{\sigma}_e^{\rm {FEM}}/\sigma^{\rm {UBS}}-1)\times 100$ plotted in $f$ versus $\ell/a$ graphs on log-log scale. In graphs (a)-(c), $\alpha = 0.5$ and in graphs (d)-(f), $\alpha = 1$. The first, second and third column belongs to $G_{\rm p}/G_{\rm m} = 0.1, 1 \, {\rm {and}}\, 10$, respectively.}
		\label{Emaps1}
	\end{center}
\end{figure}

\begin{figure}[H]
	\begin{center}
		\includegraphics[width=0.95\textwidth]{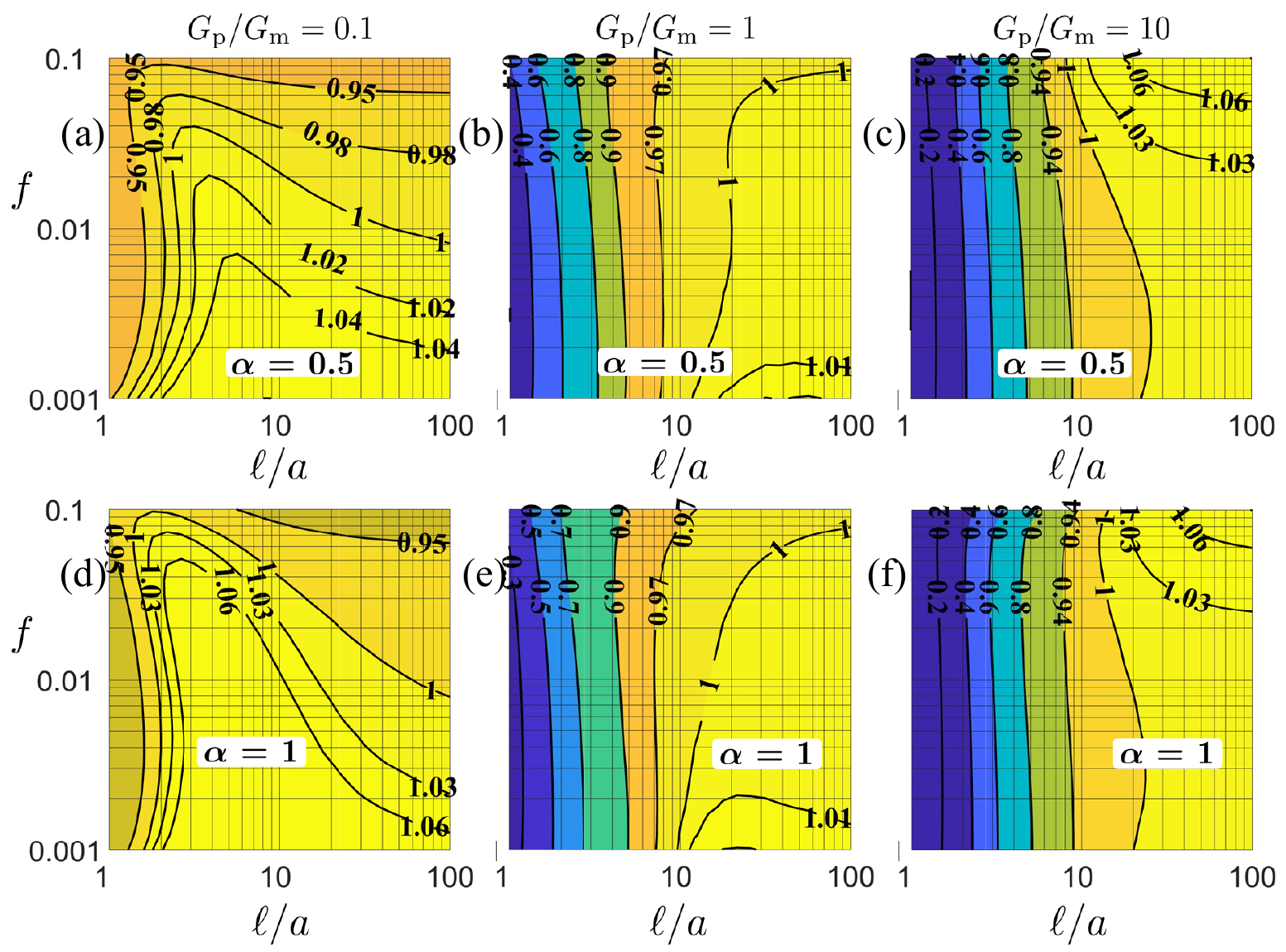}
		\caption{Ratio of the volume average effective stress-strain tangent between FEM and UBS, obtained at $\bar{\varepsilon}_{\rm e} = \bar{\varepsilon}_{\rm e*}$, plotted in $f$ versus $\ell/a$ graphs on log-log scale. In graphs (a)-(c), $\alpha = 0.5$ and in graphs (d)-(f), $\alpha = 1$. The first, second and third column belongs to $G_{\rm p}/G_{\rm m} = 0.1, 1 \, {\rm {and}}\, 10$, respectively.}
		\label{Emaps2}
	\end{center}
\end{figure}

The accuracy of the UBS predictions are presented in Figure \ref{Emaps1} as iso-contours of the relative error (in percent) of the UBS predictions plotted in $\ell/a$ versus $f$ graphs on log-log scale. Thus, the relative error is defined as $[\sigma_{\rm e}^{\rm UBS}/ \sigma_{\rm e}^{\rm FEM} - 1]\times 100$, which was evaluated at $\bar{\varepsilon}_{\rm e} = \bar{\varepsilon}_{\rm e*}$. Error maps for material systems with $\alpha = 0.5$ are shown in Figs. \ref{Emaps1}(a)-(c). As can be seen, the accuracy of the UBS predictions are within $\pm$1\% for materials over a broad range in the parametric space, especially for under-matched systems with $G_{\rm p} < G_{\rm m}$. For matched and over-matched systems ($G_{\rm p} \ge G_{\rm m}$), the accuracy of USB predictions are somewhat impaired at lower ratios of $\ell/a$ when $f$ becomes high. The trends are the same for materials with micro-hard interfaces as noted in Figs. \ref{Emaps1}(d)-(f), alltough the regions in which the relative error is within $\pm$1\% are sligthly reduced.
Insight into the decreasing accuracy noticed in regions with lower ratios of $\ell/a$ in Fig. \ref{Emaps1} may be understood from the observation made in Fig. \ref{fig5}. Even if the UBS is highly accurate at the onset of plastic yielding, as also noted in \cite{faleskog2021analytical}, it will cease to be valid at sufficiently high strain levels where the particle induced strain hardening saturates, as seen in Fig. \ref{fig5}. The strain level marking the transition to saturation depends on $\ell/a$ and $G_{\rm p}/G_{\rm m}$ as discussed above and further elaborated on in Section \ref{Sect.Saturation}. The possible impact of this phenomenon on the error maps in Fig. \ref{Emaps1} can be ascertained by plotting the hardening slope evaluated at $\bar{\varepsilon_{\rm e}} = \bar{\varepsilon}_{\rm e*}$ from FEM calculations, ${\rm d}\bar{\sigma}_{\rm e}^{\rm FEM}/{\rm d}\bar{\varepsilon}_{\rm e}$, normalised by the linear hardening slope obtained from UBS according to Eq. \eref{eqn:UBS_slope}. If this quotient, becomes significantly smaller than one, the hardening has began to saturate. The regions with a relative error significantly larger than one seen in Fig. \ref{Emaps1} are indeed associated with a quotient much less than one, as can be concluded from Fig. \ref{Emaps2}, where this quotient is plotted.

\subsection{3D-validation: Variation of interface strength and shear modulus with particle size} \label{sec4.5}
So far, materials with monodisperse particle distributions have been analysed, but particles in real materials typically exhibit some variation in size. This aspect was addressed by use of the 3D cube model in Fig. \ref{fig3}(b). It was assumed that particles varied according to a geometric series as decribed in Section 4.1, with a ratio between the largest and smallest radius of $a_8/a_1 = 3.16$. Hence, the variation in particle size is significant, and the effective particle radius is according to Eq. \eref{eqn:EffectiveVals}, $\bar{a} = 2.405a_1$ (the geometric mean value is $1.907a_1$). Two sets of analyses were carried out. In the first set, the interface strength depends on particle size, whereas the shear modulus does not. In the second set, the interface strength is independent on particle size and the shear modulus vary with particle size.

\begin{figure}[H]
	\begin{center}
		\subfloat[]{\includegraphics[width=0.4\textwidth]{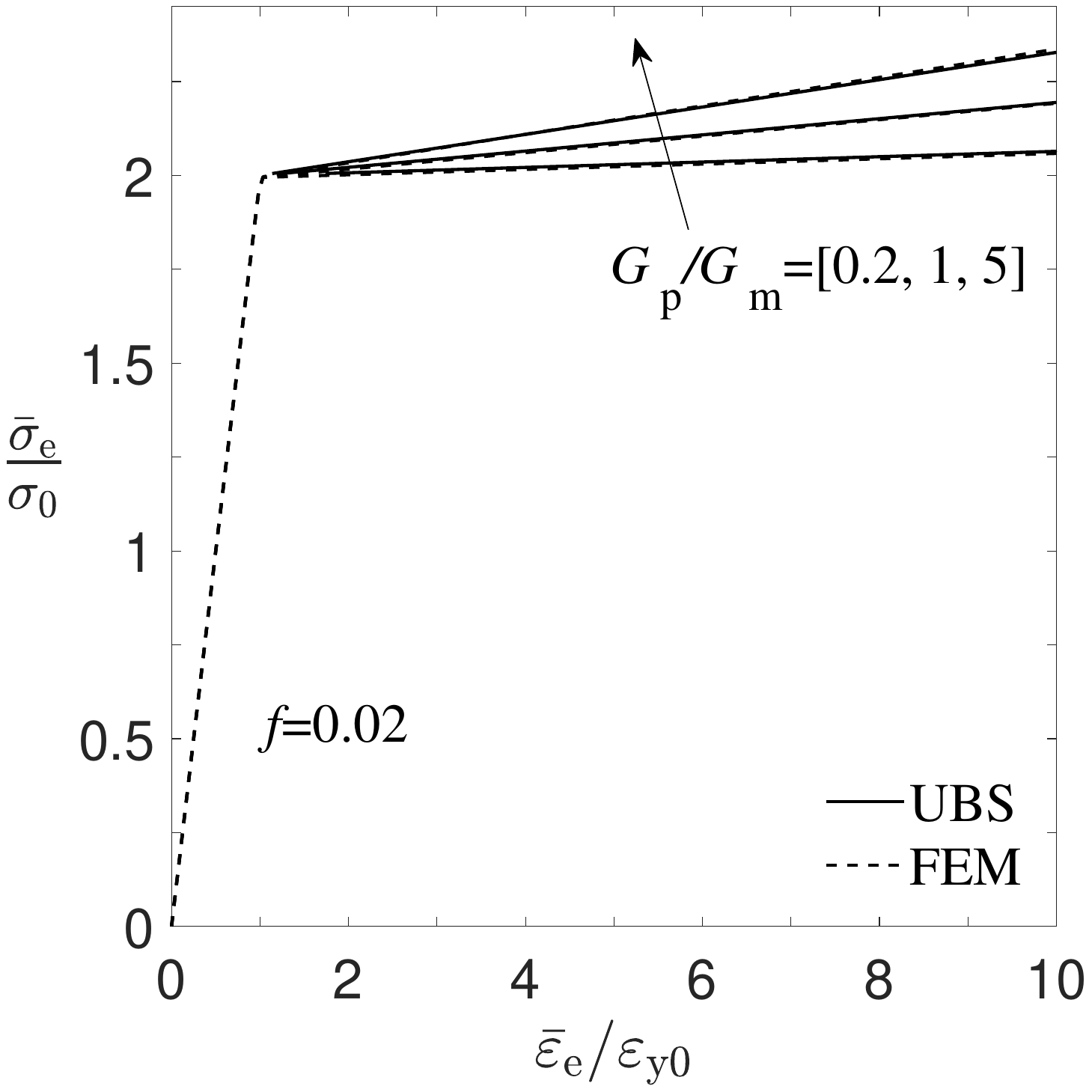}} \hspace{1 cm}
		\subfloat[]{\includegraphics[width=0.4\textwidth]{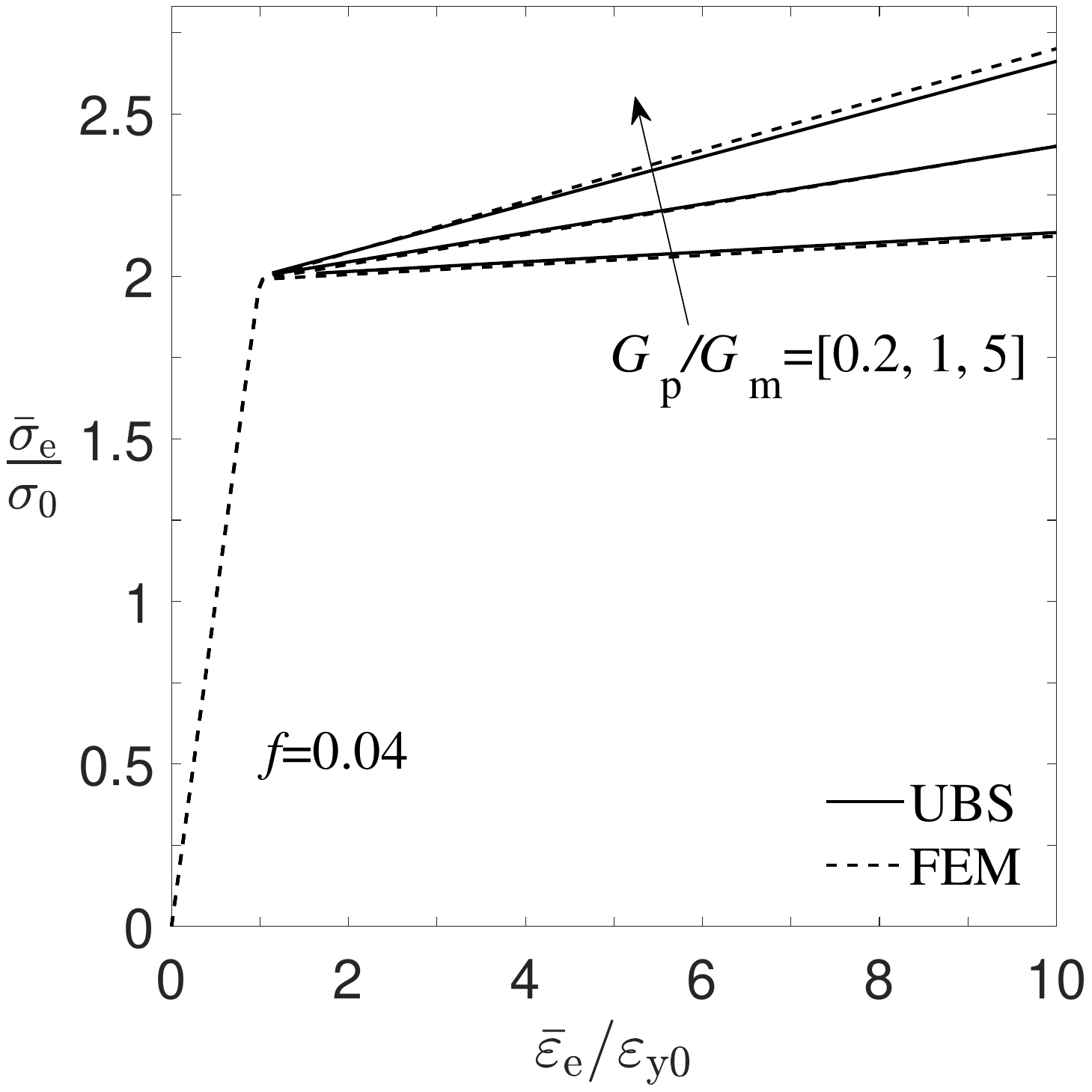}}\\
		\subfloat[]{\includegraphics[width=0.4\textwidth]{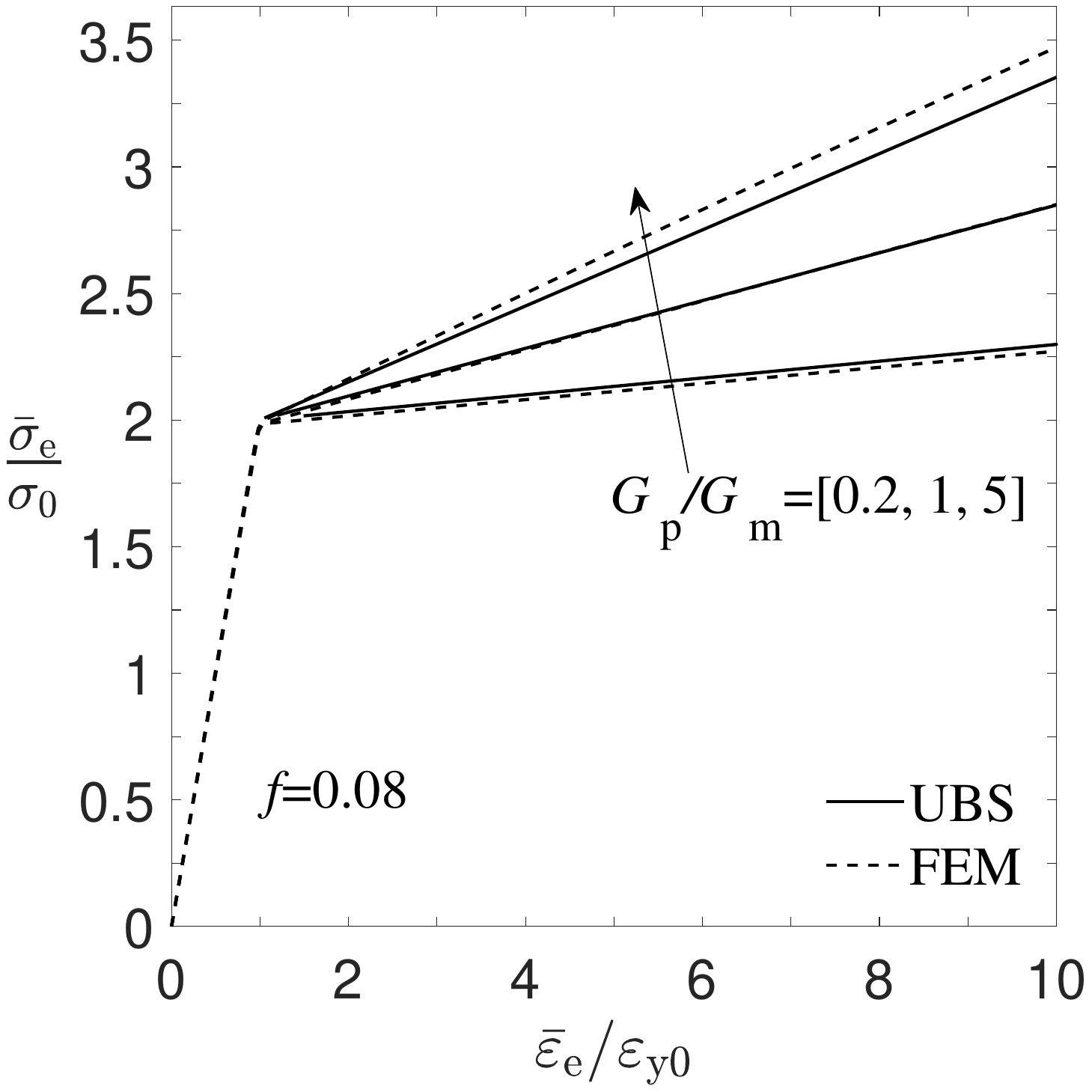}}
		\caption{Comparison of normalized volume average effective stress-strain curves resulting from UBS and FEM analysis of the 3D cubic model, for materials with a perfectly plastic matrix and interfaces with linearly varying strenth $\alpha(a_1) = 0.98, ..., \alpha(a_8)=0.31$. Particle volume fractions; (a) 2\%, (b) 4\%, and (c) 8\%.}
		\label{fig10}
	\end{center}
\end{figure}

In the first, a linear variation of interface strength with particle size was considered, defined as $\alpha_i = \alpha_1 [1-(a_i-a_1)/a_8]$, where $\alpha_1 = 0.98$. Moreover, interfaces were assumed to be independent on the accumulated plastic strain ($\omega = 1$). The effective interface strength of the composite is then given by Eq. \eref{eqn:EffectiveVals} as $\bar{\alpha} = 0.545$. Results from the the first set are shown in Fig. \ref{fig10}, where effective stress-strain curves from UBS (Eqs. \eref{eqn:UBS}-\eref{eqn:EffectiveVals},\eref{eqn:UBS_Effectve_se_eps}) are compared with the outcome from the FEM analyses. Three graphs are presented, pertaining to $f = {0.02, 0.04, 0.08}$. In each graph, three different shear mudulii are considered for the particles: $G_{\rm p}/G_{\rm m} = \{0.2, 1, 5\}$. As noticed, the model predictions agree very well with the FEM results for the lower $f$ values, but at the highest volume fraction of particles ($f = 0.08$), some deviations are noticed for stiff particles ($G_{\rm p} = 5G_{\rm m}$).

\begin{figure}[H]
	\begin{center}
		\includegraphics[width=0.4\textwidth]{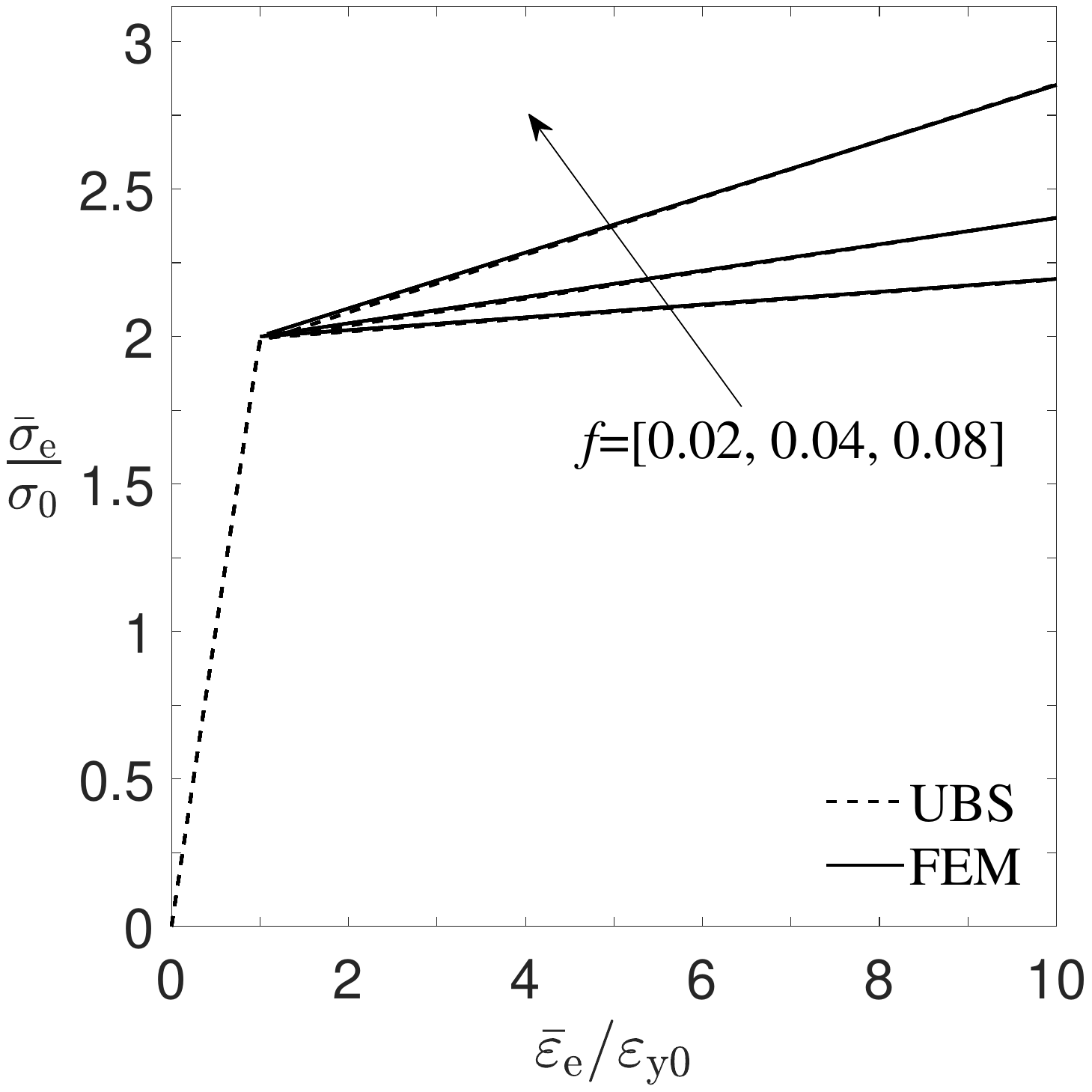}
		\caption{Comparison of normalized volume average effective stress-strain curves resulting from UBS and FEM analysis of the 3D cubic model, for $f=\{0.02, 0.04, 0.08\}$. The matrix is perfectly plastic and the shear modulus vary with particle size.}
		\label{fig11}
	\end{center}
\end{figure}

In the second set, the interface strength of all particles was chosen to be equal to the effective value resulting in set 1, \ie, $\alpha = 0.545$, and the shear modulus was taken to vary linearly with particle size as $G_{{\rm p}i} = [(1-\xi_i)G_{\rm p1}+ \xi_i G_{\rm p8}]$, where $\xi_i=(a_i-a_1)/(a_8-a_1)$. Ratios $G_{\rm p1}/G_{\rm p8}$ equal to $1/4$, $1/2$, $2$ and $4$ were considered, and the relation to the matrix shear modulus was chosen such that $\bar{\Gamma} = 1$ by use of \eref{eqn:EshelbyParam} and \eref{eqn:EffectiveVals}. As both $\bar{\Gamma}$ and $\bar{G}$ are evaluated in the same manner, cf. Eq. \eref{eqn:EffectiveVals}, both strengthening and strain hardening are expected to be independent of ratio $G_{\rm p1}/G_{\rm p8}$ under this circumstance, and only depend on $f$. Moreover, with $\bar{\Gamma}=1$, Eq. \eref{eqn:EffG_composite} yields the result $G_{\rm{eff}}=G_{\rm m}$. The results can be observed in Fig. \ref{fig11}, where effective stress-strain curves are plotted for volume fractions $f = {0.02, 0.04, 0.08}$, and clearly show that the hardening slopes are independent of $G_{\rm p1}/G_{\rm p8}$. Note that each $f$ value is represented by four curves pertaining to UBS and four curves belonging to the FEM analysis, and that all curves collapse on top of each other.

%

\subsection{Saturation of particle induced strain hardening}  \label{Sect.Saturation}
\noindent The results shown in Fig. \ref{fig5}(a) strongly suggests that in the current model, additional hardening due to particles that remains elastic will saturate at a certain level of overall plastic straining. This behaviour of hardening is in line with what is commonly observed in alloys containing hard second phase particles, \eg \cite{fribourg2011microstructure,lloyd1994particle,myhr2010combined,fazeli2008modeling}. Qualitatively, it resembles stages II and III in classic work hardening theory for polycrystals.

\begin{figure}[H]
	\begin{center}
		\includegraphics[width=0.4\textwidth]{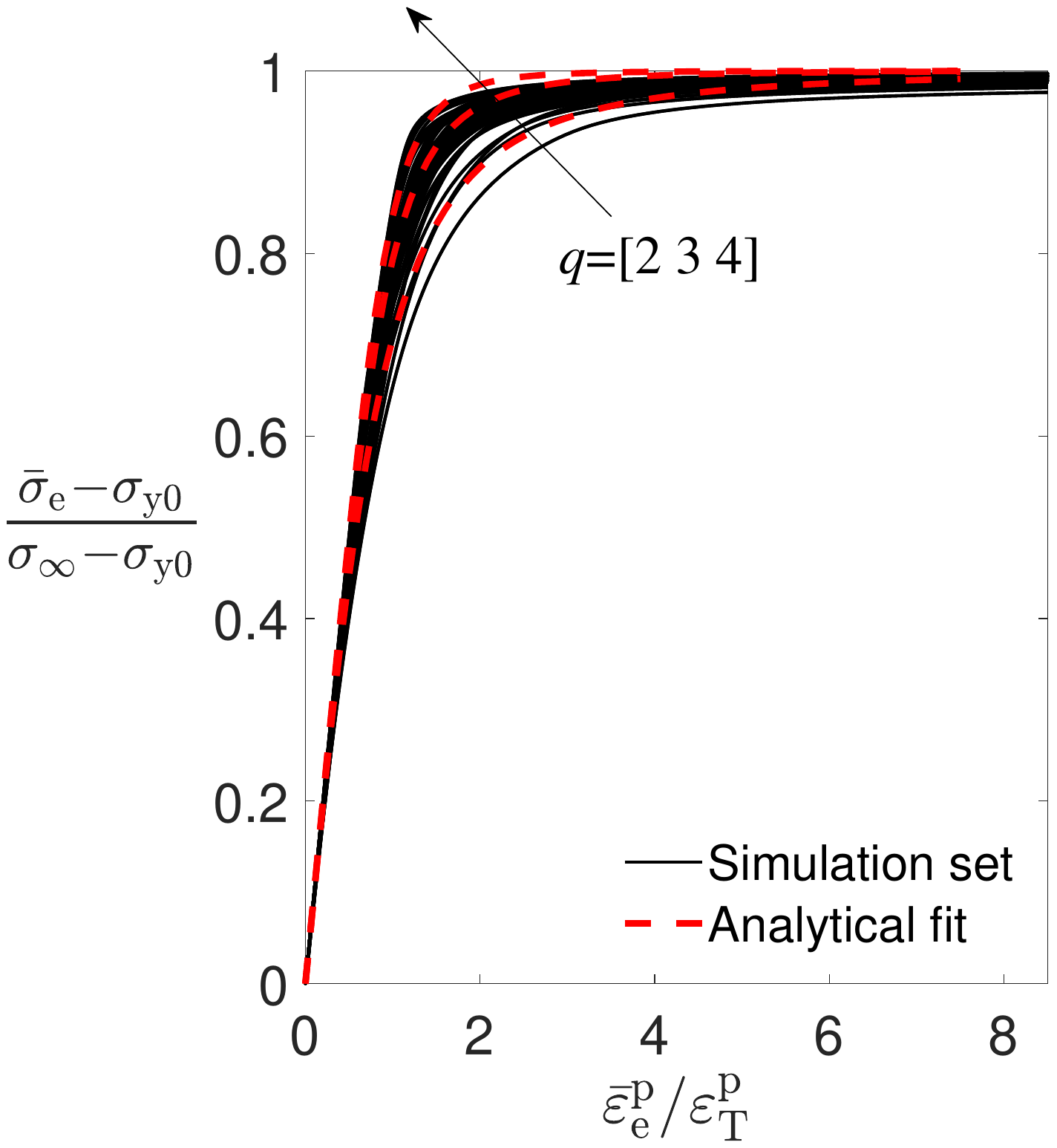}
		\caption{Normalized strain hardening plotted vs. effective plastic strain normalized the transition strain introduced in Eq. \eref{transitionfcn}. Dashed lines in colour red represents Eq. \eref{transitionfcn} evaluated for $q=2,3,4$.}
		\label{fig12}
	\end{center}
\end{figure}

As is noteworthy, the saturation of stress observed in Fig. \ref{fig5}(b) is also followed by a saturation of plastic strain at the particle/matrix interface, see Fig. \ref{fig5}(c). Physically, this can be tied to the process of plastic relaxation, by which, the accumulation rate of dislocations surrounding particles is reduced, see \eg \cite{myhr2010combined,fribourg2011microstructure}. The transition from linear hardening to saturation will here be addressed in a purely phenomenological manner. As such, the following function is introduced:
\begin{equation} \label{transitionfcn}
	\phi_{\rm T} = \big( 1+ (\bar{\varepsilon}_e^p / \varepsilon_T )^q   \big)^{-1/q}.
\end{equation}
In Eq. (\ref{transitionfcn}), $\varepsilon_T$ corresponds to the intersection between the linear slope in region I, Eqs. \eref{eqn:UBS}-\eref{eqn:UBS_peeq}, and a numerical estimate of the saturation stress level $\sigma_{\infty}$ (see Fig. \ref{fig5}(b)). Moreover, $q$ is an exponent in the range 2 to 4.
The relevance of this function can be seen in Fig. \ref{fig12}, where a comparison is made between Eq. (\ref{transitionfcn}) and the results from a large amount of numerical simulations using the axi-symmetric model with an isotropic distribution of particles of radius $a$.
\noindent Using the simulation results underpinning Fig. \ref{fig12}, it is also possible to find an explicit expression for $\varepsilon_T$ on the following form:
\begin{equation}
	\label{eqn:epsT}
	\varepsilon_T\approx K\varepsilon_{0}\frac{\ell}{a}(1+\frac{G_{\rm m}}{G_{\rm p}}),
\end{equation}
where $\varepsilon_{0} = \sigma_0/E_{\rm m}$ refers to the yield strain of the matrix material. For spherical particles distributed as shown in Fig. \ref{fig3}(a) with $H=R$, $K \approx  15$. Note that the simple relation \eref{eqn:epsT} is qualitatively supported by the trend observed in Fig. \ref{Emaps2}, where the decrease in the tangent of the effective stress-strain curves obtained from FEM seems to scale with $\ell/a$ with a negligible influence of $f$. Furtermore, in \cite{faleskog2021analytical}, a perturbation analysis based on non-dimensional forms of the governing equations \eref{eqn:kinematic}-\eref{eqn:intfstr} was carried out. The analysis was based on the assumption of $f \ll 1$ and $a/\ell \ll 1$, following the notation in the present paper. A closed form solution for the initial yield stress was in this way possible to derive. If the corresponding analysis is extended to strain hardening, it can be concluded that the perturbation assumptions break down for a plastic strain of the order of $\varepsilon_0 \ell/a$. This is confirmed by the estimate expressed in Eq. \eref{eqn:epsT}.

\section{Strengthening contribution from particles subjected to shearing}
The strengthening mechanism addressed so far is associated with dislocations that by-pass particles at the onset of macroscopic plastic straining. However, if the size of a particle located on a slip plane is sufficiently small it will be sheared by dislocations instead. In general this shearing mechanism will primarily affect the yield strength of the material. The obstacle strength from a sheared particle depends on size in some manner, and several models have been proposed for this dependency, see discussions in \cite{ardell1985precipitation} and \cite{reppich1993materials}. To account for this additional strength contribution, the model proposed in \cite{deschamps1998influence} will be employed and incorporated into the upper bound solution.

A distribution of particle size may be represented by a discrete or continuous random variable described by an appropriate distribution function. Assume that particles will be sheared if belonging to a population of $N_{\rm s}$ particles with a radius less than or equal to $a_{\rm c}$. The remaining particles will be by-passed by dislocations and belongs to a population of $N_{\rm b}$ particles. The total number of particles in $V$ is then $N_{\rm p} = N_{\rm s} + N_{\rm b}$.

In \cite{deschamps1998influence}, it is assumed that the obstacle strength of a sheared particle depends linearly on the particle size. The contribution to strengthening then hinges on assumptions made for mean particle spacing, which in turn depends on obstacle strength. Three models for the contribution to yield strength will be considered here; a mean spacing proposed by \cite{friedel1964} (weak obstacles):
\begin{equation} \label{eqn:ShearFriedel}
	\sigma_{\rm s} = C_{\rm F} \frac{a_{\rm s}^{3/2}}{a_{0}} f^{1/2},
\end{equation}
a mean spacing according to the Mott-Labusch model, see \cite{labusch1970statistical} and \cite{neuhauser2006solid} (strong obstacles):
\begin{equation} \label{eqn:ShearLabusch}
	\sigma_{\rm s} = C_{\rm L} \frac{a_{\rm s}^{4/3}}{a_{0}} f^{2/3},
\end{equation}
and finally, a mean spacing given by \cite{kocks1975models} (strong obstacles): \begin{equation} \label{eqn:ShearKocks}
	\sigma_{\rm s} = C_{\rm K} \frac{a_{\rm s}}{a_{0}} f^{1/2}.
\end{equation}
In Eqs. \eref{eqn:ShearFriedel}-\eref{eqn:ShearKocks}, $a_0$ denotes the expected mean particle size of all the $N_{\rm p}$ particles, and $a_s$ refers to a representative size of the sheared particles in population $N_{\rm s}$. These quatities are evaluated as
\begin{equation} \label{eqn:shear_radius}
	a_0 = \frac{1}{N_{\rm p}} \sum_{i=1}^{N_{\rm p}} a_i = \int_{0}^{\infty} a \cdot{\rm pdf}(a) {\rm d}a, \quad a_{\rm s} = \frac{1}{N} \sum_{i=1}^{N_{\rm s}} a_i = \int_{0}^{a_{\rm c}} a \cdot{\rm pdf}(a) {\rm d}a.
\end{equation}

The contribution from sheared particles to the overall yield strength will be added to the matrix material in an ad hoc manner. Eq. \eref{eqn:UBS} modified to include sheared particles is then taken as
\begin{equation} \label{eqn:UBSwithShear}
	\sigma_{\rm e} \le \frac{(1-f_{\rm b})}{(1-\bar{\Gamma}f_{\rm b})} \left[ \sigma_{\rm m}(\varepsilon_{\rm e}^{\rm p}) + \sigma_{\rm s} \right] + \frac{3 f_{\rm b} \sigma_0 \bar{\alpha} \omega}{(1-\bar{\Gamma}f_{\rm b})}\frac{\ell}{\bar{a}} + \frac{3 \bar{G} f_{\rm b} }{(1-\bar{\Gamma}f_{\rm b})} \varepsilon_{\rm e}^{\rm p}.
\end{equation}
Recall that $\varepsilon_{\rm e}^{\rm p} = \bar{\varepsilon}_{\rm e}^{\rm p}/(1-f)$, and note that $f_{\rm b}$ replaces $f$ in Eq. \eref{eqn:UBS} and that the effective properties are evaluated in the same manner as described in Section 3 by replacing $N_{\rm p}$ with $N_{\rm b}$. Eq. \eref{eqn:UBSwithShear} serves an extension of Eq. \eref{eqn:UBS} where the strengthening effect of shearable particles is accounted for. Formally the model has five parameters: $\bar{\alpha}\ell$, $a_c$, $C$, $c$, $\varepsilon_{\Gamma}$ that need to be calibrated from experimental results. However, $C$ should not be viewed as a free parameter. For a material with a monodispersion of particles with volume fraction $f$, the contribution from shearable particles ($\sigma_{\rm s}$) should, at the onset of macroscopic yield, be equal to the contribution from impenetrable particles when their size equals the critical radius $a_{\rm c}$. This reflects the notion that the obstacle strength from shearable and non-shearable particles are assumed to be the same at the critical radius $a_{\rm c}$, cf. \cite{deschamps1998influence}. In the current model, this implies that 
\begin{equation}
	\label{eqn:DetermineC}
	\sigma_s=\frac{3 f \sigma_0}{(1-\bar{\Gamma}f)}\frac{\ell}{a_c},
\end{equation}
where $C$ enters the left hand side of Eq. \eref{eqn:DetermineC} in a manner dependent on the choice of $\sigma_{\rm s}$ from Eqs. \eref{eqn:ShearFriedel}-\eref{eqn:ShearKocks}. In practice, materials typically contain particles exhibbiting a size distribution, $f$ in Eq. \eref{eqn:DetermineC} must then be chosen judiciously.

\section{Experimental validation of upper bound solution extended with shearing particles}
\begin{figure}[H]
	\begin{center}
		\subfloat[]{\includegraphics[width=0.4\textwidth]{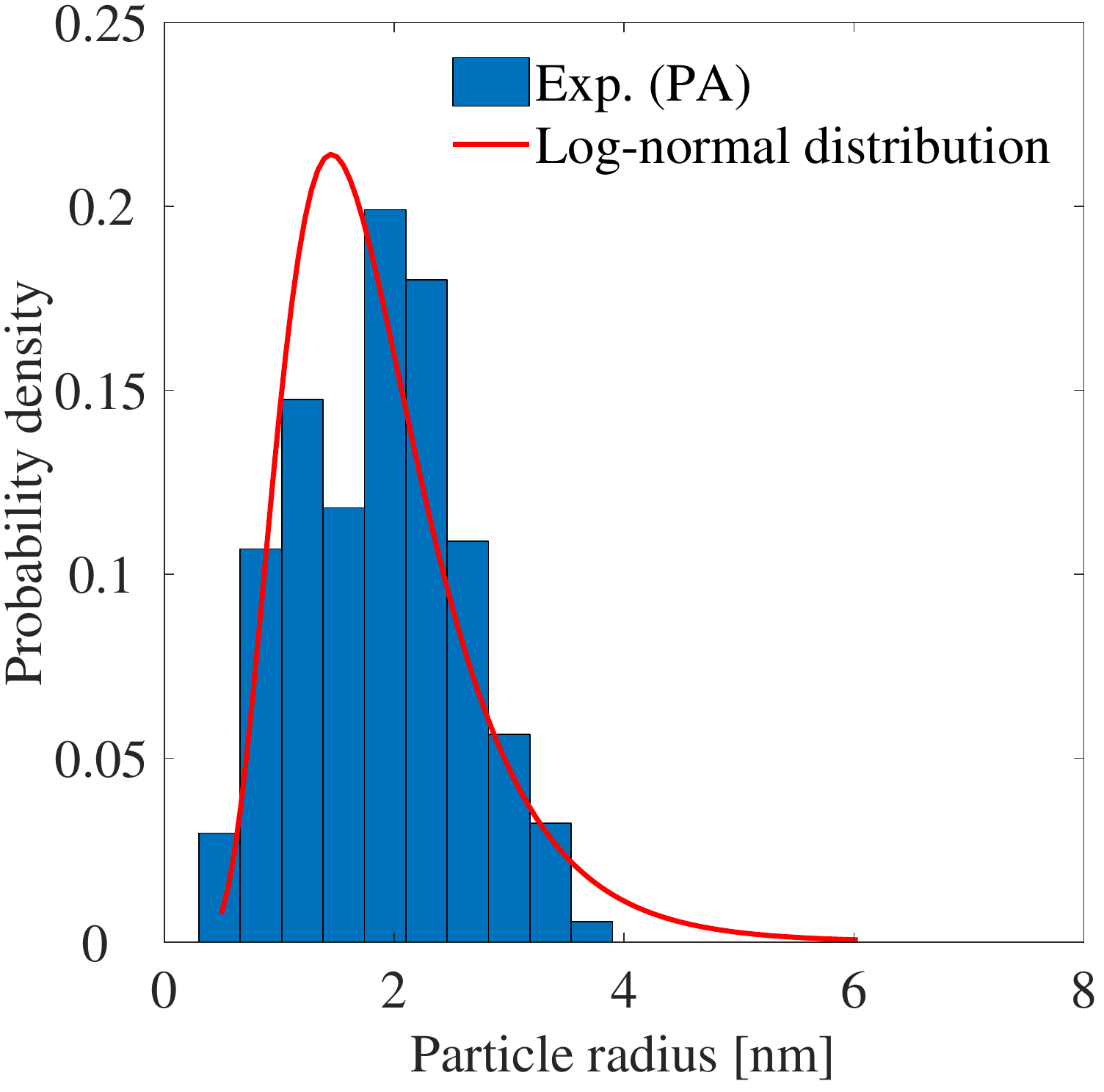}} \hspace{1cm}
		\subfloat[]{\includegraphics[width=0.4\textwidth]{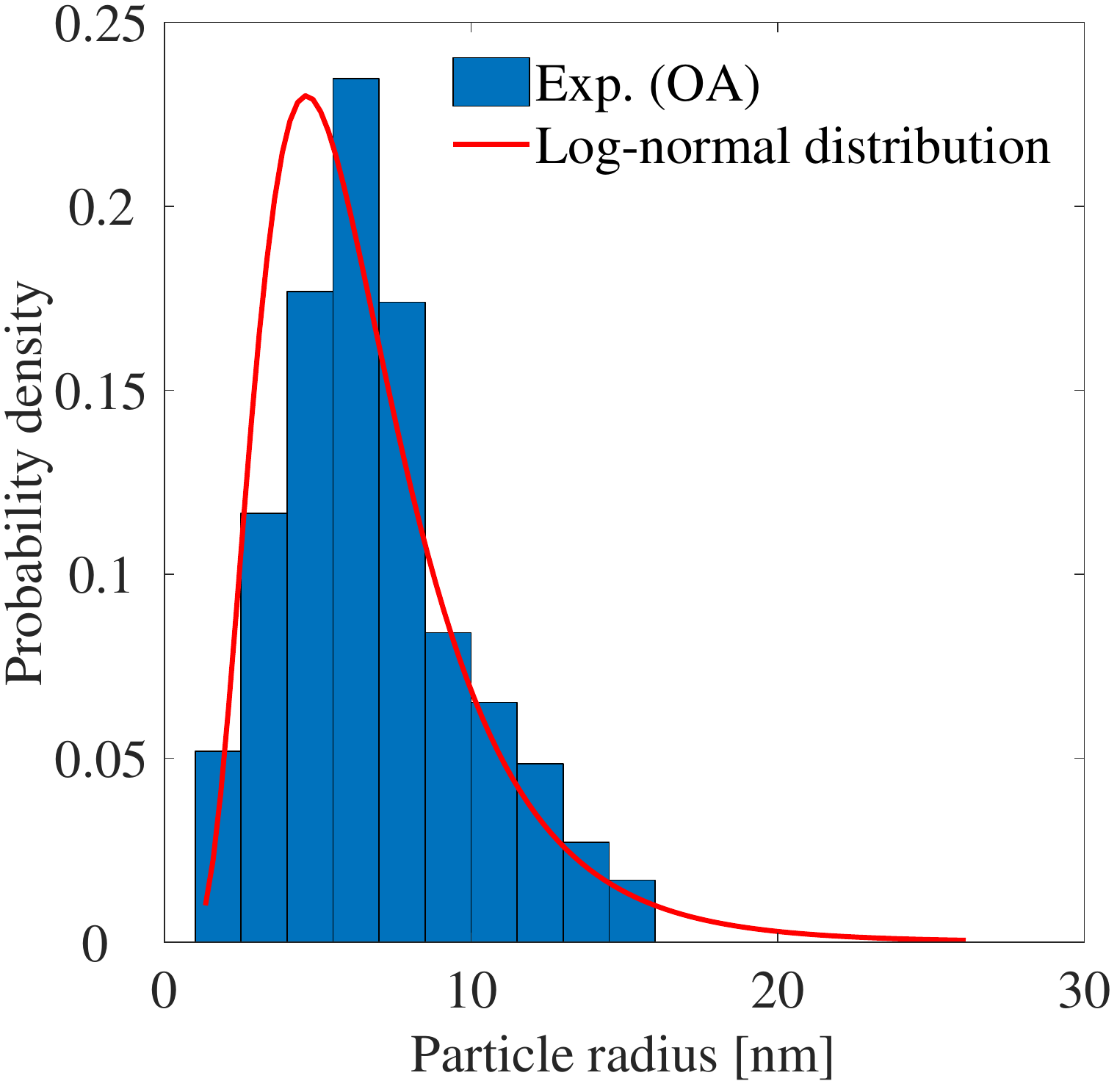}}
		\caption{Log-normal probability distributions of precipitate radii fitted against the data from \cite{fazeli2008role} on an age hardened $Al-2.8wt\%Mg-0.16wt\%Sc$ alloy. (a) Peak aged sample (b) Over aged sample.}
		\label{fig13}
	\end{center}
\end{figure}

\begin{figure}[H]
	\begin{center}
		\subfloat[]{\includegraphics[width=0.4\textwidth]{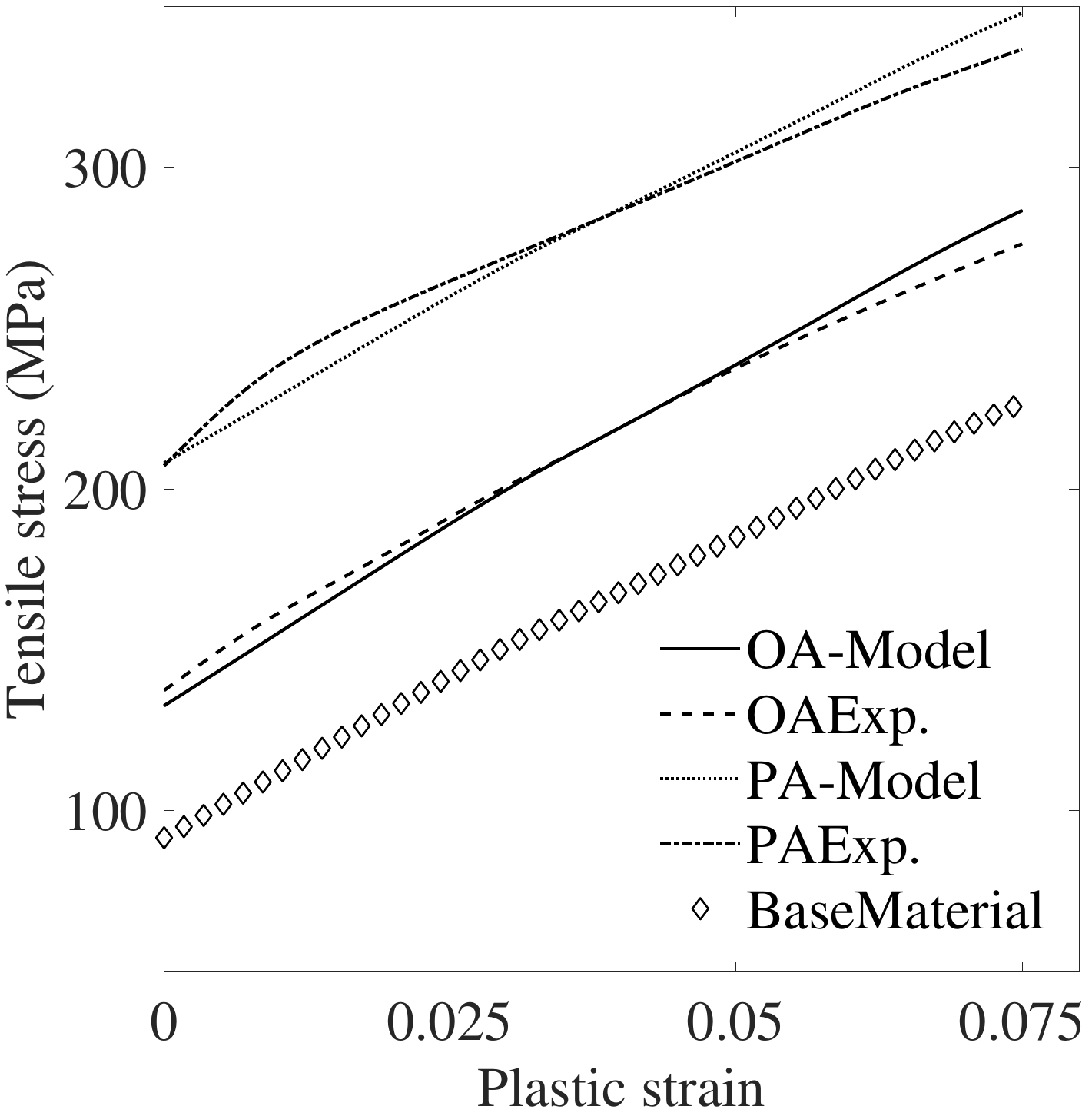}} \hspace{1cm}
		\subfloat[]{\includegraphics[width=0.4\textwidth]{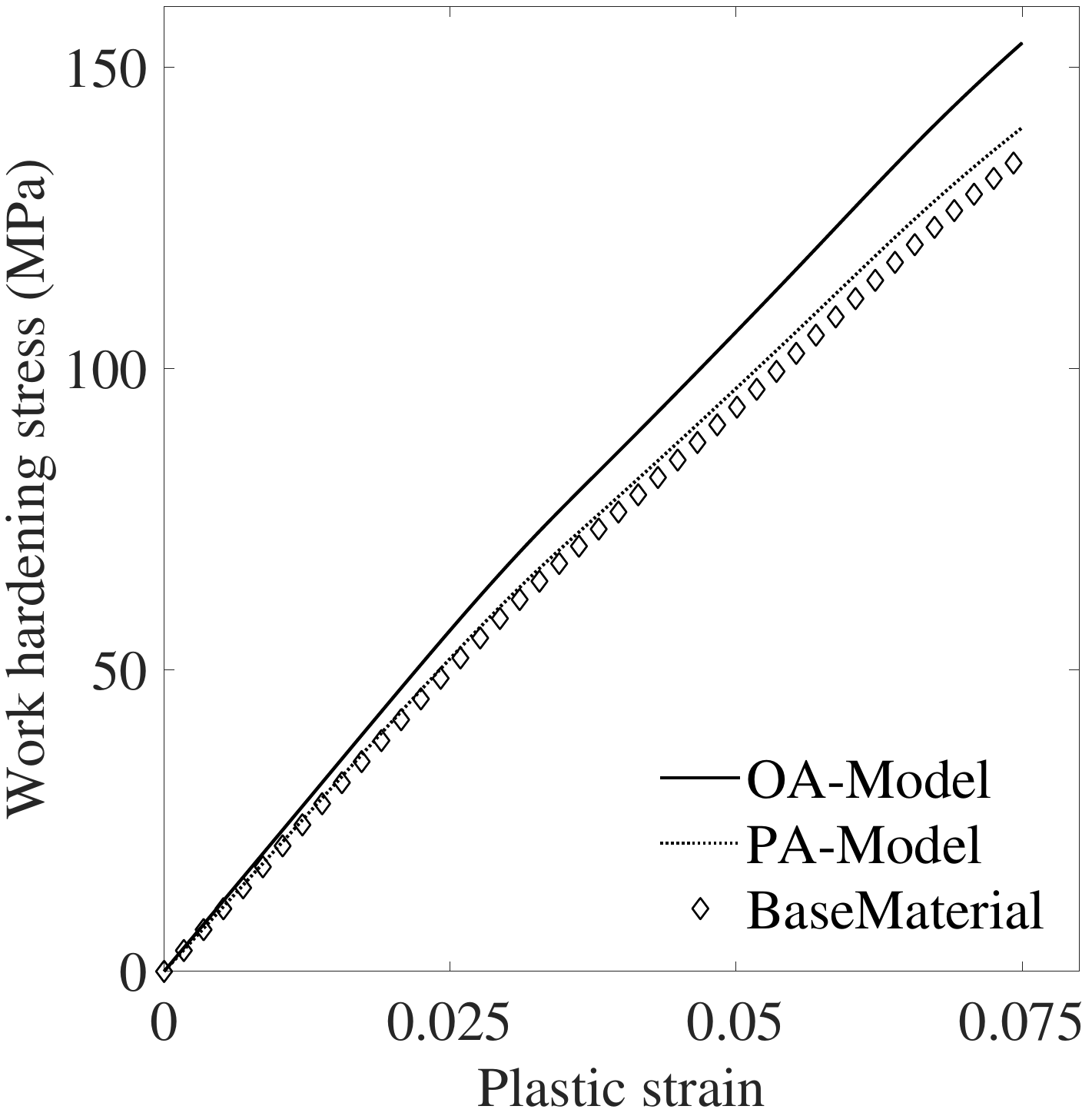}}
		\caption{Comparison between UBS-model and experimental data from \cite{fazeli2008modeling,fazeli2008role}. a) UBS-model fitted on experimetal data b) Increase in work hardening relative to base material as predicted by the model.}
		\label{fig14}
	\end{center}
\end{figure}

\noindent To verify the proposed model, a comparison with uni-axial tensile data from \cite{fazeli2008modeling} on an aged hardened $Al-2.8wt\%Mg-0.16wt\%Sc$ alloy was performed. The tensile data corresponds to a peak hardened sample (PA) with a volume fraction $f_{\rm PA}=0.45\%$, of spherical $Al_3Sc$ precipitates with average radius $1.8$ nm, and an over aged sample (OA) with a precipitate volume fraction $f_{\rm OA}=0.37\%$ and average radius equal to $6.4$ nm. Flow properties for the $Al$ matrix material was taken from \cite{jobba2015flow}. The tensile tests performed in \cite{fazeli2008modeling} and \cite{jobba2015flow} were conducted at temperature 77 K. This choice of test temperature was chosen to minimize additional growth of precipitates during straining (dynamic strain ageing), cf. \cite{fazeli2008modeling}. All input parameters used for the $Al-2.8wt\%Mg-0.16wt\%Sc$ alloy are summarized in Table \ref{T1}.
\begin{table} [H]
	\begin{center}		
		\begin{tabular}{|c|c|c|}
			\hline
			$E_m$ &$ 75$ GPa & Youngs modulus of matrix material  \\
			\hline
			$E_p $&$ 165$ GPa  & Youngs modulus of particle \\
			\hline
			$\nu_m $&$ 0.34 $ & Poisson ratio of matrix material \\
			\hline
			$\nu_p $&$ 0.2 $ & Poisson ratio of particle \\
			\hline
			$f_{OA} $&$ 0.37\% $ & Volume fraction of particles (Over aged sample) \\
			\hline
			$f_{PA} $&$ 0.45\% $ & Volume fraction of particles (Peak aged sample) \\
			\hline
			$\sigma_{0}$ & $90$ MPa  & Yield stress of matrix material \\
			\hline
		\end{tabular}\caption{Key input parameters used for the model comparison against experimental data.}		
		\label{T1}
	\end{center}
\end{table}
The precipitate size distributions with corresponding log-normal fits, as illustrated in Fig. \ref{fig13}, were created from the data in \cite{fazeli2008role} using the Matlab functions\footnote{Copyright 1990-2015 The MathWorks, Inc.} \texttt{\textbf{histfit}} and \texttt{\textbf{fitdist}}. The log-normal probability density fits were then used instead of the raw data when calculating quantities dependent on precipitate size. Moreover, the yield stress contribution from shearable particles was included using Eq. \eref{eqn:ShearKocks}, \textit{i.e.} the model based on Kocks statistics.
Illustrated in Fig. \ref{fig14} is the outcome of model comparison against experimental data. Formally, the model  has four tunable parameters ($\alpha \ell$, $a_c$, $c$, $\varepsilon_{\Gamma}$), however, with the choice $\omega=1$, \ie,  a weak (negligible) logarithmic influence on the interfacial energy, the tunable parameters reduce to only $\alpha l$ and $a_c$. The model was fitted to the experimental data using the Matlab function \texttt{\textbf{lsqcurvefit}}. Since the proposed model can only operate within a small strain regime, the fitting algorithm was carried out between 0.2\% and 7.5\% plastic strain.  The resulting fitting parameters are summarized in Table \ref{T2}. As seen in Fig. \ref{fig14}(a), the model can be fitted against the experimental data quite well. Due to the low volume fractions, the work hardening contribution from particles is not as easy to see compared to the effect on yield stress. However, as illustrated in Fig. \ref{fig14}(b), where the differences in yield stress have been eliminated, a significant contribution to work hardening is produced by the model in the case of an over aged sample, reflecting the increased storage of dislocations around non shearable particles. On the contrary, the model for the peak aged sample only shows a very small deviation from the base material in terms of work hardening, suggesting that most particles are shearable. The latter is corroborated by  $a_c$ being equal to $4.07$ nm. As a comparison, the authors in \cite{fazeli2008modeling} obtained a value of $a_c$ equal to $3.7$ nm. The initially high work hardening seen in Fig. \ref{fig14}(a) for the PA sample can not be captured by the current model using a constant $\bar{\alpha}$. However, since the physical implications of a diminishing interface strength as predicted by Eq. \ref{eqn:intfstr}, is still rather unclear to the authors, a constant value, equivalent to putting $c=0$, has been adopted. Moreover, the obtained value of $\bar{\alpha}l=0.33 \mu {\rm m}$ is in line with what was found in \cite{Asgharzadeh2021a}.
\begin{table}
	\begin{center}		
		\begin{tabular}{|c|c|c|}
			\hline
			$a_c$ &$4.07$ nm & Shearing/bypassing transition radius  \\
			\hline
			$\bar{\alpha}l $&$0.33$ $\mu m$  & Effective length scale \\
			\hline
		\end{tabular}\caption{Resulting values for the model fitting parameters.}		
		\label{T2}
	\end{center}
\end{table}

\section{Summary and concluding remarks}
The SGP-model of materials reinforced with impenetrable particles from \cite{Asgharzadeh2021a} has been refined with a more physically based interface model and the upper bound solution (UBS) for yield strength derived in \cite{faleskog2021analytical} augmented to include post yield hardening. In the absence of interfacial softening, the model predicts a linear hardening contribution from non shearable particles. This result is in line with the classic continuum models in \cite{tanaka1970hardening,brown1971work}. Numerical validation of the model was conducted using a 2D axi-symmetric- and a 3D-cubic FEM model. In the 3D model, the particles were allowed to vary in both size and elastic properties. As a final validation, the UBS-model was tested against experimental data. The numerical simulations show that the UBS-model gives a highly accurate approximation to corresponding FEM solutions within a reasonably wide parameter space (see error maps in Figs. \ref{Emaps1}-\ref{Emaps2}). The reason for the breakdown of the UBS-model is twofold. Firstly, as the volume fraction increases, the particles can no longer be regarded as isolated and the Eshelby formalism loses its validity. Secondly, at a certain level of stress, the plastic strain field becomes increasingly non uniform and significant gradients throughout the matrix follows. Since this violates the underlying assumption of a uniform plastic strain, a significant deviation from the FEM solutions follows. This effect appears however at quite large plastic strains, typically an order of magnitude larger (or more) than the composite yield strain $\varepsilon_{\text{y}0}$. For lower strains, the upper bound solution shows excellent agreement with corresponding FEM solutions.

An approximate expression for the strain at which the plastic strain field starts to show significant non uniformity has been given by Eq. \eref{eqn:epsT}. Noteworthy is that as the plastic strain becomes increasingly non uniform, the macroscopic stress transitions from linear hardening until it reaches a constant saturation stress. This behaviour is very similar to what has been consistently observed in tensile tests of alloys containing hard second phase particles and can physically be tied to the process of plastic relaxation. In the current work, a purely phenomenological approach is adopted to account for stress saturation by introducing a transition function defined by Eq. \eref{transitionfcn}. To account for shearable particles, the UBS-model was augmented in an ad hoc manner by adding the shearable yield stress contribution to the matrix flow stress.

The UBS-model was finally fitted to experimental data on an $Al-2.8wt\%Mg-0.16wt\%Sc$ alloy heat treated to produce a peak aged sample containing a precipitate volume fraction of $f=0.45\%$ spherical precipitates with mean radius $\bar{r}=1.8$nm and an over aged sample with corresponding values $f=0.37\%$ and $\bar{r}=6.4$. Kocks statistics was used to account for  shearable precipitates. Very good agreement between model and experimental data was achieved with resulting fitting parameters in line with results from previous research. A more rigorous test of the UBS-model against experimental data, using samples subjected to several different ageing times and containing much larger precipitate volume fractions, will be published in an forthcoming paper.

\section*{Acknowledgments}
\noindent This work was performed within the VINN Excellence Center Hero-m, financed by VINNOVA, the Swedish Governmental Agency for Innovation Systems, Swedish industry, and KTH Royal Institute of Technology.

\begin{appendices}
	\renewcommand{\theequation}{\Alph{section}.\arabic{equation}}
	\counterwithin*{equation}{section}

\section{Free energy per unit surface area at the particle/matrix interface}\label{AppendixA}
The free energy per unit volume, $w$, of dislocation networks has been numerically investigated by \cite{BertinCai2019}. It was found that the numerical results, for a wide range of dislocation configurations, could be very well captured by the expression
\begin{equation} \label{appeq:A01}
	w = \dfrac{G b^2}{4\pi} \rho \ln(\dfrac{1}{a_0\sqrt{\rho}}).
\end{equation}
This equation is based on classical analyses (see for example \cite{BaconHull2011}) for the elastic energy of a dislocation line with an effective core radius $a_0=0.1b$ and an outer cut-off radius $R=1/\sqrt{\rho}$, where $b$ denotes the magnitude of Burgers vector and $\rho$ the dislocation density. In \eref{appeq:A01}, $G$ denotes the shear modulus.

An interface between a plastically deforming matrix material and an elastic inclusion is now considered. Plastic deformations will then lead to an accumulation of dislocations in a thin zone adjacent to the interface. The dislocation density in this zone is defined by $\rho$. If the thickness of the zone is assumed to be equal to the average distance between dislocations $1/\sqrt{\rho}$, then a surface density of dislocations $\hat{\rho}$ (total length of dislocations per unit area) can be defined as
\begin{equation} \label{appeq:A02}
	\hat{\rho} = \sqrt{\rho}.
\end{equation}
Furtermore, a free energy per unit surface area $\psi$ can be defined as,
\begin{equation} \label{appeq:A03}
	\psi = \dfrac{1}{\hat{\rho}}w
\end{equation}

Equations \eref{appeq:A01}-\eref{appeq:A03} give an expression for the free energy per unit surface area $\psi$ in terms of the surface density of dislocations $\hat{\rho}$ as
\begin{equation} \label{appeq:A04}
	\psi =  \dfrac{G b^2}{4\pi} \hat{\rho} \ln(\dfrac{10}{b\hat{\rho}}),
\end{equation}
where $a_0=0.1b$ has been utilized according to \cite{BertinCai2019}.

The surface density of dislocations is related to the accumulation of plastic strain at the interface. Here, an order of magnitude argument will be presented for this relation. An interface zone of thickness $1/{\hat{\rho}}$ is considered. The increment of plastic strain $\dot{\varepsilon}_e^p$ in the interface can be estimated as
\begin{equation} \label{appeq:A05}
	\dot{\varepsilon}_e^p \sim \dfrac{b \dot{A}_s}{V}.
\end{equation}
Here, $\dot{A}_s$ denotes the increment in area swept by accumulated dislocations in the volume $V$ of the interface zone. It can be approximately be expressed as
\begin{equation} \label{appeq:A06}
	\dot{A}_s \sim A \dot{\hat{\rho}} \dfrac{1}{\hat{\rho}}
\end{equation}
In \eref{appeq:A06}, $A$ is the surface area of the considered area element of the interface. The first term $A \dot{\hat{\rho}}$ denotes the increment of dislocation length in the considered volume $V = A / \hat{\rho}$ resulting from accumulation of dislocations in the interface zone. The last term in Eq. \eref{appeq:A06} is an estimate of the length swept by incoming dislocations. This length scales as the thickness of the interface $1/\hat{\rho}$. From Eqs. \eref{appeq:A05} and \eref{appeq:A06}, a linear relation between $\dot{\varepsilon}_e^p$ and $\dot{\hat{\rho}}$ can be derived. Hence, the relation between plastic strain at the interface and surface density of dislocations can be written as
\begin{equation} \label{appeq:A07}
	\hat{\rho} = \hat{\rho}_0 + \dfrac{k}{b} \varepsilon_{\rm e}^{\rm p}.
\end{equation}
Here, $\hat{\rho}_0$ denotes the initial surface density of dislocations which can be estimated as $\sqrt{\rho_0}$, where $\rho_0$ is the initial volume density of dislocations in the matrix material. The numerical constant $k$ will be of order one.

Equations \eref{appeq:A04} and \eref{appeq:A07} defines a relation between surface energy and accumulated plastic strain at the interface.

\section{Stress in an ellipsoidal inclusion}\label{AppendixB}
\setcounter{figure}{0}
\renewcommand\thefigure{\thesection.\arabic{figure}}  
\noindent A linear elastic ellipsoidal inhomogeneity (a particle with different elastic properties than the matrix) in an infinite solid (matrix) is considered. A uniform stress $\sigma_{ij}^\infty$ is applied at infinity that corresponds to a uniform strain $\varepsilon_{ij}^\infty$. In addition, a homogeneous plastic strain field $\varepsilon_{ij}^{\rm p}$ exists in the matrix. The plastic strain field will give rise to an eigenstrain $\bar{\varepsilon}_{ij} = -\varepsilon_{ij}^{\rm p}$, in the inhomogeneity. This relation can be understood from Fig. \ref{figB1}, which illustrates the distribution of total strains in the solid resulting from a homogeneous plastic strain field in the matrix. The problem may be split up into two sub-problems where the total strains in the real problem are obtained by superposition of the strains in the sub-problems. Note that only problem (I) will generate stresses, and from problem (I) it is clear that the eigenstrain arising in the inhomogeneity is equal to $-\varepsilon_{ij}^{\rm p}$.

\begin{figure}[!htb]
	\begin{center}
		\includegraphics[width=0.9\textwidth]{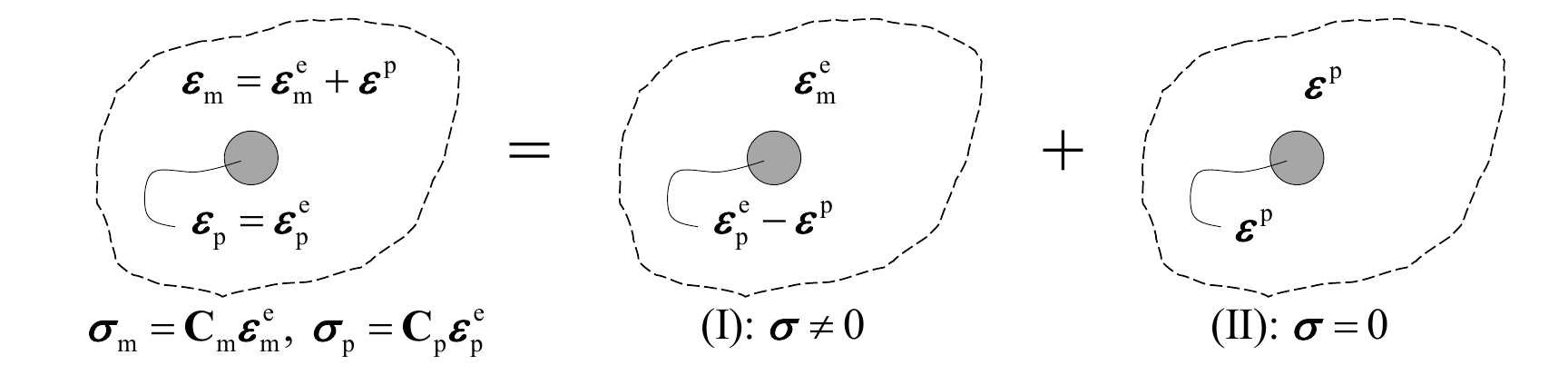}
		\caption{An infinite solid containing an elastic inclusion (particle). A constant plastic strain field exists in the matrix, $\vecsym{\varepsilon}^{\rm p}$, which causes elastic stresses in the particle and in a region close to the particle. In the figure, sub-indices 'm' and 'p' denote quantities belonging to the matrix and inclusion, respectively, whereas super-indices 'e' and 'p' distinguish between elastic and plastic strains. The problem can be divided into two sub-problems by superposition, where stresses arise in (I) and vanishes in (II). (I) can be solved as an Eshelby problem with a known eigenstrain $\bar{\vecsym{\varepsilon}} = - \vecsym{\varepsilon}^{\rm p}$.}
		\label{figB1}
	\end{center}
\end{figure}

The remotely applied stress and the eigenstrain will both result in a homogeneous stress disturbance in the inhomogeneity, and therefore the stress in the particle will be different from the remotely applied stress. The total stress in the inhomogeneity can be calculated by use of the equivalent inclusion method according to \cite{eshelby1957determination}, which will be presented below as outlined in \cite{Mura87}. Without loss of generality for the current application and to simplify matters, it is assumed that the principal directions of strain coincide with the principal orientations of the ellipsoidal inhomogeneity (the principal directions of $\varepsilon_{ij}^\infty$ and $\bar{\varepsilon}_{ij}$ coincide in the current problem). Therefore, the derivation below will then be given on matrix form in a coordinate system aligned with the principal directions. Thus, the total stress in the particle can be written on the compact form
\begin{equation} \label{appeq:B01}
	\vecsym{\sigma} = \mathbf{C}_{\rm p} \left( \vecsym{\varepsilon}^\infty + \vecsym{\varepsilon} - \bar{\vecsym{\varepsilon}} \right) ,
\end{equation}
where $\vecsym{\sigma} = [\sigma_1, \sigma_2, \sigma_3 ]^{T}$ is the principal stress vector in the particle, $\vecsym{\varepsilon}^\infty = [\varepsilon_1^\infty, \varepsilon_2^\infty, \varepsilon_3^\infty ]^{T}$, $\vecsym{\varepsilon} = [\varepsilon_1, \varepsilon_2, \varepsilon_3 ]^{T}$ denotes the strain disturbance due to the presence of an inhomogeneity (particle), $\bar{\vecsym{\varepsilon}} = [\bar{\varepsilon}_1, \bar{\varepsilon}_2, \bar{\varepsilon}_3]^{T}$ denotes the eigenstrain due to plastic straining in the matrix material, and $\mathbf{C}_{\rm p}$ is the elastic stiffness matrix of the particle defined by elastic modulus $E_{\rm p}$ and Poissons's ratio $\nu_{\rm p}$ ($G_{\rm p} = E_{\rm p}/[2(1+\nu_{\rm p})]$).

The stress in the inhomogeneity can be simulated by an equivalent inclusion problem where the whole solid is homogeneous with elastic modulus $\mathbf{C}_{\rm m}$ everywhere, and by adding an equivalent (fictitious) eigenstrain $\vecsym{\varepsilon}^{*}$ to the known eigenstrain $\bar{\vecsym{\varepsilon}}$. Hence, 
\begin{equation} \label{appeq:B02}
	\vecsym{\sigma} = \mathbf{C}_{\rm m} \left( \vecsym{\varepsilon}^\infty + \vecsym{\varepsilon} - \bar{\vecsym{\varepsilon}} - \vecsym{\varepsilon}^{*} \right).
\end{equation}
The stiffness matrix $\mathbf{C}_{\rm m}$ is defined by elastic modulus $E_{\rm m}$ and Poissons's ratio $\nu_{\rm m}$ ($G_{\rm m} = E_{\rm m}/[2(1+\nu_{\rm m})]$), and represents the properties of the matrix material. Since,  $\bar{\vecsym{\varepsilon}}$ is uniform in the particle and $\vecsym{\varepsilon}^{\infty}$ is uniform in the solid, the sum of the eigenstrains are also uniform in the particle and related to the strain disturbance as
\begin{equation} \label{appeq:B03}
	\vecsym{\varepsilon} = \mathbf{S} \left( \vecsym{\bar{\varepsilon}} + \vecsym{\varepsilon}^* \right).
\end{equation}
Here, $\mathbf{S}$ is the Eshelby tensor given on $3\times3$ matrix form.
Equivalencey of stresses and strains in the two problems requires that
\begin{equation} \label{appeq:B04}
	\mathbf{C}_{\rm p} \left( \vecsym{\varepsilon}^\infty + \vecsym{\varepsilon} - \bar{\vecsym{\varepsilon}} \right) = \mathbf{C}_{\rm m} \left( \vecsym{\varepsilon}^\infty + \vecsym{\varepsilon} - \bar{\vecsym{\varepsilon}} - \vecsym{\varepsilon}^{*} \right).
\end{equation}
Substitution of \eref{appeq:B03} into \eref{appeq:B04} and solving for the equivalent eigenstrain gives after some manipulation
\begin{equation} \label{appeq:B05}
	\vecsym{\varepsilon}^{*} = \left(\mathbf{Q}-\mathbf{S}\right)^{-1}\vecsym{\varepsilon}^\infty + \left(\mathbf{Q}-\mathbf{S}\right)^{-1} \left(\mathbf{S}-\mathbf{I}\right)  \bar{\vecsym{\varepsilon}},
\end{equation}
where $\mathbf{Q} = (\mathbf{C}_{\rm m} - \mathbf{C}_{\rm p})^{-1} \mathbf{C}_{\rm m} $ has been introduced, and  $\mathbf{I}$ is a $3 \times 3$ identity matrix. Insertion of \eref{appeq:B05} into \eref{appeq:B03} then gives the strain disturbance 
\begin{equation} \label{appeq:B06}
	\vecsym{\varepsilon} =  \mathbf{S} \left(\mathbf{Q}-\mathbf{S}\right)^{-1}\vecsym{\varepsilon}^\infty + \left(\mathbf{S} + \mathbf{S} \left(\mathbf{Q}-\mathbf{S}\right)^{-1} \left(\mathbf{S}-\mathbf{I}\right) \right)\bar{\vecsym{\varepsilon}}.
\end{equation}
By the replacements $\vecsym{\varepsilon}^\infty = \mathbf{C}_{\rm m}^{-1} \vecsym{\sigma}^\infty$ and $\bar{\vecsym{\varepsilon}} = -\vecsym{\varepsilon}^{\rm p}$, the total stress in the particle can now be calculated by substitution of Eqs. \eref{appeq:B05} and \eref{appeq:B06} into Eq. \eref{appeq:B01}, which gives
\begin{equation} \label{appeq:B07}
	\vecsym{\sigma} = \underbrace{\mathbf{C}_{\rm p}\mathbf{Q}\left(\mathbf{Q}-\mathbf{S}\right)^{-1}\mathbf{C}_{\rm m}^{-1}}_{\vecsym{\Phi}_\infty} \vecsym{\sigma}^\infty + \underbrace{\mathbf{C}_{\rm p}\mathbf{Q}\left(\mathbf{Q}-\mathbf{S}\right)^{-1}\left(\mathbf{S}-\mathbf{I}\right)}_{\vecsym{\Phi}_{\rm p}} \vecsym{\varepsilon}^{\rm p}.
\end{equation}
For a spherical inclusion, with an Eshelby tensor given on matrix form as (cf. \cite{Mura87})
\begin{equation} \label{appeq:B08}
	\mathbf{S} = \begin{bmatrix}
		s_1 & s_2 & s_2\\
		s_2 & s_1 & s_2\\
		s_2 & s_2 & s_1
	\end{bmatrix},
	s_1 = \frac{7-5\nu_{\rm m}}{15(1-\nu_{\rm m})}, s_2 = \frac{5\nu_{\rm m}-1}{15(1-\nu_{\rm m})},
\end{equation}
matrices $\vecsym{\Phi}_\infty$ and $\vecsym{\Phi}_{\rm p}$ take the simple forms
\begin{equation} \label{appeq:B09}
	\vecsym{\Phi}_\infty = \begin{bmatrix}
		g_1 & g_2 & g_2\\
		g_2 & g_1 & g_2\\
		g_2 & g_2 & g_1
	\end{bmatrix},
	\vecsym{\Phi}_{\rm p} = \begin{bmatrix}
		p_1 & p_2 & p_2\\
		p_2 & p_1 & p_2\\
		p_2 & p_2 & p_1
	\end{bmatrix}.
\end{equation}
Now, it is desirable to split the resulting stress in the particle into a hydrostatic part and a deviatoric part. This can be done by means of the matrices,
\begin{equation} \label{appeq:B10}
	\mathbf{H} =  \frac{1}{3}\begin{bmatrix}
		1 & 1 & 1\\
		1 & 1 & 1\\
		1 & 1 & 1
	\end{bmatrix},
	\mathbf{D} =  \frac{1}{3}\begin{bmatrix}
		2 & -1 & -1\\
		-1 & 2 & -1\\
		-1 & -1 & 2
	\end{bmatrix},
\end{equation}
which have the properties $\mathbf{I} = \mathbf{H} + \mathbf{D}$, $\mathbf{H}\mathbf{D} = \mathbf{D}\mathbf{H}= \mathbf{0}$, $\mathbf{H}\mathbf{H} = \mathbf{H}$, and $\mathbf{D}\mathbf{D} = \mathbf{D}$. Recognizing that the mean (hydrostatic) value and the deviatoric part of a vector $\vecsym{v}$, $v_{\rm h}$ and $\vecsym{v}^{\prime}$, respectively, are brought out by the operations $\mathbf{H} \vecsym{v} = v_{\rm h} \vecsym{i}$, with $\vecsym{i} = [1, 1, 1 ]^{T}$, and $\mathbf{D}\vecsym{v} = \vecsym{v}^{\prime}$, the total stress in the particle can be recast as
\begin{equation} \label{appeq:B11}
	\begin{array}{ll}		
	\vecsym{\sigma} & = \left(\mathbf{H}+\mathbf{D}\right)\vecsym{\Phi}_\infty\left(\mathbf{H}+\mathbf{D}\right)\vecsym{\sigma}^\infty + \left(\mathbf{H}+\mathbf{D}\right)\vecsym{\Phi}_{\rm p}\left(\mathbf{H}+\mathbf{D}\right)\vecsym{\varepsilon}^{\rm p} \vspace{2mm} \\ & = \underbrace{\left( \Lambda \sigma_{\rm h}^\infty + 3 K \varepsilon^{\rm p}_{\rm h} \right)}_{\sigma_{\rm h}} \vecsym{i} + \underbrace{\Gamma \vecsym{\sigma}^{\infty \prime} + 2 G \vecsym{\varepsilon}^{\rm p \prime}}_{\vecsym{\sigma}^{\prime}} \:,
	\end{array}
\end{equation}
where
\begin{equation} \label{appeq:B12}
	\begin{array}{l}
		\Lambda  = g_1 + 2g_2 = \dfrac{3e(1-\nu_{\rm m})}{2(1-2\nu_{\rm p}) + e(1-\nu_{\rm m})},
		\vspace{2 mm} \\ \Gamma = g_1 - g_2  = \dfrac{15e(1-\nu_{\rm m}^2)}{(7 - 5\nu_{\rm m})(1 + \nu_{\rm p}) + 2e(4 - 5\nu_{\rm m})(1+\nu_{\rm m})} = \dfrac{15(1-\nu_{\rm m})g}{7 - 5\nu_{\rm m} + 2(4 - 5\nu_{\rm m})g},  
		\vspace{2 mm} \\  3K = p_1 + 2p_2 = \dfrac{-E_{\rm p}}{1-2\nu_{\rm p} + e(1+\nu_{\rm m})/2},
		\vspace{2 mm} \\ 2G  = p_1 - p_2 =  \dfrac{E_{\rm p}} {2(1+\nu_{\rm p}) + 4e(4-5\nu_{\rm m})(1+\nu_{\rm m})} = \dfrac{2G_{\rm m} \cdot  (7-5\nu_{\rm m}) g}{7-5\nu_{\rm m} + 2(4-5\nu_{\rm m})g},
		\vspace{2 mm} \\ e = E_{\rm p} / E_{\rm m}, \quad g = G_{\rm p}/G_{\rm m}.
	\end{array}
\end{equation}
Note that, $\varepsilon^{\rm p}_{\rm h} = 0$ in \eref{appeq:B11}, since $\vecsym{\varepsilon}^{\rm p}$ is deviatoric.

\end{appendices}

\bibliographystyle{model2-names}\biboptions{authoryear}
\bibliography{ReferencesPCA}

\begin{thebibliography}{64}
\expandafter\ifx\csname natexlab\endcsname\relax\def\natexlab#1{#1}\fi
\providecommand{\url}[1]{\texttt{#1}}
\providecommand{\href}[2]{#2}
\providecommand{\path}[1]{#1}
\providecommand{\DOIprefix}{doi:}
\providecommand{\ArXivprefix}{arXiv:}
\providecommand{\URLprefix}{URL: }
\providecommand{\Pubmedprefix}{pmid:}
\providecommand{\doi}[1]{\href{http://dx.doi.org/#1}{\path{#1}}}
\providecommand{\Pubmed}[1]{\href{pmid:#1}{\path{#1}}}
\providecommand{\bibinfo}[2]{#2}
\ifx\xfnm\relax \def\xfnm[#1]{\unskip,\space#1}\fi
\bibitem[{Ardell(1985)}]{ardell1985precipitation}
\bibinfo{author}{Ardell, A.}, \bibinfo{year}{1985}.
\newblock \bibinfo{title}{Precipitation hardening}.
\newblock \bibinfo{journal}{Metallurgical Transactions A} \bibinfo{volume}{16},
  \bibinfo{pages}{2131--2165}.
\bibitem[{Arsenlis et~al.(2007)Arsenlis, Cai, Tang, Rhee, Oppelstrup, Hommes,
  Pierce and Bulatov}]{arsenlis2007enabling}
\bibinfo{author}{Arsenlis, A.}, \bibinfo{author}{Cai, W.},
  \bibinfo{author}{Tang, M.}, \bibinfo{author}{Rhee, M.},
  \bibinfo{author}{Oppelstrup, T.}, \bibinfo{author}{Hommes, G.},
  \bibinfo{author}{Pierce, T.G.}, \bibinfo{author}{Bulatov, V.V.},
  \bibinfo{year}{2007}.
\newblock \bibinfo{title}{Enabling strain hardening simulations with
  dislocation dynamics}.
\newblock \bibinfo{journal}{Modelling and Simulation in Materials Science and
  Engineering} \bibinfo{volume}{15}, \bibinfo{pages}{553}.
\bibitem[{Asgharzadeh and Faleskog(2021a)}]{Asgharzadeh2021b}
\bibinfo{author}{Asgharzadeh, M.}, \bibinfo{author}{Faleskog, J.},
  \bibinfo{year}{2021}a.
\newblock \bibinfo{title}{3\uppercase{D} analysis of a strain gradient
  plasticity material reinforced by elastic particles}.
\newblock \bibinfo{note}{Preprint arXiv:2106.092290 [cond-mat.mtrl-sci]}.
\bibitem[{Asgharzadeh and Faleskog(2021b)}]{Asgharzadeh2021a}
\bibinfo{author}{Asgharzadeh, M.}, \bibinfo{author}{Faleskog, J.},
  \bibinfo{year}{2021}b.
\newblock \bibinfo{title}{A strengthening model of particle-matrix interaction
  based on an axisymmetric strain gradient plasticity analysis}.
\newblock \bibinfo{note}{Preprint arXiv:2106.08432 [cond-mat.mtrl-sci]}.
\bibitem[{Bao et~al.(1991)Bao, Hutchinson and McMeeking}]{bao1991particle}
\bibinfo{author}{Bao, G.}, \bibinfo{author}{Hutchinson, J.},
  \bibinfo{author}{McMeeking, R.}, \bibinfo{year}{1991}.
\newblock \bibinfo{title}{Particle reinforcement of ductile matrices against
  plastic flow and creep}.
\newblock \bibinfo{journal}{Acta metallurgica et materialia}
  \bibinfo{volume}{39}, \bibinfo{pages}{1871--1882}.
\bibitem[{Bertin and Cai(2018)}]{BertinCai2019}
\bibinfo{author}{Bertin, N.}, \bibinfo{author}{Cai, W.}, \bibinfo{year}{2018}.
\newblock \bibinfo{title}{Energy of periodic discrete dislocation networks}.
\newblock \bibinfo{journal}{Journal of the Mechanics and Physics of Solids}
  \bibinfo{volume}{121}, \bibinfo{pages}{133--146}.
\bibitem[{Brown and Stobbs(1971)}]{brown1971work}
\bibinfo{author}{Brown, L.}, \bibinfo{author}{Stobbs, W.},
  \bibinfo{year}{1971}.
\newblock \bibinfo{title}{The work-hardening of copper-silica}.
\newblock \bibinfo{journal}{Philosophical Magazine} \bibinfo{volume}{23},
  \bibinfo{pages}{1201--1233}.
\bibitem[{Chen and Wang(2002)}]{chen2002size}
\bibinfo{author}{Chen, S.}, \bibinfo{author}{Wang, T.}, \bibinfo{year}{2002}.
\newblock \bibinfo{title}{Size effects in the particle-reinforced metal-matrix
  composites}.
\newblock \bibinfo{journal}{Acta Mechanica} \bibinfo{volume}{157},
  \bibinfo{pages}{113--127}.
\bibitem[{Cheng et~al.(2003)Cheng, Poole, Embury and
  Lloyd}]{cheng2003influence}
\bibinfo{author}{Cheng, L.}, \bibinfo{author}{Poole, W.},
  \bibinfo{author}{Embury, J.}, \bibinfo{author}{Lloyd, D.},
  \bibinfo{year}{2003}.
\newblock \bibinfo{title}{The influence of precipitation on the work-hardening
  behavior of the aluminum alloys aa6111 and aa7030}.
\newblock \bibinfo{journal}{Metallurgical and materials transactions A}
  \bibinfo{volume}{34}, \bibinfo{pages}{2473--2481}.
\bibitem[{Christman et~al.(1989)Christman, Needleman and
  Suresh}]{christman1989experimental}
\bibinfo{author}{Christman, T.}, \bibinfo{author}{Needleman, A.},
  \bibinfo{author}{Suresh, S.}, \bibinfo{year}{1989}.
\newblock \bibinfo{title}{An experimental and numerical study of deformation in
  metal-ceramic composites}.
\newblock \bibinfo{journal}{Acta Metallurgica} \bibinfo{volume}{37},
  \bibinfo{pages}{3029--3050}.
\bibitem[{Dahlberg and Faleskog(2013)}]{dahlberg2013improved}
\bibinfo{author}{Dahlberg, C.F.}, \bibinfo{author}{Faleskog, J.},
  \bibinfo{year}{2013}.
\newblock \bibinfo{title}{An improved strain gradient plasticity formulation
  with energetic interfaces: theory and a fully implicit finite element
  formulation}.
\newblock \bibinfo{journal}{Computational Mechanics} \bibinfo{volume}{51},
  \bibinfo{pages}{641--659}.
\bibitem[{Dahlberg et~al.(2013)Dahlberg, Faleskog, Niordson and
  Legarth}]{dahlberg2013deformation}
\bibinfo{author}{Dahlberg, C.F.}, \bibinfo{author}{Faleskog, J.},
  \bibinfo{author}{Niordson, C.F.}, \bibinfo{author}{Legarth, B.N.},
  \bibinfo{year}{2013}.
\newblock \bibinfo{title}{A deformation mechanism map for polycrystals modeled
  using strain gradient plasticity and interfaces that slide and separate}.
\newblock \bibinfo{journal}{International Journal of Plasticity}
  \bibinfo{volume}{43}, \bibinfo{pages}{177--195}.
\bibitem[{Dai et~al.(1999)Dai, Ling and Bai}]{dai1999strain}
\bibinfo{author}{Dai, L.}, \bibinfo{author}{Ling, Z.}, \bibinfo{author}{Bai,
  Y.}, \bibinfo{year}{1999}.
\newblock \bibinfo{title}{A strain gradient-strengthening law for particle
  reinforced metal matrix composites}.
\newblock \bibinfo{journal}{Scripta Materialia} \bibinfo{volume}{41},
  \bibinfo{pages}{245--251}.
\bibitem[{De~Vaucorbeil et~al.(2013)De~Vaucorbeil, Poole and
  Sinclair}]{de2013superposition}
\bibinfo{author}{De~Vaucorbeil, A.}, \bibinfo{author}{Poole, W.},
  \bibinfo{author}{Sinclair, C.}, \bibinfo{year}{2013}.
\newblock \bibinfo{title}{The superposition of strengthening contributions in
  engineering alloys}.
\newblock \bibinfo{journal}{Materials Science and Engineering: A}
  \bibinfo{volume}{582}, \bibinfo{pages}{147--154}.
\bibitem[{Deschamps and Brechet(1998)}]{deschamps1998influence}
\bibinfo{author}{Deschamps, A.}, \bibinfo{author}{Brechet, Y.},
  \bibinfo{year}{1998}.
\newblock \bibinfo{title}{Influence of predeformation and ageing of an
  al--zn--mg alloy—ii. modeling of precipitation kinetics and yield stress}.
\newblock \bibinfo{journal}{Acta Materialia} \bibinfo{volume}{47},
  \bibinfo{pages}{293--305}.
\bibitem[{Eshelby(1957)}]{eshelby1957determination}
\bibinfo{author}{Eshelby, J.D.}, \bibinfo{year}{1957}.
\newblock \bibinfo{title}{The determination of the elastic field of an
  ellipsoidal inclusion, and related problems}.
\newblock \bibinfo{journal}{Proceedings of the royal society of London. Series
  A. Mathematical and physical sciences} \bibinfo{volume}{241},
  \bibinfo{pages}{376--396}.
\bibitem[{Estrin and Mecking(1984)}]{estrin1984unified}
\bibinfo{author}{Estrin, Y.}, \bibinfo{author}{Mecking, H.},
  \bibinfo{year}{1984}.
\newblock \bibinfo{title}{A unified phenomenological description of work
  hardening and creep based on one-parameter models}.
\newblock \bibinfo{journal}{Acta metallurgica} \bibinfo{volume}{32},
  \bibinfo{pages}{57--70}.
\bibitem[{Faleskog and Gudmundson(2021)}]{faleskog2021analytical}
\bibinfo{author}{Faleskog, J.}, \bibinfo{author}{Gudmundson, P.},
  \bibinfo{year}{2021}.
\newblock \bibinfo{title}{Analytical predictions of yield stress of a strain
  gradient plasticity material reinforced by small elastic particles}.
\newblock \bibinfo{journal}{Journal of the Mechanics and Physics of Solids}
  \bibinfo{volume}{157}, \bibinfo{pages}{1--26}.
\newblock \bibinfo{note}{104623}.
\bibitem[{Fazeli et~al.(2008a)Fazeli, Poole and Sinclair}]{fazeli2008modeling}
\bibinfo{author}{Fazeli, F.}, \bibinfo{author}{Poole, W.},
  \bibinfo{author}{Sinclair, C.}, \bibinfo{year}{2008}a.
\newblock \bibinfo{title}{Modeling the effect of al3sc precipitates on the
  yield stress and work hardening of an al--mg--sc alloy}.
\newblock \bibinfo{journal}{Acta Materialia} \bibinfo{volume}{56},
  \bibinfo{pages}{1909--1918}.
\bibitem[{Fazeli et~al.(2008b)Fazeli, Sinclair and Bastow}]{fazeli2008role}
\bibinfo{author}{Fazeli, F.}, \bibinfo{author}{Sinclair, C.},
  \bibinfo{author}{Bastow, T.}, \bibinfo{year}{2008}b.
\newblock \bibinfo{title}{The role of excess vacancies on precipitation
  kinetics in an al-mg-sc alloy}.
\newblock \bibinfo{journal}{Metallurgical and Materials Transactions A}
  \bibinfo{volume}{39}, \bibinfo{pages}{2297}.
\bibitem[{Fleck and Hutchinson(1993)}]{fleck1993phenomenological}
\bibinfo{author}{Fleck, N.}, \bibinfo{author}{Hutchinson, J.},
  \bibinfo{year}{1993}.
\newblock \bibinfo{title}{A phenomenological theory for strain gradient effects
  in plasticity}.
\newblock \bibinfo{journal}{Journal of the Mechanics and Physics of Solids}
  \bibinfo{volume}{41}, \bibinfo{pages}{1825--1857}.
\bibitem[{Fleck and Hutchinson(2001)}]{fleck2001reformulation}
\bibinfo{author}{Fleck, N.}, \bibinfo{author}{Hutchinson, J.},
  \bibinfo{year}{2001}.
\newblock \bibinfo{title}{A reformulation of strain gradient plasticity}.
\newblock \bibinfo{journal}{Journal of the Mechanics and Physics of Solids}
  \bibinfo{volume}{49}, \bibinfo{pages}{2245--2271}.
\bibitem[{Fleck et~al.(1994)Fleck, Muller, Ashby and
  Hutchinson}]{fleck1994strain}
\bibinfo{author}{Fleck, N.}, \bibinfo{author}{Muller, G.},
  \bibinfo{author}{Ashby, M.F.}, \bibinfo{author}{Hutchinson, J.W.},
  \bibinfo{year}{1994}.
\newblock \bibinfo{title}{Strain gradient plasticity: theory and experiment}.
\newblock \bibinfo{journal}{Acta Metallurgica et materialia}
  \bibinfo{volume}{42}, \bibinfo{pages}{475--487}.
\bibitem[{Fleck and Willis(2009a)}]{fleck2009mathematical2}
\bibinfo{author}{Fleck, N.}, \bibinfo{author}{Willis, J.},
  \bibinfo{year}{2009}a.
\newblock \bibinfo{title}{A mathematical basis for strain-gradient plasticity
  theory. part ii: Tensorial plastic multiplier}.
\newblock \bibinfo{journal}{Journal of the Mechanics and Physics of Solids}
  \bibinfo{volume}{57}, \bibinfo{pages}{1045--1057}.
\bibitem[{Fleck and Willis(2009b)}]{fleck2009mathematical}
\bibinfo{author}{Fleck, N.}, \bibinfo{author}{Willis, J.},
  \bibinfo{year}{2009}b.
\newblock \bibinfo{title}{A mathematical basis for strain-gradient plasticity
  theory—part i: Scalar plastic multiplier}.
\newblock \bibinfo{journal}{Journal of the Mechanics and Physics of Solids}
  \bibinfo{volume}{57}, \bibinfo{pages}{161--177}.
\bibitem[{Fredriksson and Gudmundson(2005)}]{fredriksson2005size}
\bibinfo{author}{Fredriksson, P.}, \bibinfo{author}{Gudmundson, P.},
  \bibinfo{year}{2005}.
\newblock \bibinfo{title}{Size-dependent yield strength of thin films}.
\newblock \bibinfo{journal}{International journal of plasticity}
  \bibinfo{volume}{21}, \bibinfo{pages}{1834--1854}.
\bibitem[{Fredriksson and Gudmundson(2007)}]{fredriksson2007modelling}
\bibinfo{author}{Fredriksson, P.}, \bibinfo{author}{Gudmundson, P.},
  \bibinfo{year}{2007}.
\newblock \bibinfo{title}{Modelling of the interface between a thin film and a
  substrate within a strain gradient plasticity framework}.
\newblock \bibinfo{journal}{Journal of the Mechanics and Physics of Solids}
  \bibinfo{volume}{55}, \bibinfo{pages}{939--955}.
\bibitem[{Fribourg et~al.(2011)Fribourg, Br{\'e}chet, Deschamps and
  Simar}]{fribourg2011microstructure}
\bibinfo{author}{Fribourg, G.}, \bibinfo{author}{Br{\'e}chet, Y.},
  \bibinfo{author}{Deschamps, A.}, \bibinfo{author}{Simar, A.},
  \bibinfo{year}{2011}.
\newblock \bibinfo{title}{Microstructure-based modelling of isotropic and
  kinematic strain hardening in a precipitation-hardened aluminium alloy}.
\newblock \bibinfo{journal}{Acta Materialia} \bibinfo{volume}{59},
  \bibinfo{pages}{3621--3635}.
\bibitem[{Friedel(1964)}]{friedel1964}
\bibinfo{author}{Friedel, J.}, \bibinfo{year}{1964}.
\newblock \bibinfo{title}{Dislocations}.
\newblock \bibinfo{publisher}{Pergamon Press, Oxford}.
\bibitem[{Gudmundson(2004)}]{gudmundson2004unified}
\bibinfo{author}{Gudmundson, P.}, \bibinfo{year}{2004}.
\newblock \bibinfo{title}{A unified treatment of strain gradient plasticity}.
\newblock \bibinfo{journal}{Journal of the Mechanics and Physics of Solids}
  \bibinfo{volume}{52}, \bibinfo{pages}{1379--1406}.
\bibitem[{Gurtin(2004)}]{gurtin2004gradient}
\bibinfo{author}{Gurtin, M.E.}, \bibinfo{year}{2004}.
\newblock \bibinfo{title}{A gradient theory of small-deformation isotropic
  plasticity that accounts for the burgers vector and for dissipation due to
  plastic spin}.
\newblock \bibinfo{journal}{Journal of the Mechanics and Physics of Solids}
  \bibinfo{volume}{52}, \bibinfo{pages}{2545--2568}.
\bibitem[{Gurtin and Anand(2005)}]{gurtin2005theory}
\bibinfo{author}{Gurtin, M.E.}, \bibinfo{author}{Anand, L.},
  \bibinfo{year}{2005}.
\newblock \bibinfo{title}{A theory of strain-gradient plasticity for isotropic,
  plastically irrotational materials. part i: Small deformations}.
\newblock \bibinfo{journal}{Journal of the Mechanics and Physics of Solids}
  \bibinfo{volume}{53}, \bibinfo{pages}{1624--1649}.
\bibitem[{Hull and Bacon(2011)}]{BaconHull2011}
\bibinfo{author}{Hull, D.}, \bibinfo{author}{Bacon, D.}, \bibinfo{year}{2011}.
\newblock \bibinfo{title}{Introduction to Dislocations}.
\newblock \bibinfo{edition}{5} ed., \bibinfo{publisher}{Butterworth-Heinemann,
  Oxford}.
\bibitem[{Hutchinson and Fleck(1997)}]{hutchinson1997strain}
\bibinfo{author}{Hutchinson, J.}, \bibinfo{author}{Fleck, N.},
  \bibinfo{year}{1997}.
\newblock \bibinfo{title}{Strain gradient plasticity}.
\newblock \bibinfo{journal}{Advances in applied mechanics}
  \bibinfo{volume}{33}, \bibinfo{pages}{295--361}.
\bibitem[{Jobba et~al.(2015)Jobba, Mishra and Niewczas}]{jobba2015flow}
\bibinfo{author}{Jobba, M.}, \bibinfo{author}{Mishra, R.},
  \bibinfo{author}{Niewczas, M.}, \bibinfo{year}{2015}.
\newblock \bibinfo{title}{Flow stress and work-hardening behaviour of al--mg
  binary alloys}.
\newblock \bibinfo{journal}{International Journal of Plasticity}
  \bibinfo{volume}{65}, \bibinfo{pages}{43--60}.
\bibitem[{Kocks et~al.(1975)Kocks, Argon and Ashby}]{kocks1975models}
\bibinfo{author}{Kocks, U.}, \bibinfo{author}{Argon, A.},
  \bibinfo{author}{Ashby, M.}, \bibinfo{year}{1975}.
\newblock \bibinfo{title}{Models for macroscopic slip}.
\newblock \bibinfo{journal}{Prog. Mater. Sci} \bibinfo{volume}{19},
  \bibinfo{pages}{171--229}.
\bibitem[{Kocks and Mecking(2003)}]{kocks2003physics}
\bibinfo{author}{Kocks, U.}, \bibinfo{author}{Mecking, H.},
  \bibinfo{year}{2003}.
\newblock \bibinfo{title}{Physics and phenomenology of strain hardening: the
  fcc case}.
\newblock \bibinfo{journal}{Progress in materials science}
  \bibinfo{volume}{48}, \bibinfo{pages}{171--273}.
\bibitem[{Koslowski et~al.(2002)Koslowski, Cuitino and
  Ortiz}]{koslowski2002phase}
\bibinfo{author}{Koslowski, M.}, \bibinfo{author}{Cuitino, A.M.},
  \bibinfo{author}{Ortiz, M.}, \bibinfo{year}{2002}.
\newblock \bibinfo{title}{A phase-field theory of dislocation dynamics, strain
  hardening and hysteresis in ductile single crystals}.
\newblock \bibinfo{journal}{Journal of the Mechanics and Physics of Solids}
  \bibinfo{volume}{50}, \bibinfo{pages}{2597--2635}.
\bibitem[{Kouzeli and Mortensen(2002)}]{kouzeli2002size}
\bibinfo{author}{Kouzeli, M.}, \bibinfo{author}{Mortensen, A.},
  \bibinfo{year}{2002}.
\newblock \bibinfo{title}{Size dependent strengthening in particle reinforced
  aluminium}.
\newblock \bibinfo{journal}{Acta Materialia} \bibinfo{volume}{50},
  \bibinfo{pages}{39--51}.
\bibitem[{Labusch(1970)}]{labusch1970statistical}
\bibinfo{author}{Labusch, R.}, \bibinfo{year}{1970}.
\newblock \bibinfo{title}{A statistical theory of solid solution hardening}.
\newblock \bibinfo{journal}{physica status solidi (b)} \bibinfo{volume}{41},
  \bibinfo{pages}{659--669}.
\bibitem[{Liu et~al.(2003)Liu, Dai and Yang}]{liu2003strain}
\bibinfo{author}{Liu, L.}, \bibinfo{author}{Dai, L.}, \bibinfo{author}{Yang,
  G.}, \bibinfo{year}{2003}.
\newblock \bibinfo{title}{Strain gradient effects on deformation strengthening
  behavior of particle reinforced metal matrix composites}.
\newblock \bibinfo{journal}{Materials Science and Engineering: A}
  \bibinfo{volume}{345}, \bibinfo{pages}{190--196}.
\bibitem[{Lloyd(1994)}]{lloyd1994particle}
\bibinfo{author}{Lloyd, D.}, \bibinfo{year}{1994}.
\newblock \bibinfo{title}{Particle reinforced aluminium and magnesium matrix
  composites}.
\newblock \bibinfo{journal}{International Materials Reviews}
  \bibinfo{volume}{39}, \bibinfo{pages}{1--23}.
\bibitem[{Mecking and Kocks(1981)}]{mecking1981kinetics}
\bibinfo{author}{Mecking, H.}, \bibinfo{author}{Kocks, U.},
  \bibinfo{year}{1981}.
\newblock \bibinfo{title}{Kinetics of flow and strain-hardening}.
\newblock \bibinfo{journal}{Acta metallurgica} \bibinfo{volume}{29},
  \bibinfo{pages}{1865--1875}.
\bibitem[{Mott and Nabarro(1940)}]{mott1940attempt}
\bibinfo{author}{Mott, N.}, \bibinfo{author}{Nabarro, F.N.},
  \bibinfo{year}{1940}.
\newblock \bibinfo{title}{An attempt to estimate the degree of precipitation
  hardening, with a simple model}.
\newblock \bibinfo{journal}{Proceedings of the Physical Society}
  \bibinfo{volume}{52}, \bibinfo{pages}{86}.
\bibitem[{Mura(1987)}]{Mura87}
\bibinfo{author}{Mura, T.}, \bibinfo{year}{1987}.
\newblock \bibinfo{title}{Micromechanics of Defects in Solids}.
\newblock \bibinfo{edition}{2} ed., \bibinfo{publisher}{Kluwer Academic
  Publishers}.
\bibitem[{Myhr et~al.(2010)Myhr, Grong and Pedersen}]{myhr2010combined}
\bibinfo{author}{Myhr, O.R.}, \bibinfo{author}{Grong, {\O}.},
  \bibinfo{author}{Pedersen, K.O.}, \bibinfo{year}{2010}.
\newblock \bibinfo{title}{A combined precipitation, yield strength, and work
  hardening model for al-mg-si alloys}.
\newblock \bibinfo{journal}{Metallurgical and Materials Transactions A}
  \bibinfo{volume}{41}, \bibinfo{pages}{2276--2289}.
\bibitem[{Neuh{\"a}user and Schwink(2006)}]{neuhauser2006solid}
\bibinfo{author}{Neuh{\"a}user, H.}, \bibinfo{author}{Schwink, C.},
  \bibinfo{year}{2006}.
\newblock \bibinfo{title}{Solid solution strengthening}.
\newblock \bibinfo{journal}{Materials science and technology} .
\bibitem[{Orowan(1948)}]{orowan1948discussion}
\bibinfo{author}{Orowan, E.}, \bibinfo{year}{1948}.
\newblock \bibinfo{title}{Discussion on internal stresses}, in:
  \bibinfo{booktitle}{Symposium on internal stresses in metals and alloys},
  \bibinfo{organization}{Institute of Metals London}. pp.
  \bibinfo{pages}{451--453}.
\bibitem[{Polizzotto(2010)}]{polizzotto2010strain}
\bibinfo{author}{Polizzotto, C.}, \bibinfo{year}{2010}.
\newblock \bibinfo{title}{Strain gradient plasticity, strengthening effects and
  plastic limit analysis}.
\newblock \bibinfo{journal}{International journal of solids and structures}
  \bibinfo{volume}{47}, \bibinfo{pages}{100--112}.
\bibitem[{Qu et~al.(2005)Qu, Siegmund, Huang, Wu, Zhang and
  Hwang}]{qu2005study}
\bibinfo{author}{Qu, S.}, \bibinfo{author}{Siegmund, T.},
  \bibinfo{author}{Huang, Y.}, \bibinfo{author}{Wu, P.},
  \bibinfo{author}{Zhang, F.}, \bibinfo{author}{Hwang, K.},
  \bibinfo{year}{2005}.
\newblock \bibinfo{title}{A study of particle size effect and interface
  fracture in aluminum alloy composite via an extended conventional theory of
  mechanism-based strain-gradient plasticity}.
\newblock \bibinfo{journal}{Composites Science and Technology}
  \bibinfo{volume}{65}, \bibinfo{pages}{1244--1253}.
\bibitem[{Queyreau et~al.(2010)Queyreau, Monnet and
  Devincre}]{queyreau2010orowan}
\bibinfo{author}{Queyreau, S.}, \bibinfo{author}{Monnet, G.},
  \bibinfo{author}{Devincre, B.}, \bibinfo{year}{2010}.
\newblock \bibinfo{title}{Orowan strengthening and forest hardening
  superposition examined by dislocation dynamics simulations}.
\newblock \bibinfo{journal}{Acta Materialia} \bibinfo{volume}{58},
  \bibinfo{pages}{5586--5595}.
\bibitem[{Reddy and Sysala(2020)}]{reddy2020bounds}
\bibinfo{author}{Reddy, B.}, \bibinfo{author}{Sysala, S.},
  \bibinfo{year}{2020}.
\newblock \bibinfo{title}{Bounds on the elastic threshold for problems of
  dissipative strain-gradient plasticity}.
\newblock \bibinfo{journal}{Journal of the Mechanics and Physics of Solids}
  \bibinfo{volume}{143}, \bibinfo{pages}{104089}.
\bibitem[{Reppich(1993)}]{reppich1993materials}
\bibinfo{author}{Reppich, B.}, \bibinfo{year}{1993}.
\newblock \bibinfo{title}{Materials science and technology}.
\newblock \bibinfo{journal}{Particle strengthening} \bibinfo{volume}{6},
  \bibinfo{pages}{311--357}.
\bibitem[{Russell and Brown(1972)}]{russell1972dispersion}
\bibinfo{author}{Russell, K.C.}, \bibinfo{author}{Brown, L.},
  \bibinfo{year}{1972}.
\newblock \bibinfo{title}{A dispersion strengthening model based on differing
  elastic moduli applied to the iron-copper system}.
\newblock \bibinfo{journal}{Acta Metallurgica} \bibinfo{volume}{20},
  \bibinfo{pages}{969--974}.
\bibitem[{Santos-G{\"u}emes et~al.(2020)Santos-G{\"u}emes, Bell{\'o}n,
  Esteban-Manzanares, Segurado, Capolungo and LLorca}]{santos2020multiscale}
\bibinfo{author}{Santos-G{\"u}emes, R.}, \bibinfo{author}{Bell{\'o}n, B.},
  \bibinfo{author}{Esteban-Manzanares, G.}, \bibinfo{author}{Segurado, J.},
  \bibinfo{author}{Capolungo, L.}, \bibinfo{author}{LLorca, J.},
  \bibinfo{year}{2020}.
\newblock \bibinfo{title}{Multiscale modelling of precipitation hardening in
  al--cu alloys: Dislocation dynamics simulations and experimental validation}.
\newblock \bibinfo{journal}{Acta Materialia} \bibinfo{volume}{188},
  \bibinfo{pages}{475--485}.
\bibitem[{Tanaka and Mori(1970)}]{tanaka1970hardening}
\bibinfo{author}{Tanaka, K.}, \bibinfo{author}{Mori, T.}, \bibinfo{year}{1970}.
\newblock \bibinfo{title}{The hardening of crystals by non-deforming particles
  and fibres}.
\newblock \bibinfo{journal}{Acta Metallurgica} \bibinfo{volume}{18},
  \bibinfo{pages}{931--941}.
\bibitem[{Vattr{\'e} et~al.(2009)Vattr{\'e}, Devincre and
  Roos}]{vattre2009dislocation}
\bibinfo{author}{Vattr{\'e}, A.}, \bibinfo{author}{Devincre, B.},
  \bibinfo{author}{Roos, A.}, \bibinfo{year}{2009}.
\newblock \bibinfo{title}{Dislocation dynamics simulations of precipitation
  hardening in ni-based superalloys with high $\gamma\prime$ volume fraction}.
\newblock \bibinfo{journal}{Intermetallics} \bibinfo{volume}{17},
  \bibinfo{pages}{988--994}.
\bibitem[{Voyiadjis and Deliktas(2009)}]{voyiadjis2009formulation}
\bibinfo{author}{Voyiadjis, G.Z.}, \bibinfo{author}{Deliktas, B.},
  \bibinfo{year}{2009}.
\newblock \bibinfo{title}{Formulation of strain gradient plasticity with
  interface energy in a consistent thermodynamic framework}.
\newblock \bibinfo{journal}{International Journal of Plasticity}
  \bibinfo{volume}{25}, \bibinfo{pages}{1997--2024}.
\bibitem[{Voyiadjis and Song(2019)}]{voyiadjis2019strain}
\bibinfo{author}{Voyiadjis, G.Z.}, \bibinfo{author}{Song, Y.},
  \bibinfo{year}{2019}.
\newblock \bibinfo{title}{Strain gradient continuum plasticity theories:
  theoretical, numerical and experimental investigations}.
\newblock \bibinfo{journal}{International Journal of Plasticity}
  \bibinfo{volume}{121}, \bibinfo{pages}{21--75}.
\bibitem[{Xue et~al.(2002)Xue, Huang and Li}]{xue2002particle}
\bibinfo{author}{Xue, Z.}, \bibinfo{author}{Huang, Y.}, \bibinfo{author}{Li,
  M.}, \bibinfo{year}{2002}.
\newblock \bibinfo{title}{Particle size effect in metallic materials: a study
  by the theory of mechanism-based strain gradient plasticity}.
\newblock \bibinfo{journal}{Acta Materialia} \bibinfo{volume}{50},
  \bibinfo{pages}{149--160}.
\bibitem[{Yueguang(2001)}]{yueguang2001particulate}
\bibinfo{author}{Yueguang, W.}, \bibinfo{year}{2001}.
\newblock \bibinfo{title}{Particulate size effects in the particle-reinforced
  metal-matrix composites}.
\newblock \bibinfo{journal}{Acta Mechanica Sinica} \bibinfo{volume}{17},
  \bibinfo{pages}{45}.
\bibitem[{Zbib et~al.(2000)Zbib, De~La~Rubia, Rhee and Hirth}]{zbib20003d}
\bibinfo{author}{Zbib, H.M.}, \bibinfo{author}{De~La~Rubia, T.D.},
  \bibinfo{author}{Rhee, M.}, \bibinfo{author}{Hirth, J.P.},
  \bibinfo{year}{2000}.
\newblock \bibinfo{title}{3d dislocation dynamics: stress--strain behavior and
  hardening mechanisms in fcc and bcc metals}.
\newblock \bibinfo{journal}{Journal of Nuclear Materials}
  \bibinfo{volume}{276}, \bibinfo{pages}{154--165}.
\bibitem[{Zhang et~al.(2007)Zhang, Huang, Hwang, Qu and Liu}]{zhang2007three}
\bibinfo{author}{Zhang, F.}, \bibinfo{author}{Huang, Y.},
  \bibinfo{author}{Hwang, K.}, \bibinfo{author}{Qu, S.}, \bibinfo{author}{Liu,
  C.}, \bibinfo{year}{2007}.
\newblock \bibinfo{title}{A three-dimensional strain gradient plasticity
  analysis of particle size effect in composite materials}.
\newblock \bibinfo{journal}{Materials and Manufacturing Processes}
  \bibinfo{volume}{22}, \bibinfo{pages}{140--148}.
\bibitem[{Zhu et~al.(1997)Zhu, Zbib and Aifantis}]{zhu1997strain}
\bibinfo{author}{Zhu, H.T.}, \bibinfo{author}{Zbib, H.},
  \bibinfo{author}{Aifantis, E.}, \bibinfo{year}{1997}.
\newblock \bibinfo{title}{Strain gradients and continuum modeling of size
  effect in metal matrix composites}.
\newblock \bibinfo{journal}{Acta Mechanica} \bibinfo{volume}{121},
  \bibinfo{pages}{165--176}.

\end{thebibliography}
\end{document}